\documentclass[12pt]{article}
\usepackage[utf8]{inputenc}
\usepackage[T1]{fontenc}
\usepackage{enumitem}
\usepackage[english]{babel}
\usepackage{lmodern, csquotes, eurosym}
\usepackage[colorlinks=true, citecolor=blue, linkcolor=blue, urlcolor=blue, pdfborder={0 0 0}]{hyperref}
\usepackage[round]{natbib}
\usepackage{tikz}
\usepackage{caption}
\usepackage{subcaption}
\usepackage{mathtools}
\usepackage{comment}
\usepackage{booktabs}  
\usepackage{siunitx}
\usepackage{pgfplots}
\usepgfplotslibrary{fillbetween}
\pgfplotsset{compat=1.18}

\usepackage{amssymb}
\usepackage{amsfonts}
\usepackage{geometry}
\usepackage{graphicx}

\newtheorem{proposition}{Proposition}

\title{Barriers to AI Adoption: Image Concerns at Work\thanks{I am especially grateful for the guidance and support of Yuval Salant, Benjamin Golub, Alex Imas, Daniel Martin, Alvaro Sandroni, and Jörg Spenkuch. I also thank Effi Benmelech, Meghan Busse, Joshua Dean, John Horton, Alexander Jakobsen, Rafael Jiménez-Durán, Diego Jiménez-Hernández, Andreas Kraft, Annie Liang, Ryan Oprea, Devin Pope, Mike Powell, Luis Rayo, Avner Strulov-Shlain, and audiences at Advances with Field Experiments Conference, Booth Behavioral Lab, and Kellogg Theory and Strategy Seminars for their valuable feedback. Refine.ink provided helpful comments during the writing stage. The experiment was pre-registered on aspredicted.org (\#239005 and \#242197). The IRB ID at Northwestern University is STU00223689.}}
\author{David Almog\thanks{Kellogg School of Management, Northwestern University,  \href{mailto:david.almog@kellogg.northwestern.edu}{david.almog@kellogg.northwestern.edu}.} \\ \textit{Job Market Paper} }
\date{This version: \today\\ [1ex]
  \href{https://drive.google.com/file/d/1a2AzcwAbFq-DbwL8qZzPdXNJpF4DaJ1e/view?usp=sharing}{\textcolor{blue}{Click here for latest version}}}
\medskip

\begin{document}

\maketitle

\begin{abstract}
Concerns about how workers are perceived can deter effective collaboration with artificial intelligence (AI). In a field experiment on a large online labor market, I hired 450 U.S.-based remote workers to complete an image-categorization job assisted by AI recommendations. Workers were incentivized by the prospect of a contract extension based on an HR evaluator’s feedback. I find that workers adopt AI recommendations at lower rates when their reliance on AI is visible to the evaluator, resulting in a measurable decline in task performance. The effects are present despite a conservative design in which workers know that the evaluator is explicitly instructed to assess expected accuracy on the same AI-assisted task. This reduction in AI reliance persists even when the evaluator is reassured about workers’ strong performance history on the platform, underscoring how difficult these concerns are to alleviate. Leveraging the platform’s public feedback feature, I introduce a novel incentive-compatible elicitation method showing that workers fear heavy reliance on AI signals a lack of confidence in their own judgment, a trait they view as essential when collaborating with AI.
\end{abstract}


\newpage 

\section{Introduction}

The rapid deployment of artificial intelligence (AI) in organizations promises large productivity gains, but because AI use is often visible or trackable, it also introduces a new source of image concerns. Concerns about how individuals are perceived by others influence many everyday choices.\footnote{See \cite{Bursztyn2017} for a review of the recent literature on social image concerns.} In the workplace, these concerns are particularly salient, as workers care not only about performing well but also about being viewed favorably by colleagues, managers, and clients \citep{Ellingsen2008,Bandiera2005}. Despite the scale of AI diffusion, we still know little about how image concerns affect AI adoption. In principle, these concerns could push behavior in either direction: heavy reliance on AI might signal adaptability and technological savvy, but it could also be interpreted as a lack of effort, poor judgment, or limited competence.

While AI systems continue to improve their predictive capabilities \citep{Agrawal2019}, humans remain central in many decision-making domains. This persistence reflects multiple forces: productivity synergies, social preferences rooted in tradition or ethical considerations, and labor market frictions that prevent full automation. In many settings, humans make the final call with AI recommendations in hand. This arrangement preserves human authority while offering the accuracy of machines. Yet in practice, decisions in these AI-human collaboration systems often fall short of what could be achieved if AI recommendations were used more effectively.

Evidence from hiring \citep{Hoffman2018}, radiology \citep{Agarwal2023}, and pretrial courts \citep{Angelova2025} suggests that professionals under-use AI recommendations, leaving substantial value on the table. A critical step toward realizing these gains is to understand why people hesitate to follow algorithmic advice. Existing studies, including the work in these domains, document a range of cognitive and information-aggregation biases, but none can causally establish whether image concerns also contribute. In this paper, I use a field experiment to exogenously vary the observability of AI reliance on a large online labor platform, in order to provide the first causal evidence that image concerns shape workers’ AI adoption. I show that (i) image concerns can deter the use of AI at work, leading to performance losses; (ii) workers fear that relying on AI signals a lack of confidence in their own judgment; and (iii) overcoming such concerns is particularly difficult.

Consider an HR committee where other members can readily see that one colleague almost always follows the hiring algorithm’s recommendations; a hospital where supervisors notice that a radiologist rarely departs from the AI’s diagnosis; or a pretrial court where a judge’s low override rate of algorithmic risk scores is visible to peers and administrators. In each case, heavy reliance on AI can yield excellent performance given the accuracy of these systems. Yet it may be hard for observers to avoid forming a negative impression, interpreting extensive reliance on AI as a sign of limited effort, skill, or judgment.

Studying this question with observational data is extremely challenging. Any attempt to measure how visibility shapes AI adoption would require clean variation in whether workers’ AI use is observable, without simultaneously changing how they perceive the technology or altering the career incentives they face. \cite{Angelova2024} show that when judges face greater public scrutiny, their use of algorithmic recommendations systematically changes. While this result suggest that image concerns may be at play, the study lacks the controlled environment necessary to isolate this mechanism, leaving open the question that this paper directly addresses.\footnote{Incentives and observability move together in that paper, as public scrutiny is closely tied to the timing of a reelection.} To overcome this challenge, I turn to a field experiment on one of the largest online labor market platforms, \emph{Upwork}. This setting provides both the control needed to isolate image concerns and the realism of observing workers in their natural work environment, engaged in the kinds of short-term data annotation jobs they routinely complete on the platform.

For the field experiment, I hired 450 freelance workers on Upwork with prior experience in data annotation to complete an image-categorization task assisted by AI recommendations.\footnote{Data annotation continues to rely on human input, although it is increasingly assisted by AI.} Workers were informed that their performance would be evaluated by an HR specialist, with top performers eligible for higher-paid contract extensions. By randomly varying whether HR evaluators could observe workers’ AI reliance, I isolate the effect of image concerns on workers’ use of AI. In the control group, evaluators saw only each worker’s accuracy, whereas in the treatment group they also saw how often the worker changed their answers to match the AI’s recommendations. The HR specialist’s evaluations guided which workers were invited to return for higher-paid extensions, making these assessments directly relevant to participants’ incentives. Importantly, workers were explicitly told that HR specialists were instructed to evaluate candidates based on expected accuracy in another session of the same AI-assisted categorization task, leaving no uncertainty about the evaluation criteria or the nature of the task.

The main finding is that workers reduced their reliance on AI when their this was observed by an evaluator. In the control group, workers switched their initial answer to the AI recommendation in roughly 30.5\% of cases, whereas making AI reliance visible lowered this rate to 26.2\%, a reduction of about 14\% (statistically significant at the 1\% level). The implication for performance is not obvious. In principle, workers could have compensated by exerting greater effort on their initial choices or by exercising better judgment about when to adopt AI recommendations. Neither of these channels played a meaningful role. Treated workers did spend 2 seconds more (a 10\% increase in consideration time, a natural proxy for effort) on their initial choices, yet their initial accuracy remained unchanged. Conditioning on whether they switched their answer to the AI recommendation or not, their accuracy did not improve, ruling out the possibility that they became more selective in deciding when to adopt the AI recommendation. Instead, the reduction in AI reliance translated directly into lower performance, with accuracy falling from 79.1\% to 76.4\%, a decline of 3.4\% (also significant at the 1\% level). These effects are consistent across demographic groups and levels of platform experience, indicating that the phenomenon is general rather than driven by specific subpopulations.

To better understand the mechanism behind these results, I develop a novel incentive-compatible elicitation method using platform feedback to measure what workers fear their AI use signals. After completing the categorization task, workers selected which trait—effort, skill, or confidence in their judgment—they wanted emphasized in the public feedback we provided, which appeared on their profile and remained visible to future employers on the platform. In the control group, this choice simply revealed which trait workers most wished to emphasize. In the treatment group, however, workers were told that the feedback would also show whether their AI use was above or below average and that their feedback choice would allow them to highlight a trait alongside this AI-use measure. This variation identifies how disclosing AI use in feedback, which makes image concerns about AI reliance more salient, changes which trait workers prefer to include in their feedback. The results show that treated workers shifted sharply toward emphasizing confidence in their judgment: the share selecting this trait more than doubled relative to the control group (a 117\% increase). Follow-up questionnaire evidence supports this interpretation, indicating that while confidence in judgment is generally viewed as the least important trait to signal, in AI-assisted work it emerges as the most important, surpassing both effort and skill.

Finally, I provide three pieces of evidence that these image concerns are difficult to overcome. I first highlight features of the experimental design that were deliberately chosen to mute many of the channels through which image concerns might normally operate. Workers know that HR evaluators are instructed to focus exclusively on expected accuracy in the same AI-assisted task, eliminating ambiguity about the evaluation criteria and the scope of the task. Even under these tightly controlled conditions, workers reduce their reliance on AI in ways that lower performance. This suggests that the estimated effects likely understate the role of image concerns in real workplaces, where evaluation criteria are less clearly defined and tasks are more varied.

I then ask whether reducing informational frictions can attenuate these concerns. I implement an intervention that reassures HR specialists about workers’ quality and track record on the platform and makes workers aware of this. Nonetheless, this attempt to close the information gap, akin to repeated interactions with the same evaluator, does not mitigate the image-driven reduction in AI reliance and performance.

I also examine where these beliefs come from by placing workers in the evaluator role. In the second job, rehired workers evaluate other returning workers, mimicking the role of HR: they observe both accuracy and AI reliance from the first job and rate these profiles. Their pay depends on the second job accuracy of the worker with whom they are ultimately paired, with a higher chance of being paired with a worker they rated more favorably, so that monetary incentives are tied solely to expected accuracy. Nevertheless, workers penalize AI use: on average, adopting three additional AI recommendations is penalized slightly more than a single incorrect answer. This pattern suggests that workers’ beliefs about being penalized for using AI reflect their own behavior as evaluators, rather than merely their beliefs about how others would assess AI use.

The rest of the paper is organized as follows. Section \ref{sec:lit} reviews the relevant literature. Section \ref{sec:ED} describes the experimental design and explains the key design choices. Section \ref{sec:CF} offers a brief framework that formalizes the worker’s decision problem and derives testable implications. Section \ref{sec:Results} presents the main findings: making AI use observable reduces workers’ reliance on it, with measurable costs for performance. I then show how observability undermines effective collaboration and provide evidence on the mechanism driving image concerns. Section \ref{hard} documents the robustness of these image-driven responses, highlighting the challenges in overcoming them. Finally, Section \ref{sec:conclusion} offers a brief discussion.

\subsection{Related Literature}\label{sec:lit}

Image concerns shape highly consequential decisions across many domains, including voting \citep{Dellavigna2017}, credit consumption \citep{Bursztyn2018}, educational investment \citep{Bursztyn2019}, political contributions \citep{Perez-Truglia2017}, preventive health take-up \citep{Karing2024, Jee2024}, and, more closely related to this paper, workplace behavior \citep{Mas2009}. The current paper extends this literature by documenting causal evidence that image concerns influence behavior in a new and increasingly important workplace domain: AI-human collaboration.

Beyond this, the paper contributes to the study of stigmatized behaviors \citep{Golub2019,Celhay2025, Friedrichsen2018}, a domain closely tied to social image concerns. Recent evidence suggests that social norms are beginning to emerge against the use of AI. For instance, \cite{Yang2025} use a vignette study to show that physicians evaluate peers who rely on generative AI less favorably, perceiving them as having weaker clinical skills. Along similar lines, \cite{Reif2025} provide experimental evidence of workplace penalties for AI users. Even in a low-stakes context such as a university survey, \cite{Ling2025} find that social desirability bias leads people to under-report their use of AI. This paper aligns with these findings by showing that AI use carries negative connotations in the workplace, and it does so through an incentive-compatible design implemented in a real labor-market setting. 

The paper also contributes to the rapidly growing literature on AI–human collaboration. I focus on recommendation-based collaboration, which is common in labor settings such as radiology, lending, and pretrial adjudication. This format is particularly useful for studying image concerns because it has been in place far longer than newer forms such as generative AI (so social norms and user familiarity are more settled, reducing confounds from novelty), and because AI use in this context is more transparent and easier to identify.\footnote{One of the few settings where generative AI use can be reliably tracked is examined by \cite{Goldberg2025}, who study its equilibrium effects in a creative goods marketplace.}

Field evidence shows that high-stakes professional environments have struggled to realize the potential of AI recommendations \citep{Agarwal2023, Angelova2025, Hoffman2018, Stevenson2024}.\footnote{Large language models, another collaboration format particularly well-suited for writing and articulation tasks, have shown promising results \citep{Brynjolfsson2023, Noy2023, Otis2024, Peng2023}.} A common pattern across these studies is the systematic under-utilization of AI recommendations, often referred to as \emph{algorithmic aversion} \citep{Dietvorst2015}. The dominant explanations emphasize overconfidence and related deviations from Bayesian updating.\footnote{Using a controlled experimental environment, \cite{Caplin2025} shows that calibration and belief formation play a key role in determining the gains from working with AI recommendations.} This paper shows that image concerns, beyond belief formation, can also deter AI use, even at the expense of performance. This perspective aligns with recent research demonstrating that AI can shape preferences directly \citep{McLaughlin2024,Albright2024,Almog2025}. Moreover, in a companion paper \citep{AlmogNIIC}, I use the same task with a Prolific sample to isolate non-instrumental image concerns.\footnote{Sometimes referred to as \textit{hedonic} concerns, which can involve embarrassment, stress, or emotional discomfort.} While the present study examines behavior under predominantly monetary incentives tied to career concerns, the companion paper uses a setting and incentive structure designed to test for non-instrumental motives, which represent another manifestation of image concerns. I show that even when these perceptions carry no monetary consequences, concerns about how one is perceived lead participants to reduce their adoption of AI recommendations, thereby lowering their own chances of earning a performance-based bonus.

In addition, the paper contributes to the labor literature using online platforms such as Upwork to study market frictions \citep{Pallais2014, Stanton2016,Horton2017,Barach2021,He2021}. Methodologically, I introduce a new incentive-compatible mechanism that allows workers to select features of the public feedback they receive after job completion, offering a new tool for studying mechanisms in platform environments.

Finally, the paper advances research on technology adoption and digitization. Adoption often faces challenges such as organizational frictions \citep{Atkin2017} or a lack of familiarity with new tools \citep{AlmogCI}, and digitalization can amplify these barriers. As \cite{Goldfarb2012} emphasize, new technologies frequently enable firms to collect novel forms of information. AI recommendations, digital by design, create a durable and auditable trace, making them particularly vulnerable to image concerns \citep{Goldfarb2019}. This stands in contrast to advice exchanged in a meeting with a co-worker, where attribution is diffuse and harder to corroborate. Consistent with this logic, \cite{Houeix2025} show that data observability slowed the adoption of digital payments in the Senegalese taxi industry. I document a similar pattern: observability also deters AI adoption, though here the mechanism is image concerns rather than contract enforcement.

\section{Field Experiment Design}\label{sec:ED}

\begin{figure}[h!]
\centering
\caption{Upwork Job Posting}
\fbox{\includegraphics[width=6.5in]{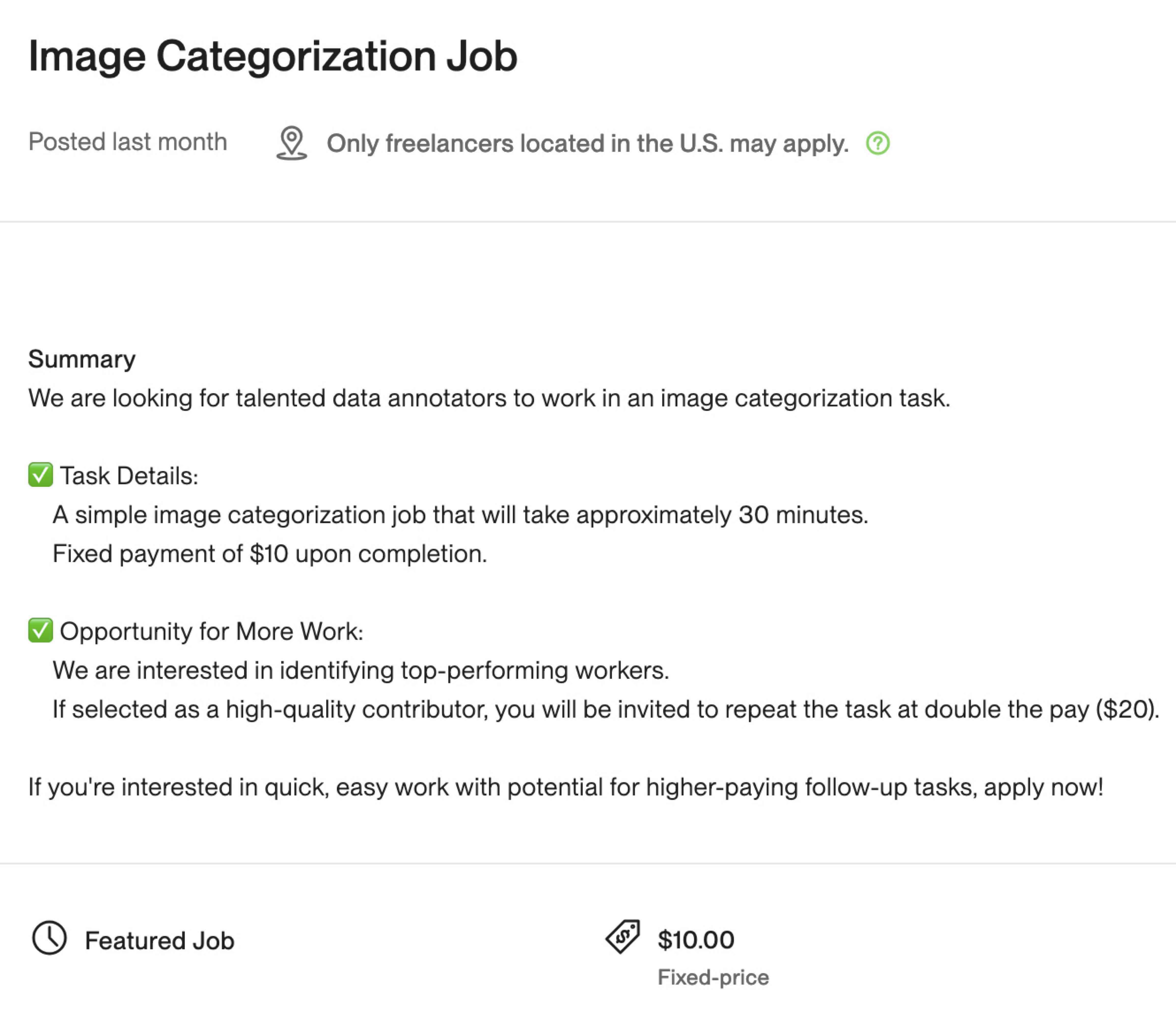}}
\label{fig:Post}
\end{figure}

I hired 450 workers\footnote{After debriefing, one worker elected to withdraw their data, leaving 449 workers for research purposes, without affecting any of the paper’s results.} over a two-week period in July 2025 by posting a 30-minute image categorization job on Upwork under the registered employer name \textit{BuildingAI}, a company that advertises AI annotation solutions on its webpage.\footnote{Screenshots of the \textit{BuildingAI} webpage are available in Appendix \ref{sec:Webpage}.} Upwork is a leading online marketplace that connects employers with independent contractors for a wide range of remote tasks. The job offered a fixed payment of \$10 upon completion and highlighted the firm’s interest in identifying top-performing workers, who would have the opportunity to be invited back to repeat the task at double the pay rate. Figure \ref{fig:Post} shows how the job posting appeared on Upwork.

During the instructions, workers were informed that the firm would extend second job offers to 30\% of participants, with decisions guided by feedback from an HR specialist reviewing their anonymized participation profiles. Workers were randomly assigned to one of three experimental conditions, which varied the information available to the HR specialist at the time of evaluation. Importantly, each worker was also made aware of the specific information that the HR specialist would have in their respective condition.

There is a \textit{Private} condition, which serves as the study’s control. In this condition, AI reliance, defined as the frequency with which workers changed their answers to match the AI’s recommendation, remains private, and HR specialists observe only task accuracy. Throughout the paper, accuracy will serve as the main performance benchmark and is defined as the percentage of correct answers after considering the AI recommendation. Workers in the \textit{Private} condition were informed of the evaluation process as follows:

\begin{quote}
    \emph{“The HR specialist will only see each worker's percentage of final correct answers (after considering the AI recommendation).”}
\end{quote}

The first treatment group is the \textit{Public} condition. In this condition, HR specialists observe not only workers’ accuracy but also their AI reliance. Workers in the \textit{Public} condition were informed of the evaluation process as follows:

\begin{quote}
    \emph{“The HR specialist will only see each worker's percentage of final correct answers (after considering the AI recommendation) and how often each worker changed their answer after seeing the AI recommendation.”}
\end{quote}

The second treatment group is the \textit{Public with Information} condition. As in the \textit{Public} condition, HR specialists observe both workers’ accuracy and their AI reliance. In addition, only workers in this condition are informed that the HR specialist has been assured of the rigorous pre-screening process and the strong track record of all candidates on the platform. This is not new information to the workers, as they were already told on the initial screen that they had been selected based on their strong Upwork history. The intervention therefore manipulates only workers’ beliefs about what the HR specialist knows. The exact message provided exclusively to workers in the \textit{Public with Information} condition was:

\begin{quote}
    \emph{“We also assured the HR specialist that all candidates had been carefully pre-screened and had a solid track record of positive experiences on Upwork, minimizing the risk of unnecessary concerns about worker quality.”}
\end{quote}

In all experimental conditions, workers were told that HR specialists were explicitly instructed to evaluate candidates based on who was expected to perform most accurately if invited back for another session of the same AI-assisted categorization task. The instructions highlighted accuracy as the decisive factor in achieving a higher score.

At the end of the categorization task, without prior knowledge of this component, workers participated in a novel, incentive-compatible exercise designed to reveal what they fear high AI reliance may signal to employers. All workers were told that our firm typically leaves public feedback for employees, intended to be informative for future employers on the platform. They were informed that the feedback would include a brief description of the completed task, in this case, image categorization with AI assistance.  

Workers were then given the opportunity to select one of three statements they wished to emphasize at the end of their feedback:\footnote{These three options reflected the main mechanisms identified in a pilot survey as reasons why workers may fear using AI when it is observable.}  
\begin{enumerate}[label=(\roman*)]
    \item You are a hard worker and put in strong effort.
    \item You are highly skilled and capable in these tasks.
    \item You have confidence in your own judgment.
\end{enumerate}

In the treatment groups, a few modifications were introduced. Workers were told that their public feedback would also indicate whether they relied more or less on the AI recommendations than the average worker for the same job. They were further reminded that heavy reliance on AI might raise concerns for future employers, and that their selected sentence could help address those concerns if they were identified as high AI users.\footnote{Northwestern’s human subjects committee was concerned that feedback evaluations disclosing AI reliance could harm workers. In response, treatment group workers were asked a follow-up question on whether they wanted the AI reliance information included in their feedback. They were informed that if at least one worker objected, the information would be removed for everyone for fairness considerations. Because about 30\% of the workers requested its removal, the information was ultimately excluded.}

For the control group, this exercise captures the baseline distribution of preferences over signaling effort, skill, or confidence in judgment. For the treatment groups, it captures preferences over these same signals under the potential threat of being exposed as a high AI user. Taken together, this allows us to illustrate, at the aggregate level, what workers fear their AI use may signal to employers.  

After all workers completed the first job, an experienced HR specialist hired on Upwork was familiarized with the task and proceeded to assign each worker a score from 0 to 100, adhering strictly to the advertised information in their condition (observing only accuracy for the control group, and both accuracy and AI reliance for the treatment groups). Based on a rehiring rule that depended on both score and chance (described in the next section), we offered contract extensions to 140 workers, of which 135 accepted within one week.\footnote{The 96.4\% acceptance rate reinforces that the rehiring incentives were strong and that workers on the platform are committed to longer-horizon jobs.}

For rehired workers originally assigned to a treatment group, the second job began with an evaluation exercise. They assessed 20 profiles of other returning workers based solely on accuracy (correct answers) and AI reliance rates, replicating the HR specialist’s evaluation in the first job. This component was designed to identify whether changes in AI reliance induced by the treatment were driven solely by anticipated external evaluation or also reflected self-projection into the evaluator’s role. After this, all rehired workers completed a short questionnaire followed by a 10-round version of the categorization task. Control-group workers did not perform the evaluation task, since AI reliance had not been visible in their prior environment and introducing it at this stage would have been unnatural.

The evaluation exercise was incentive compatible: workers were told that they would be matched with one of the evaluated workers, and that the higher a worker ranked in their ordered score ranking, the greater the chance of being paired with them. Each worker could earn up to an additional \$5, receiving \$0.25 for every correct answer of their matched partner and \$0.25 for each of their own correct answers. This design provided incentives both to evaluate carefully and to perform as well as possible themselves. 

\begin{figure}[h!]
\centering
\caption{Experiment Design}
\includegraphics[width=6in]{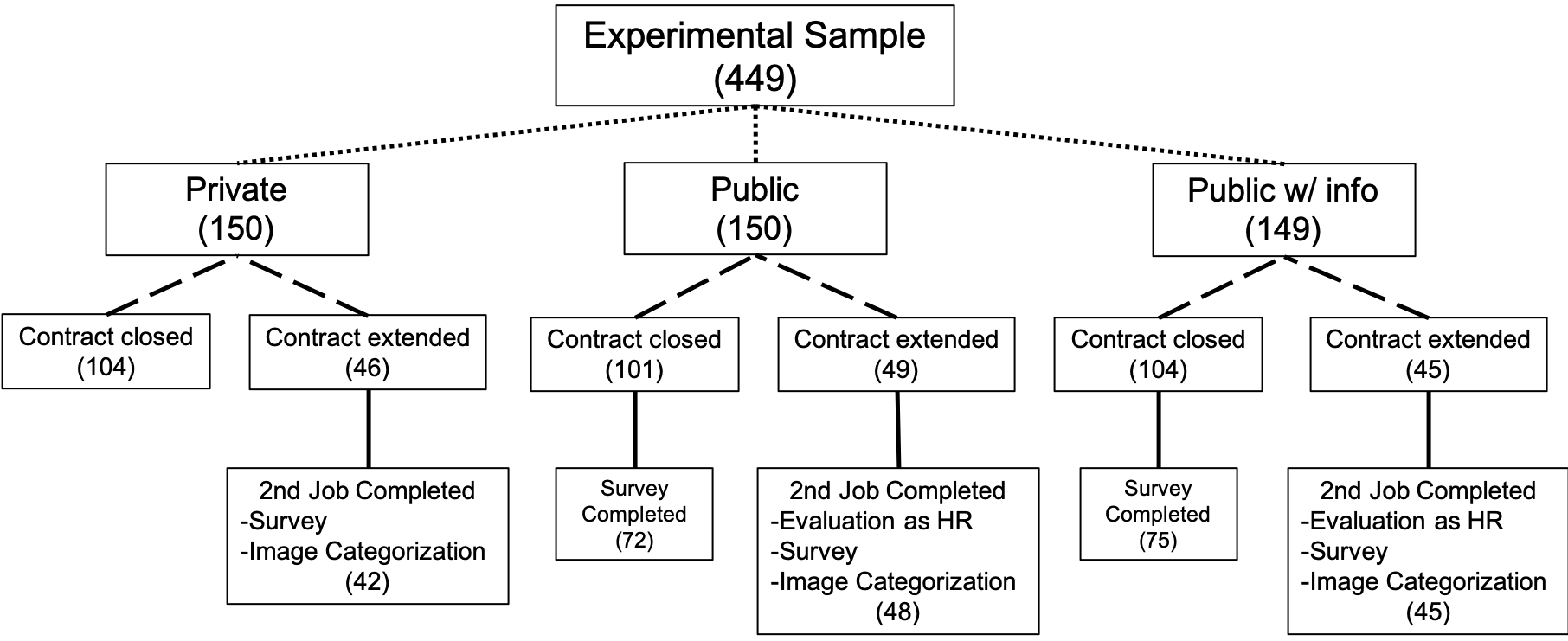}
\label{fig:ED}

    \vspace{0.5em}
    \begin{minipage}{0.9\textwidth} 
        \footnotesize \textit{Notes}: Short dashed lines denote random assignment, while long dashed lines denote quasi-random assignment based on the rehiring probability rule. The number of workers in each node is reported in parentheses.
    \end{minipage}
\end{figure}

Finally, workers from the treatment groups who were not selected for a contract extension were given an opportunity to earn \$5 by completing the short questionnaire component of the second job, of which 147 out of 205 agreed. Because most of the questions were relevant only for treated workers, they were the ones invited to complete the questionnaire. Figure \ref{fig:ED} presents a diagram of the experimental design.

\subsection{Upwork and Sample Selection}
Upwork (formerly oDesk) has been widely used in field experiments because it combines real-market conditions with strong experimental control (\cite{Horton2011} discuss in detail the virtues of using online labor markets for conducting experiments). For example, \cite{Pallais2014} examined hiring inefficiencies caused by information frictions, and \cite{Coffman2023} registered as an employer to study gender differences in job applications. A key advantage of Upwork for this study is its ability to foster authentic employer–employee relationships: the platform enables the use of actual contract extensions as incentives. Recruited workers perform data annotation professionally or as a side job and maintain reputations for doing that through ratings and public reviews. This provides a unique opportunity to study how workers in their natural workplace environment respond to an exogenous shock that makes their AI use observable to someone with influence over their career prospects.

To participate, workers had to satisfy three criteria. First, they had to be located in the United States, situating the findings in a region with substantial AI exposure while reducing concerns about language barriers or misinterpreted instructions. Second, workers needed prior platform experience in related fields such as data annotation or data entry. Because prior work has shown that new workers face substantial barriers \citep{Pallais2014, Stanton2016}, I restricted the sample to workers with previous experience; 81\% had already earned over \$100 on the platform. Finally, workers had to be registered as freelancers rather than part of an agency, ensuring direct business relationships without intermediaries.

\subsection{Image Categorization Task}
\begin{figure}[h!]
\centering
\caption{Task interface showing the initial choice and AI recommendation stages.}
\begin{subfigure}[b]{\textwidth}
    \centering
    \includegraphics[width=2.5in]{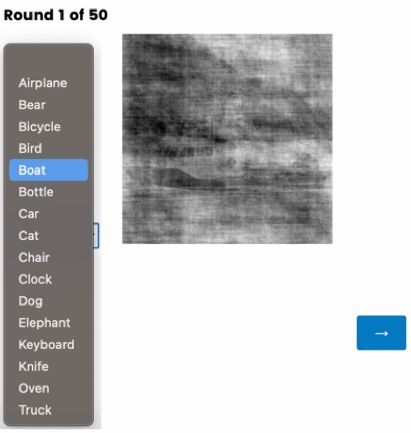}
    \caption*{Initial Choice}
\end{subfigure}

\vspace{0.5cm}

\begin{subfigure}[b]{0.49\textwidth}
    \centering
    \includegraphics[width=\textwidth]{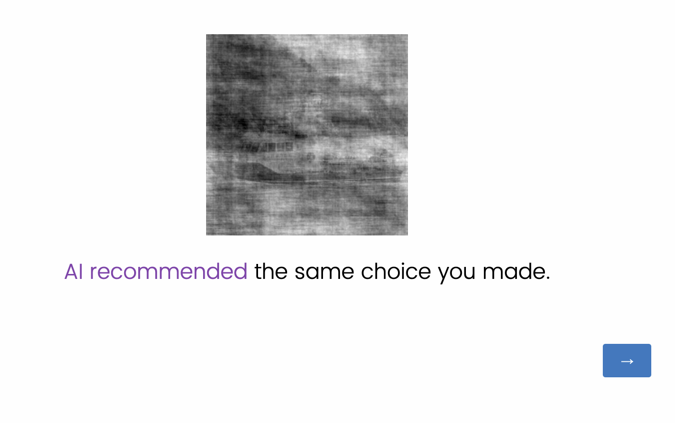}
    \caption*{When the recommendation agrees}
\end{subfigure}
\hfill
\begin{subfigure}[b]{0.49\textwidth}
    \centering
    \includegraphics[width=\textwidth]{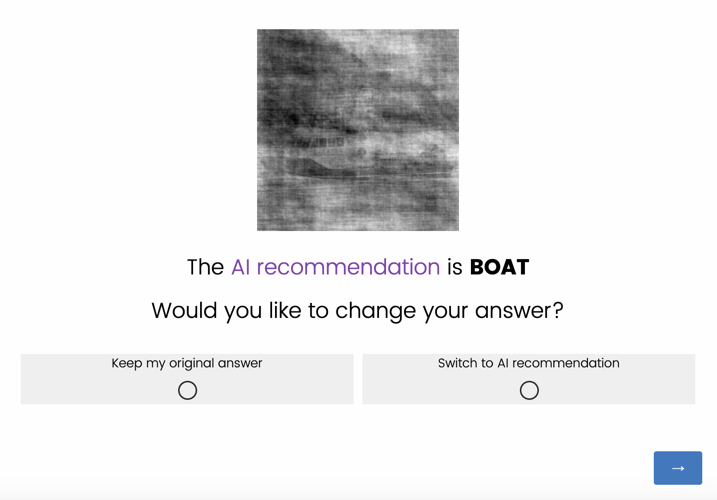}
    \caption*{When the recommendation differs}
\end{subfigure}
\label{fig:TaskInterface}
\end{figure}

The first job consisted of 50 rounds of an image categorization task with AI recommendation assistance. In each round, the workers observed a blurred image with the objective of selecting the correct category from a list of 16 options: airplane, bear, bicycle, bird, boat, bottle, car, cat, chair, clock, dog, elephant, keyboard, knife, oven, and truck. After making their initial choice, workers were shown an AI recommendation. If their choice matched the AI recommendation, they were notified of the agreement and proceeded to the next image. If the recommendation differed, they could revise their answer and switch to the AI’s recommendation. Workers were incentivized to put effort into their initial choice because, once the AI recommendation was revealed, they could only choose between their original answer and the AI’s recommendation; all other options were no longer available. The dynamics of this two-stage decision process are illustrated in Figure \ref{fig:TaskInterface}. The images and AI model were obtained from \cite{Steyvers2022}, and Appendix \ref{sec:TaskAppendix} provides additional technical details on the task construction.

Although AI has made significant progress in image categorization, human input remains essential in supervised learning processes such as data annotation. This task is particularly well-suited for studying AI–human collaboration, as it encompasses: (i) images that appear simple to humans but are misclassified by the AI; (ii) images that the AI classifies correctly but are extremely difficult for humans, making blind delegation to the AI the most effective strategy; and (iii) cases where humans often classify correctly, but those who do not may be guided to the correct answer after reconsidering the image from the perspective suggested by the AI.\footnote{See Figure \ref{fig:AppendixImages} for examples.}

Additional advantages of this task are that it is a well-established and popular category on online labor platforms such as Upwork,\footnote{Upwork has a job category named ``AI Data Annotation and Labeling.''} it is easy to explain and requires little to no training (especially for platform users with prior annotation experience), and it offers excellent experimental control. The dataset further provides ground-truth labels, presents each image at varying noise levels to manipulate difficulty, has been validated as an efficient tool for studying AI–human collaboration, and \cite{Steyvers2022} also made available high-quality AI recommendations.

\subsection{Design Choices}

In this section, we discuss several of the main choices we faced while designing the experiment.

\textit{Between-subjects.} Workers are assigned to a single experimental condition to avoid experimenter demand or priming effects. While a within-subjects design could provide insight into the distribution of treatment effects, which is particularly relevant here since effects may vary in direction, the risk of eliciting mechanical responses to treatment is substantial.

\textit{Recommendation timing.} Providing the recommendation after an initial choice allows us to clearly identify AI's impact on the final decision. This approach avoids ambiguity in cases where the choice and recommendation align, but it’s uncertain whether the choice was influenced by the AI or made independently. This design has two key advantages: first, it enables precise measurement of how often recommendations are adopted; second, it conveys to workers that their choices to adopt recommendations are being reliably tracked. In some work environments, identifying which employees rely more heavily on recommendations can take considerable time (e.g., days or even months), but this study lacks the timeframe for extended observation, which would likely be necessary if recommendations were provided before making an initial choice.

\textit{Information provided on AI recommendations.} Workers were informed that the recommendations came from an AI model trained on the same task and that it was accurate 85\% of the time, a statement that holds for both the full dataset and the subset of images used in this experiment. They were also told that AI might perform well on images that humans find difficult but could occasionally miss images that appear easy to the human eye. This statement reflects a genuine feature of the original study by \cite{Steyvers2022}. Prior research \citep{Dietvorst2015,Dreyfuss2025} shows that people may overreact to a single poor algorithmic recommendation, so this disclaimer was intended to reduce the likelihood that participants would lose trust in AI based on an isolated error. Such reactions represent an interesting phenomenon that is orthogonal to our research question, and given the short horizon of our study, we found it preferable to avoid them.

\textit{Incentives.} Workers receive a fixed \$10 payment for completing the job. However, the most important aspect of the incentive scheme is the opportunity to be rehired for a similar task at double the pay rate (\$20). The rehiring opportunity allows workers to derive instrumental value from third-party beliefs, here represented by the evaluation of an HR specialist. This approach mirrors a realistic aspect of the labor market, where employees care about how others perceive them, as these perceptions can ultimately affect hiring, promotion, or layoff decisions.

\textit{Rehiring decisions.} As advertised to workers, second job offers were made using feedback from a real HR specialist hired on the same platform. Rehiring decisions depended on both score and luck, with the probability of rehiring for each worker determined by the following formula:

\[
P(\text{Rehire}) = 0.2 + (\text{Decile} \times 0.02), \quad \text{Decile} = 1, \dots, 10
\]

Deciles were determined by workers’ scores within their experimental condition. The rehiring rule served two purposes. First, it rewarded higher scores. Although workers were never told explicitly that higher scores would lead to better outcomes, it was natural to interpret the scores in this way, and our design relies on this perception. Second, it ensured that the pool for the second job was more representative, rather than consisting only of top performers. This design makes the second-job results more generalizable.

\textit{Treatment design.} The variation across experimental conditions was designed to differ only in the information available to the HR specialist at the moment of evaluation, and workers were informed openly and transparently about what information the HR would have at their disposal. Across all conditions, participants were also made aware that the HR specialist was instructed to evaluate solely on which workers were expected to perform most accurately if brought back for another session of the same AI-assisted categorization task.

\section{Conceptual Framework}\label{sec:CF}
I introduce a model of AI-assisted categorization that mirrors the workflow of the experiment. In each round, the worker first makes an unaided choice and holds a private confidence \(p\) in the selected category; the AI then provides a calibrated recommendation with fixed confidence \(\kappa\). When the recommendation differs from the worker’s initial choice, the worker decides whether to keep or switch their initial choice, anticipating an evaluator who always observes final accuracy \(a\) and may also observe AI reliance \(r\) (the share of switches). The worker chooses when to adopt the AI recommendations to maximize the evaluator’s score. We focus on how behavior shifts when AI reliance becomes observable. Visibility can induce a trade-off between accuracy and AI adoption, which ultimately depends on how the worker expects adopting AI will influence the evaluator’s score. The model formalizes this environment and yields testable predictions for (i) willingness to adopt AI recommendations and (ii) overall task performance.

\subsection{A Model of Worker Behavior under Observable AI Assistance}

Consider a categorization task that consists of \(n\) distinct categories (in the experiment, \(n=16\)). The true category, or \emph{state}, is denoted by \(\omega \in \{\omega_1, \ldots, \omega_n\}\), and the worker chooses a category \(y \in \{y_1, \ldots, y_n\}\). An action is correct when \(y = y_i\) for the corresponding state \(\omega = \omega_i\); that is, when the true category is chosen.

A worker has baseline type \(\theta\in[1/n,1]\) summarizing her average
\emph{unaided} accuracy. Before seeing the AI recommendation, she selects an initial category \(y^0\) from a private signal. The AI then recommends a category \(\hat y\) with calibrated confidence \(\kappa\in[0,1]\), interpreted as the probability that \(\hat y\) is correct.\footnote{An algorithm is said to be well-calibrated if its predicted probabilities align with empirical outcome frequencies.} Conditional on the state \(\omega\), the worker’s private signal and the AI’s recommendation are independent.\footnote{In the experiment, workers were prompted to treat the AI’s recommendation as independent of their own signal. The qualitative predictions of the model remain unchanged if this assumption is relaxed (e.g., under a single-crossing property)}

Let \(y\) denote the worker’s final label after considering the AI recommendation. We track two reduced-form objects across tasks:
\[
a \;\equiv\; \Pr(y=\omega)\quad\text{(expected accuracy)},\qquad
r \;\equiv\; \Pr\big(y=\hat y \neq y^0\big)\quad\text{(AI reliance)}.
\]
Thus \(a\) is average accuracy after considering the AI recommendation, and \(r\) is the share of tasks on which the worker \emph{changes} her initial answer to the AI’s recommendation (only relevant when \(y^0\neq \hat y\)).

The worker decides whether to switch to the AI’s recommendation in order to maximize the score assigned by an evaluator who observes certain variables. In the control regime the evaluator’s score depends only on accuracy,
\[
S_c \;=\; S_c(a),
\]
whereas in the treatment regime it also depends on AI reliance,
\[
S_t \;=\; S_t(a,r).
\]
\paragraph{Assumption (Accuracy is rewarded).}
The evaluator’s score is strictly increasing in accuracy:
\[
\frac{d S_c(a)}{d a}\;> 0
\quad\text{and}\quad
\frac{\partial S_t(a,r)}{\partial a}\;> 0
\qquad\forall\, a\in(0,1),\; r\in[0,1].
\]

\subsection{Decision Rule and Threshold Optimality}

I deliberately take a reduced-form approach to belief formation: I do not model priors, signals, or an updating rule. All we require is that, after choosing a category \(y^0\), the worker can summarize her information by a scalar
\[
p \;\equiv\; \Pr(y^0=\omega\mid \text{worker's information}) \in \big[1/n,\,1\big],
\]
interpreted as the probability that her initial choice of category is correct.\footnote{A simple microfoundation is Bayesian: the worker begins with a prior over categories, observes a private signal \(s\), forms posteriors \(q_j(s)=\Pr(\omega=\omega_j \mid s)\), chooses \(y^0=\arg\max_j q_j(s)\), and sets \(p=\max_j q_j(s)\). Our analysis relies only on the scalar \(p\) and abstracts from the underlying structure that generates it. For some results it is convenient (but not required) to assume \(p\sim F\) with density \(f_0\) and \(\mathbb{E}[p]=\theta\).}

When \(y^0\neq \hat y\) (disagreement), the worker chooses \emph{one source} to follow:
either keep \(y^0\) or switch to \(\hat y\). Under calibration, the expected accuracy from
keeping \(y^0\) equals \(p\); from following the AI equals \(\kappa\).
(Agreement does not affect the rule because both sources induce the same category.)

A simple and natural decision rule is a cutoff policy: the worker adopts the AI’s recommendation whenever her own confidence \(p\) is below a threshold \(\tau \in [1/n, 1]\), and otherwise keeps her initial choice \(y^0\).

\begin{proposition}[Threshold optimality]\label{prop1}

\emph{(i) Scoring when AI reliance is not observed.} If the worker maximizes \(S_c(a)\) with \(S_c'(a)> 0\),
the accuracy-maximizing policy is the cutoff rule with
\[
\tau^\ast=\kappa, \qquad\text{i.e., switch to AI iff } p\le \kappa.
\]
\emph{(ii) Scoring with observable AI reliance.} Suppose the worker maximizes
\(S_t(a,r)\), where \(S_t\) is continuously differentiable and \(\partial S_t/\partial a>0\).
Then an optimal policy is again a cutoff rule:
\[
\tau^\ast \;=\; \kappa - \lambda, \qquad
\lambda \;\equiv\; -\,\frac{\partial S_t/\partial r}{\partial S_t/\partial a}\;\Bigg|_{(a,r)}
\]
Here \(\lambda\) is the evaluator’s \emph{shadow penalty} (in
accuracy units) for an additional unit of reliance. If reliance is penalized \((\partial S_t/\partial r<0)\) then \(\tau^\ast<\kappa\); if reliance is rewarded \((\partial S_t/\partial r>0)\) then \(\tau^\ast>\kappa\).
\end{proposition}

\begin{figure}[h!]
    \centering
    \caption{Threshold Decision Rule.}
    \begin{tikzpicture}
    \begin{axis}[
        width=12cm,
        height=8cm,
        xmin=0, xmax=1,
        ymin=0, ymax=1.2,
        axis lines=left,
        xlabel={Initial Choice Confidence},
        ylabel={Probability Distribution $(f_{\theta})$},
        samples=200,
        domain=0:1,
        ticklabel style={/pgf/number format/fixed},
        ytick=\empty,              
        xtick={0,0.6,0.85,1},      
        xticklabels={$1/n$,$\kappa-\lambda$,$\kappa$,$1$},
    ]
        \def\tau{0.85}
        \def\taulow{0.6}

        \addplot[name path=curve, thick]
            {0.5*exp(-((x-0.55)^2)/(2*0.10^2))
             +0.25*exp(-((x-0.25)^2)/(2*0.08^2))
             +0.25*exp(-((x-0.85)^2)/(2*0.08^2))};

        \addplot[name path=axis, draw=none] coordinates {(0,0) (1,0)};

        \addplot[fill=blue!15, draw=none]
            fill between[of=curve and axis, soft clip={domain=0:\taulow}];

        \addplot[fill=green!15, draw=none]
            fill between[of=curve and axis, soft clip={domain=\taulow:\tau}];

        \addplot[fill=red!15, draw=none]
            fill between[of=curve and axis, soft clip={domain=\tau:1}];

        \addplot[dashed] coordinates {(\tau,0) (\tau,1.1)};

        \addplot[densely dotted] coordinates {(\taulow,0) (\taulow,1.1)};

        \draw[->, thick] (axis cs:\tau-0.01,0.96) -- (axis cs:\taulow+0.01,0.96);

        \node[anchor=north, align=center, font=\footnotesize]
            at (axis cs:0.30,0.75)
            {Adopt AI\\always};

        \node[anchor=north, align=center, font=\footnotesize]
            at (axis cs:0.725,0.75)
            {Adopt AI\\only when\\not visible};

        \node[anchor=north, align=center, font=\footnotesize]
            at (axis cs:0.925,0.75)
            {Override\\AI\\always};

    \end{axis}
    \end{tikzpicture}

    \label{fig:posterior_threshold}

        \vspace{0.5em}
    \begin{minipage}{0.9\textwidth} 
        \footnotesize \textit{Notes}: The distribution of private confidence $p$ is partitioned by two different thresholds that dictate whether to adopt AI. When $p$ is below the threshold, the worker follows the AI recommendation; when $p$ is above it, keeps her original choice. The higher threshold $\kappa$ represents the optimal rule when AI reliance is not observable. When reliance becomes observable and evaluators penalize it, the threshold shifts left to $\kappa-\lambda$. The shaded green region between $\kappa-\lambda$ and $\kappa$ highlights initial confidence levels for which adopting the AI recommendation would improve accuracy, but the worker instead overrides whenever reliance is visible.
    \end{minipage}
\end{figure}
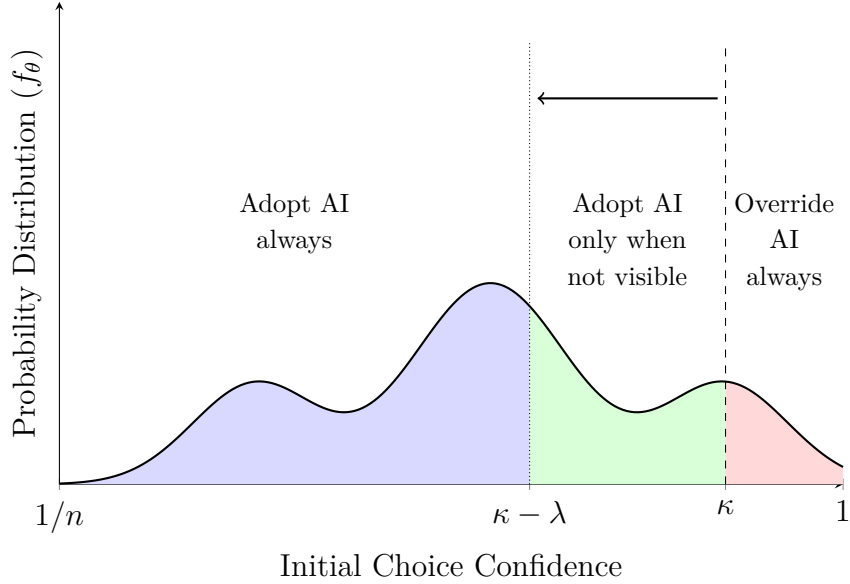

When AI reliance is penalized, the worker is willing to ignore AI recommendations for any private confidence level \(p \in (\kappa-\lambda, \kappa)\), a range in which accuracy would improve by following the AI recommendation, but not enough to offset the expected penalty for relying on AI. Conversely, when AI reliance is rewarded, the worker 
is willing to adopt AI recommendations when her private confidence level \(p \in (\kappa, \kappa-\lambda)\), a range in which accuracy would decrease by following the AI recommendation, but the reward for adopting AI more than compensates for the loss in accuracy. Figure~\ref{fig:posterior_threshold}
illustrates how the threshold rule dictates whether to adopt or override the AI recommendation as a function of initial choice confidence.

\subsection{Solution and Testable Predictions}

Assume the private confidence \(p\) admits a cdf \(F\) with density \(f\) on \([1/n,1]\) and \(\mathbb{E}[p]=\theta\). Then the closed form solution is: 
\[
\;r^\ast(\lambda)=\left[F(\kappa-\lambda)\right] \, D\;
\qquad\text{and}\qquad
\;a^\ast(\lambda)=\left[\theta+\int_{1/n}^{\kappa-\lambda}(\kappa-p)\,f(p)\,dp\;\right]\, D \,+\, (\theta\kappa)\,(1-D).
\]

where $D$ denotes the probability of disagreement $(y^0 \neq \hat y)$, and the bracketed terms represent, respectively, the reliance rate and the accuracy rate conditional on disagreement.

\paragraph{Model predictions.}
Moving from the control benchmark to a regime in which evaluators observe AI reliance, and assuming that the evaluator’s score \(S_t(a,r)\) is strictly increasing in accuracy \((\partial S_t/\partial a>0)\), the model implies:

\begin{enumerate}
\item \textbf{AI adoption.} Visibility shifts optimal AI reliance in the direction of the perceived incentive on \(r\):
\begin{itemize}
\item If evaluators penalize reliance \((\partial S_t/\partial r<0)\), then \(r^\ast\) decreases.
\item If evaluators reward reliance \((\partial S_t/\partial r>0)\), then \(r^\ast\) increases.
\end{itemize}

\item \textbf{Accuracy (strict comparative static, \(\lambda\neq 0\)).}
For any nonzero visibility weight \(\lambda\) (so \(\partial S_t/\partial r\neq 0\)) with \(f_\theta(\cdot)>0\),
\[
a^\ast(\lambda) \; < \; a^\ast(0).
\]
That is, accuracy is strictly lower under any nonzero visibility-induced distortion of the scoring rule. Appendix \ref{subsec:invU} shows that accuracy can be represented as an inverse-U-shaped function of AI reliance, attaining a maximum at \(\tau=\kappa\) (the optimal AI reliance level under the control regime).
\end{enumerate}

\noindent\emph{Implication (penalty case).} If workers anticipate a penalty on AI reliance \((\partial S_t/\partial r<0)\), then making reliance observable yields (i) lower AI reliance \(r^\ast\) and (ii) lower accuracy \(a^\ast\) relative to the control.

\section{Experimental Results}\label{sec:Results}

Table \ref{tab:Descriptive} provides a summary of workers’ characteristics. Gender and ethnicity were self-reported by the workers, while the remaining variables, encompassing platform history and previous education, were collected from their public profiles. Two-sided t-tests indicate that treatment assignment was balanced across the available observable characteristics, with no variable showing a statistically significant difference at the 10\% level. 

\begin{table}[h!]
    \centering
    \caption{Descriptive Statistics and Balance Check.}
    \label{tab:Descriptive}
    \sisetup{table-format=1.3}  
    \begin{tabular}{lSSSSSSSS}
        \toprule
        &  & {Private} & & \multicolumn{2}{c}{Public} &  & \multicolumn{2}{c}{Public w/ info} \\
        \cmidrule(lr){3-3} \cmidrule(lr){5-6} \cmidrule(lr){8-9}
        & & {Mean (sd.)} &  & {Mean (sd.)} & {t-stat}  & & {Mean (sd.)} & {t-stat}   \\
        \midrule
        Female & & {0.69 (0.46)}  & & {0.68 (0.47)} & {0.25} &  & {0.67 (0.47)} & {0.41}  \\
        Minority & & {0.43 (0.50)}  & & {0.46 (0.50)} & {-0.58} &  & {0.50 (0.50)} & {-1.22}  \\
        Previous Earnings & & {8,262 (30,292)}  & & {7,958 (19,757)} & {0.10} & & {7,635 (24,275)} & {0.20}  \\
        Previous Jobs & & {15.2 (17.1)}  & & {17.9 (20.1)} & {-1.24} &  & {15.5 (20)} & {-0.11}  \\
        Hourly fee & & {20.3 (9.6)}  & & {21.1 (9.3)} & {-0.75} &  & {21.4 (14.7)} & {-0.77} \\
        Badge & & {0.35 (0.48)}  & & {0.39 (0.49)} & {-0.72} &  & {0.40 (0.49)} & {-1.00}  \\
        Undergraduate degree & & {0.65 (0.48)}  & & {0.66 (0.48)} & {-0.24} &  & {0.70 (0.46)} & {-1.07}
        \\
        \vspace{0.8em}
        Graduate degree & & {0.2 (0.40)}  & & {0.27 (0.44)} & {-1.36} &  & {0.27 (0.44)} & {-1.40}
        \\
        Observations & & {150}  & & \multicolumn{2}{c}{150} &  & \multicolumn{2}{c}{149}
        \\
        \bottomrule
    \end{tabular}
    
    \vspace{0.5em}
    \begin{minipage}{0.9\textwidth} 
        \footnotesize \textit{Notes}: \textit{Minority} is a dummy variable equal to one for workers identifying with any non-white ethnicity. \textit{Previous Jobs} was winsorized at the 95th percentile, with values above this percentile set to 95th level. \textit{Badge} is a dummy variable equal to one for workers holding any of the platform badges: ‘Rising Talent’, ‘Top Performer’, or ‘Top Performer Plus’. The t-statistics reported correspond to two-sided t-tests comparing the private (control) group to either of the public (treatment) groups.
        \\
        * \(p<0.10\), ** \(p<0.05\), *** \(p<0.01\).
    \end{minipage}
\end{table}

\subsection{Treatment Effects}
For clarity, I pool the two \textit{public} treatments in the main analysis. The conditions produced very similar effects, with no statistically significant differences across any outcome variable.\footnote{Table \ref{tab:Results_Separate} reports results with the treatments separated, showing nearly identical and statistically significant estimates when each treatment is compared separately to the control group.} Section \ref{hard} discusses in greater detail the implications of not finding differences between the two \textit{public} treatments. Unless otherwise noted, the control group refers to the \textit{private} condition, and the treatment group combines the \textit{public} and \textit{public with information} conditions.

The empirical analysis relies on the following simple regression framework:

\begin{equation}\label{eqn:EmpiricalSpecificationMain}
Y_{ij}= \alpha + \beta{T_i} + \gamma X_j + \epsilon_{ij}
\end{equation}

Where $Y_{ij}$ denotes the outcome variable for worker $i$ on image $j$. The treatment indicator $T_i \in \{0,1\}$ equals 1 if worker $i$ was assigned to either of the \textit{Public} treatment groups, and 0 otherwise. $X_j$ represents image fixed effects. The coefficient of interest, $\beta$, captures the average treatment effect of being assigned to a condition in which AI reliance was visible to the HR specialist, relative to the control group, where AI reliance was not visible. All specifications include image fixed effects, and standard errors are clustered at the worker level.\footnote{Results are robust to excluding image fixed effects.} Treatment effects for the six pre-registered outcome variables are reported in Table \ref{tab:Results}. 

\begin{table}[h!]
    \centering
    \caption{Treatment Effects.}
    \label{tab:Results}
    \sisetup{table-format=1.3}  
    \begin{tabular}{lSSSSSSSSS}
        \toprule
        & \multicolumn{2}{c}{AI recommendation reliance} & & \multicolumn{2}{c}{Correct answer} & & \multicolumn{2}{c}{Response time} \\
        \cmidrule(lr){2-3} \cmidrule(lr){5-6} \cmidrule(lr){8-9}
        & {All} & {Conditional} & & {Initial} & {Final} & & {Initial} & {Rec. stage}  \\
        & {(1)} & {(2)} & & {(3)} & {(4)} & & {(5)} & {(6)}   \\
        \midrule
        \textbf{Treatment} & \textbf{-0.043}*** & \textbf{-0.080}*** & &\textbf{0.008} & \textbf{-0.027}***  &  & \textbf{2.10}* & \textbf{-0.72} \\\vspace{1em}
        & {(0.014)} & {(0.022)} & & {(0.012)} & {(0.008)}  & & {(1.10)}  & {(0.57)}   \\

        Constant & 0.305 & 0.640 & & 0.553 & 0.791  & & {21.3} & {10.1} \\
        Observations & {22,398} &  {10,554} &  & {22,398} & {22,398} & &{22,398} &  {10,554} \\
        \bottomrule
    \end{tabular}
    
    \vspace{0.5em}
    \begin{minipage}{0.9\textwidth} 
        \footnotesize \textit{Notes}: Each dependent variable is regressed on a treatment group indicator (equal to 1 if the worker was assigned to any of the \textit{public} treatment groups), with image-specific fixed effects included. Standard errors are clustered at the worker level.\\
        * \(p<0.10\), ** \(p<0.05\), *** \(p<0.01\).
    \end{minipage}
\end{table}

The first main result is that assignment to the \textit{Public} treatment reduced workers’ reliance on AI recommendations. Columns 1 and 2 of Table \ref{tab:Results} report the treatment effects on AI reliance.\footnote{Column 1 includes all rounds (the unconditional reliance rate), which is also the measure observed by HR evaluators and therefore the outcome most relevant for workers’ image concerns. Column 2 restricts to only rounds in which workers had the opportunity to switch from their initial answer to the AI recommendation. When the AI recommendation matched a worker’s initial choice, this does not count as relying on AI.} Across both measures, we observe a statistically significant decrease in AI reliance. Using Column 1 as the main benchmark, workers in the treatment groups followed the AI recommendation 4.3 percentage points (pp) less often, significant at the 1\% level. This effect represents a 14\% reduction relative to the 30.5\% AI reliance rate observed in the control group, where workers had no concern about an HR specialist observing their AI use. In Column 2, which focuses only on rounds where workers faced a genuine choice between their own answer and the AI recommendation, the estimated effect is naturally larger at 8 pp.

Making AI adoption visible to HR specialists reduced workers’ reliance on AI, which in turn led to worse performance in the categorization task, even when that performance itself was observable to HR. Column 4 of Table \ref{tab:Results} reports the treatment effects on the percentage of correct answers after considering the AI recommendation, our measure of job performance. Workers in the \textit{Public} treatment, by ignoring more often AI recommendations, experienced a 2.7 pp decline in accuracy, significant at the 1\% level. Relative to the 79.1\% accuracy rate observed in the control group, this corresponds to a 3.4\% performance reduction. Figure \ref{fig:RoundsTE} shows that the treatment effects on AI reliance and performance remain stable across the 50 rounds, indicating a consistent change in behavior throughout the task. 

These patterns are consistent with the two model predictions in Section~\ref{sec:CF}, which formalizes a worker’s decision in a categorization task, including the option to revise an initial answer after considering an AI recommendation. The model predicts that when AI reliance becomes visible and workers expect a penalty for relying on the AI, visibility reduces reliance and leads to lower accuracy relative to the control condition.

A reduction in performance is not guaranteed following the decrease in AI reliance due to its visibility, as several orthogonal channels not captured by the model may have sustained or even improved accuracy. One plausible scenario is that workers exerted greater effort in their initial choice in anticipation of relying less on AI.\footnote{\cite{Almog2025} provide field evidence that public overruling of human decisions by AI can trigger effort adjustments, while \cite{Agarwal2025} show how such effort responses can be incorporated into the design of human–AI collaboration systems.} Column 5 of Table \ref{tab:Results} reports a 10\% increase in response time, a natural proxy for effort (significant at the 10\% level). On average, workers in the \textit{Public} treatment took about two additional seconds per image to make their initial selection. However, column 3 shows that these extra seconds did not have an impact on performance, as initial accuracy was unaffected. Thus, helping to reasonably rule out a meaningful distortion in performance derived from workers’ initial choice behavior. 

Column 5 of Table \ref{tab:Results} reports a 10\% increase in response time, a natural proxy for effort (significant at the 10\% level). On average, workers in the \textit{Public} treatment took about two additional seconds per image to make their initial selection. However, as shown in column 3, these extra seconds did not improve performance, as initial accuracy remained unchanged. This helps reasonably rule out meaningful distortions in performance stemming from workers’ initial choice behavior.

Another potential mechanism is that treated workers became more discerning in their AI use, forgoing incorrect recommendations more often. However, the evidence does not support this interpretation. First, Column 6 shows a negative but statistically insignificant effect on response times at the recommendation stage, making it unlikely that treated workers engaged in more careful deliberation. Second, and more importantly, Figure \ref{fig:ByFollow} disaggregates accuracy by treatment status and by whether workers followed the recommendation. If anything, their judgment worsened: conditional on not following recommendations, their accuracy declined.

I also conducted a counterfactual exercise to test whether the induced reduction in AI reliance reflected a deliberate change in engagement rather than merely a mechanical decrease. Specifically, I simulated a scenario in which, for every round where control workers followed the AI recommendation, I randomly forced 14\% of those choices to be overridden, thereby reducing their AI reliance to the rate observed in the treatment group. Figure \ref{fig:Random} shows that, under this adjustment, the average accuracy gains for control workers declined from 23.8pp to 20.4pp, virtually identical to the gains among treated workers (20.3pp).

Taken together, these findings show that the decline in AI reliance when it was visible did not reflect improved judgment. Instead, it was comparable to randomly overriding additional AI recommendations, providing clear evidence that workers in the treatment group did not improve the quality of their decisions about when to rely on AI.

A recurring concern in the adoption of AI is its potential to exacerbate existing inequalities or generate new ones across different groups in society. To address this, I extend the empirical strategy to examine heterogeneous effects, testing whether workers respond differently to having their AI reliance made visible depending on their gender, ethnicity, education, or platform experience. Figure \ref{fig:Het_dem} shows no evidence that the main findings are driven by any particular demographic group, such as gender, ethnicity, or education. Likewise, Figure \ref{fig:Het_platform} illustrates that the treatment effects are consistent across the distribution of workers’ prior platform experience, whether measured by earnings or by completed jobs. Altogether, these results suggest that workers’ responses to the visibility of their AI use reflect a population-wide pattern, rather than being concentrated in specific subgroups.

\subsection{Scope for AI-human Collaboration}
\begin{figure}[htbp]
\centering
\caption{Accuracy Distributions}
\begin{subfigure}[b]{.49\textwidth}
\centering
\includegraphics[width=3.1in]{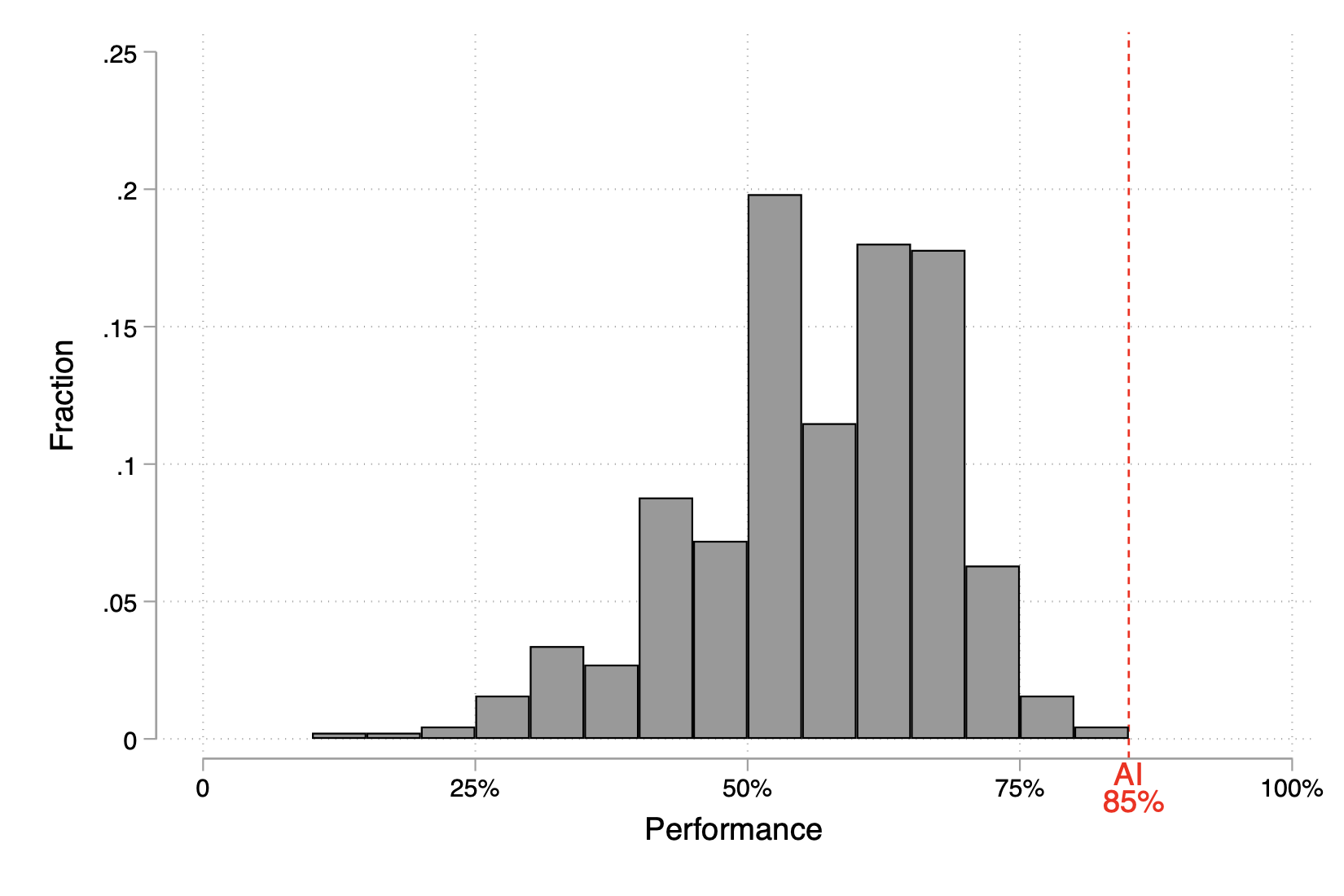}
\caption{Initial Accuracy (before AI).}
\label{fig:BeforeDist}
\end{subfigure}
\begin{subfigure}[b]{.49\textwidth}
\centering
\includegraphics[width=3.1in]{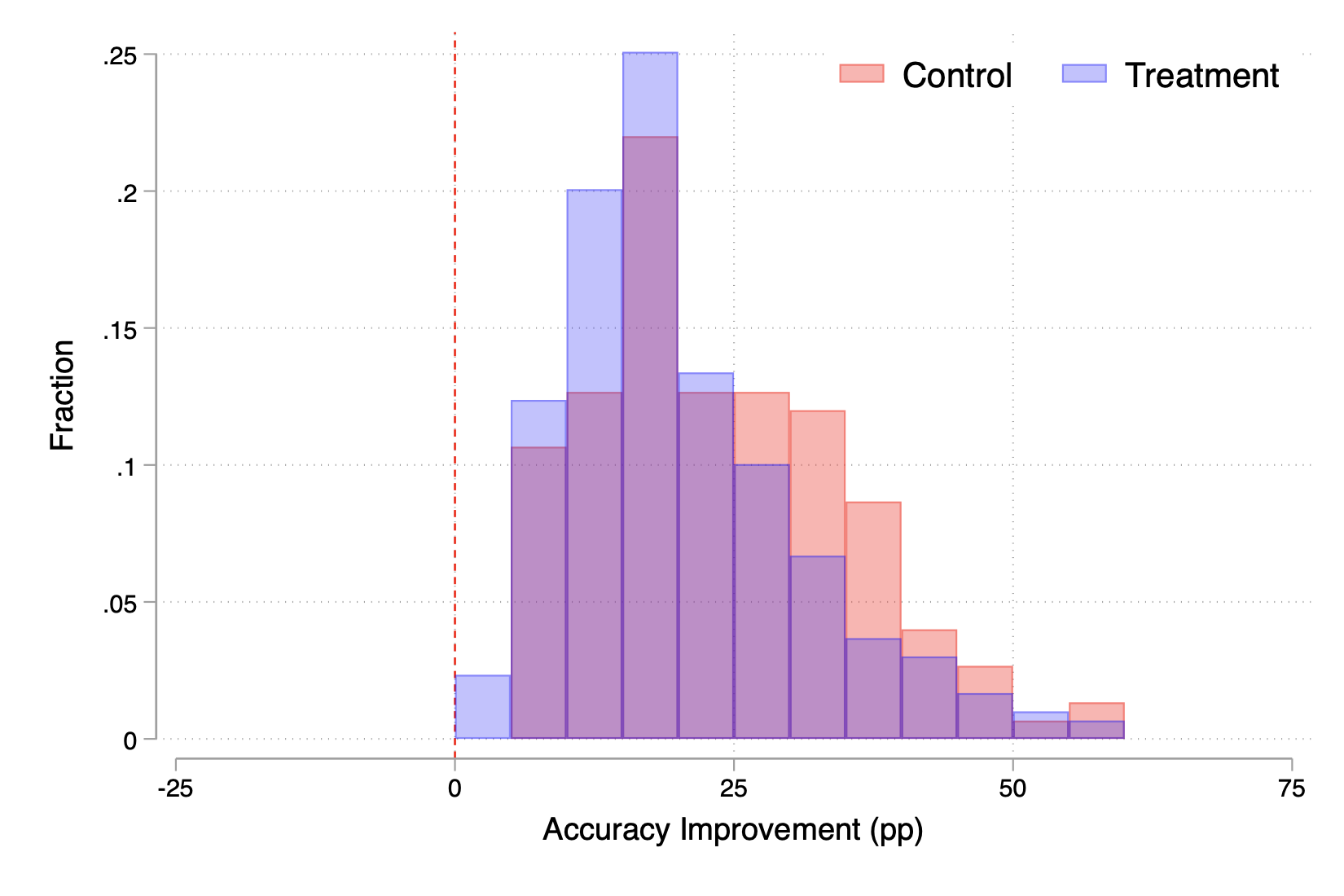}
\caption{Accuracy Gains from AI Recommendations.}
\label{fig:GainsDist}
\end{subfigure}

    \vspace{0.5em}
    \begin{minipage}{0.9\textwidth} 
        \footnotesize \textit{Notes}: Panel (a) shows workers’ accuracy before receiving AI recommendations, with the red dashed vertical line indicating the AI’s accuracy. Panel (b) displays the distribution of accuracy improvements from incorporating AI recommendations relative to workers’ initial answers, distinguishing control workers (red) from treatment workers (blue).
    \end{minipage}
\label{fig:PerfDist}
\end{figure}

The potential for AI–human collaboration lies in what humans and AI can achieve together. Although no worker in this setting was individually more accurate than the AI (Figure \ref{fig:BeforeDist}), this does not imply the absence of collaborative gains. Figure \ref{fig:GainsDist} shows the distribution of performance improvements per worker after considering the AI recommendations, highlighting two key findings: (i) every worker benefited from AI assistance, and (ii) the distribution of performance gains is more skewed to the right in the control group than in the treatment group, consistent with the larger improvements when AI reliance was private.

A natural benchmark for defining successful collaboration is whether workers, with AI assistance, surpass the AI’s standalone accuracy of 85\%. Figure \ref{fig:Share} shows that the share of workers meeting the benchmark falls from 24.7\% to 18.5\% when AI reliance becomes visible to the HR specialist. Put differently, one out of every four workers who would otherwise qualify as a successful collaborator is lost once AI reliance becomes visible.

\subsection{Mechanisms}
Understanding why workers view AI reliance as damaging to their image is crucial, since addressing this perception is key to recouping the productivity losses it generates. To investigate the underlying mechanism, I developed a novel incentive-compatible method that leverages a core feature of Upwork, \textit{public feedback}, which is also common across many online platforms.

After completing the categorization task, all workers were informed that they would receive feedback intended to be informative for future employers on the platform. The feedback described the AI-assisted categorization task they had performed, ensuring that it provided sufficient context on its own. Workers were then given the opportunity to choose among three positive but mutually exclusive traits to emphasize in their feedback. Specifically, they could select a statement highlighting their effort (signaling that they are hard workers), their skills in this type of task, or their confidence in their own judgment.\footnote{These three traits emerged as the most common in a prior survey and are also prominent in public discourse. Respondents who used the open-ended option typically referred to one or more of these traits, often in overlapping ways. To avoid ambiguity, I restricted the choice to only these three, which ensured clarity and comparability across responses.} For workers in the control group, their choice of statement reflects baseline preferences over these traits. By contrast, workers in the treatment group were told that the feedback would also report whether they used more or less AI than the average worker. They were directed to view their selected statement as an opportunity to highlight a trait that might otherwise be overlooked if their feedback identified them as high AI users. Thus, for treated workers, the distribution of chosen statements reflects preferences under a scenario where image concerns about being identified as a high AI user are salient.

\begin{figure}[h!]
\centering
\caption{Feedback Preference}
\includegraphics[width=5in]{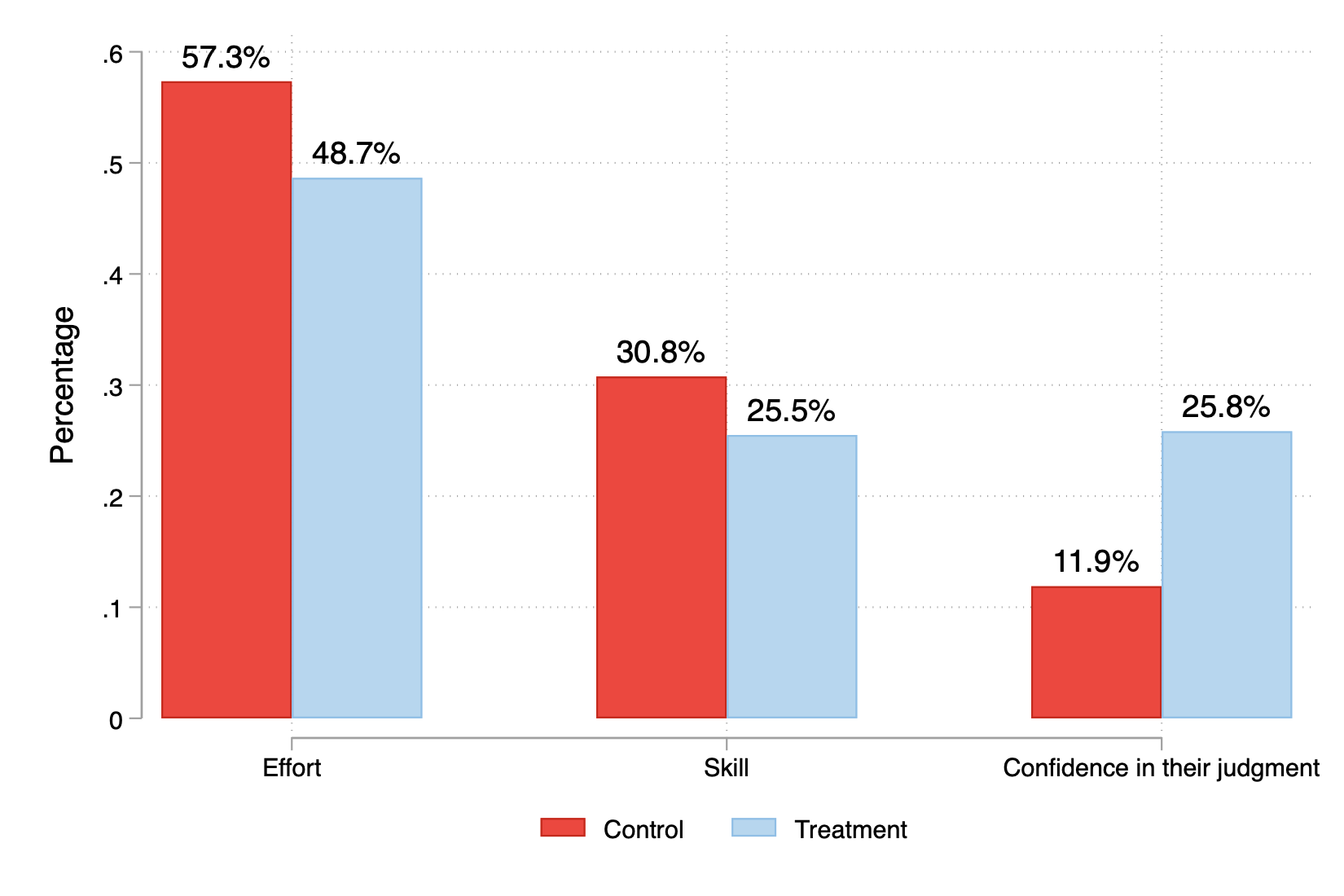}
\label{fig:Feedback}
\end{figure}

Figure \ref{fig:Feedback} presents the distribution of preferred statements by treatment status. In the control group, the majority of workers (57.3\%) preferred to signal effort, followed by skill (30.8\%), with confidence in judgment ranking last (11.9\%). In the treatment groups, while signaling effort remained the most common choice (48.7\%), both effort and skill declined as preferences shifted toward signaling confidence in judgment (25.8\%), representing a 117\% increase relative to the control group. These results provide to two main insights. First, many workers consistently value being perceived as hard-working, regardless of whether AI reliance is disclosed, which is reasonable given that monitoring remote workers is challenging and effort tends to extrapolate well to other tasks. Second, signaling confidence in their own judgment, which is relatively unpopular when AI reliance is not disclosed in the feedback, becomes much more relevant once workers worry that heavy AI reliance may be visible to future employers.

\begin{figure}[h!]
\centering
\caption{Self-reported Importance of Signals.}
\includegraphics[width=5in]{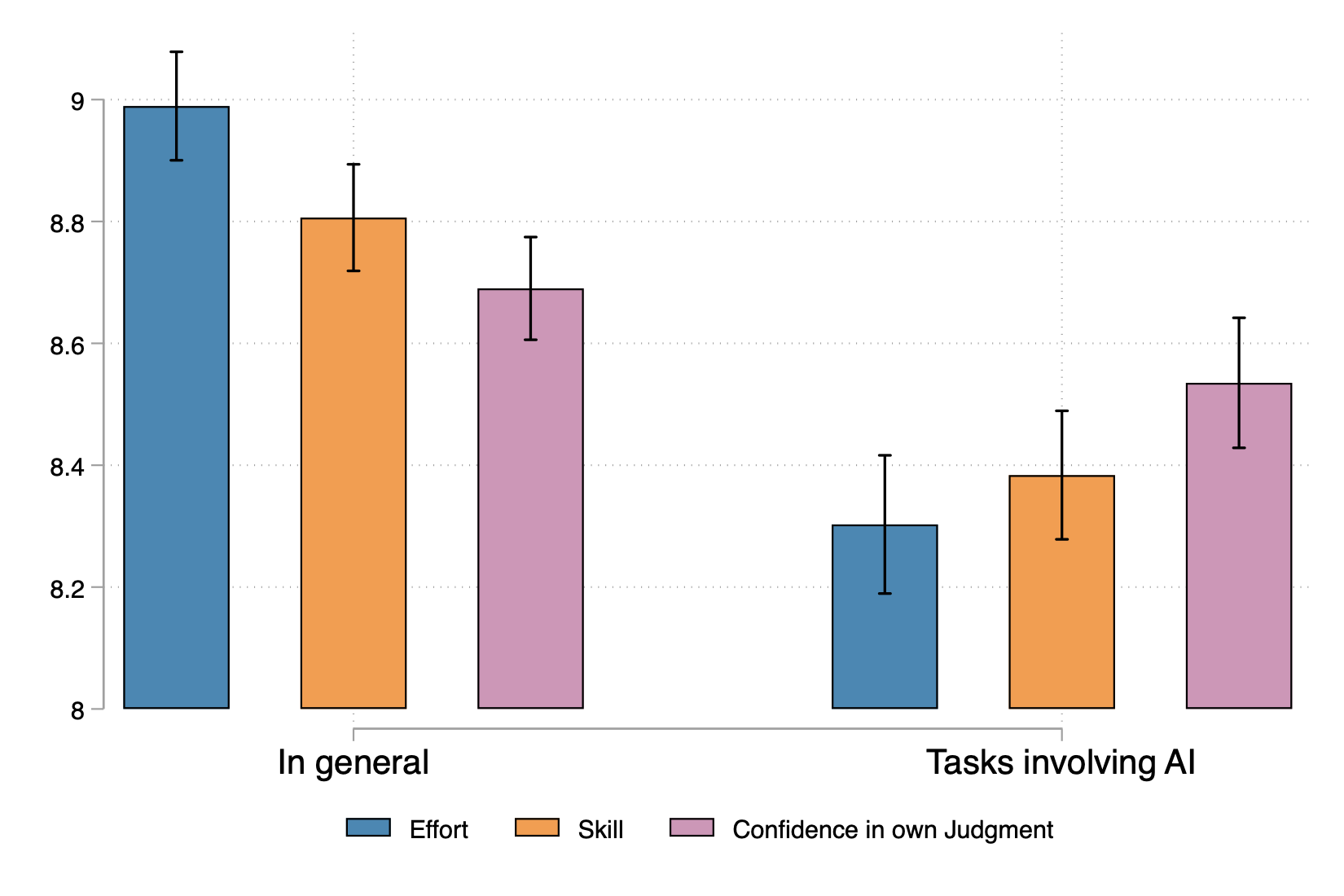}
\label{fig:Sliders}
\end{figure}

To the best of my knowledge, this is the first use of public feedback to elicit preferences in an incentive-compatible way. To validate these findings, I also included two more direct, though non-incentivized, questions in the second job questionnaire. I asked 284 returning workers to report, using a slider ranging from 1 (not important at all) to 10 (extremely important), how important it is to signal each of the same three positive traits to employers on Upwork. The question was asked twice: first in a general setting, and then specifically for tasks involving the use of AI. Figure \ref{fig:Sliders} presents the results. The first question, situated in a general setting, replicates the ranking observed in the control group: effort first, followed by skill, and lastly confidence in judgment. When the second question referred to tasks involving AI, the ranking reversed, consistent with the shift we observe in treated workers’ feedback preferences toward signaling confidence in their judgment. These questionnaire results are reassuring, suggesting that the incentive-compatible elicitation method was well understood by the workers and reflective of underlying preferences for these signals.

\section{Discussion: How Hard Is It to Overcome Image Concerns?}\label{hard}

This section presents evidence on why overcoming image concerns is difficult. I begin with the evaluation-process instructions, which are a good starting point for illustrating the strength of the results, since they were constructed to mute potential channels through which image concerns could influence outcomes. Workers were explicitly told that the HR specialist would evaluate them based on which workers were expected to perform most accurately if they were to return to the same AI-assisted image categorization task. This instruction highlights two components that work against the observed treatment effects.

First, while task accuracy is a central criterion in many jobs, evaluations are often shaped by additional factors. In our evaluation process, however, we instructed evaluators to focus solely on expected accuracy, thereby minimizing the potential influence of image concerns.

Second, in typical jobs workers often face a wide array of tasks or uncertainty about how the displayed task may evolve over time. In such situations, image concerns could become more relevant, since appearing skilled might extrapolate better to other domains or future tasks. Here, by insisting that the evaluation was tied specifically to predicting accuracy in the \emph{same} AI-assisted categorization task, we removed the possibility of image concerns playing a broader role.

This design suggests that the observed results likely understate the role of image concerns, which may be amplified when either the evaluation criteria or the task itself are less well defined. By structuring the experiment in this way, I show that image concerns emerge even in the absence of such considerations. The subsections that follow provide further evidence on why overcoming image concerns may be particularly challenging.

\subsection{Null Effect of the Information Intervention}

Earlier in the paper, I justified pooling the two treatment conditions in which AI reliance was public, as the information intervention had no effect on the main outcomes: response times, accuracy, and AI reliance.\footnote{Table \ref{tab:Results_Separate} reports $p$-values from Wald tests of equality between the \textit{Public} and \textit{Public with Information} treatments. All $p$-values are high, providing no evidence of distinct effects across any of the outcome variables of interest.}

The intervention was motivated by the idea that AI reliance might reveal something about a worker’s type in a context where the HR specialist had very limited information about the workers. In recruitment, HR specialists are often assumed to weigh traits beyond raw performance, particularly in remote work settings. Attributes such as reliability, integrity, work ethic, accountability, communication, and punctuality are central when direct supervision is limited, and these may plausibly correlate with reliance on AI.

To assess whether image concerns were driven by information gaps, workers in the information condition were told that the HR specialist had already been assured of their strong track record on the platform and that there was no reason to doubt their quality. As noted in the pre-analysis plan, the expectation was that the information intervention would partially mitigate the treatment effects resulting from making AI reliance visible, leading to levels of AI reliance and accuracy that would fall between those of the control and the standard public-treatment groups. Instead, their behavior was indistinguishable from the latter, indicating that the additional reassurance did not mitigate image concerns.

In longer relationships, workers may worry less about signaling through AI reliance, since more information about their type accumulates naturally. The experiment, by contrast, captures a short-term work environment, so the information intervention was designed to approximate such a scenario by artificially closing the information gap. Yet the results suggest that even when workers are confident their quality is recognized, image concerns persist.

Questionnaire responses help provide intuition for the null effect of the information intervention. Workers frequently noted that the reassurance raised the HR specialist’s expectations, which they sought to meet by using less AI.

As an additional check, I examine whether workers read the instructions carefully. Figure \ref{fig:Time_Instructions} shows that average reading times for the screen containing the treatment variation were longer in the treatment conditions, and longest in the \textit{Public with Information} condition, which included an additional paragraph. Reading times for the other instruction screens did not differ significantly across conditions.

\subsection{Origins of Beliefs: Workers Also Penalize AI Use}
The reduction in AI reliance reflects workers’ beliefs about the HR manager’s evaluation criteria. Specifically, workers anticipate that HR managers will assign lower scores to those who rely more heavily on AI. One possible explanation is that workers project their own beliefs onto the HR manager; another is that workers’ beliefs about the HR manager’s actions diverge from what they themselves would do. The latter possibility opens the door for a misperception equilibrium à la \cite{Saudi2020}, who present evidence that simple information interventions can be effective in addressing such distortions. In what follows, I present evidence consistent with the idea that workers were projecting how they themselves would evaluate others when choosing to rely less on AI when visible.

To better understand where workers’ beliefs about HR evaluation come from, I asked returning workers from the treatment groups to rank 20 real profiles of other workers who were also selected for the second job.\footnote{The 20 profiles combined accuracy levels of 60, 70, 80, and 90 with AI use ranging from 10 to 60 in increments of 10 (Not all combinations were feasible).} Each profile displayed the worker’s accuracy rate and AI usage from the first job, and participants were monetarily incentivized to score them based on the number of correct answers those 20 workers achieved in the second job.

\begin{figure}[h!]
\centering
\caption{Accuracy/AI Reliance Pairs Ordered by Average Ranking.}
\includegraphics[width=5in]{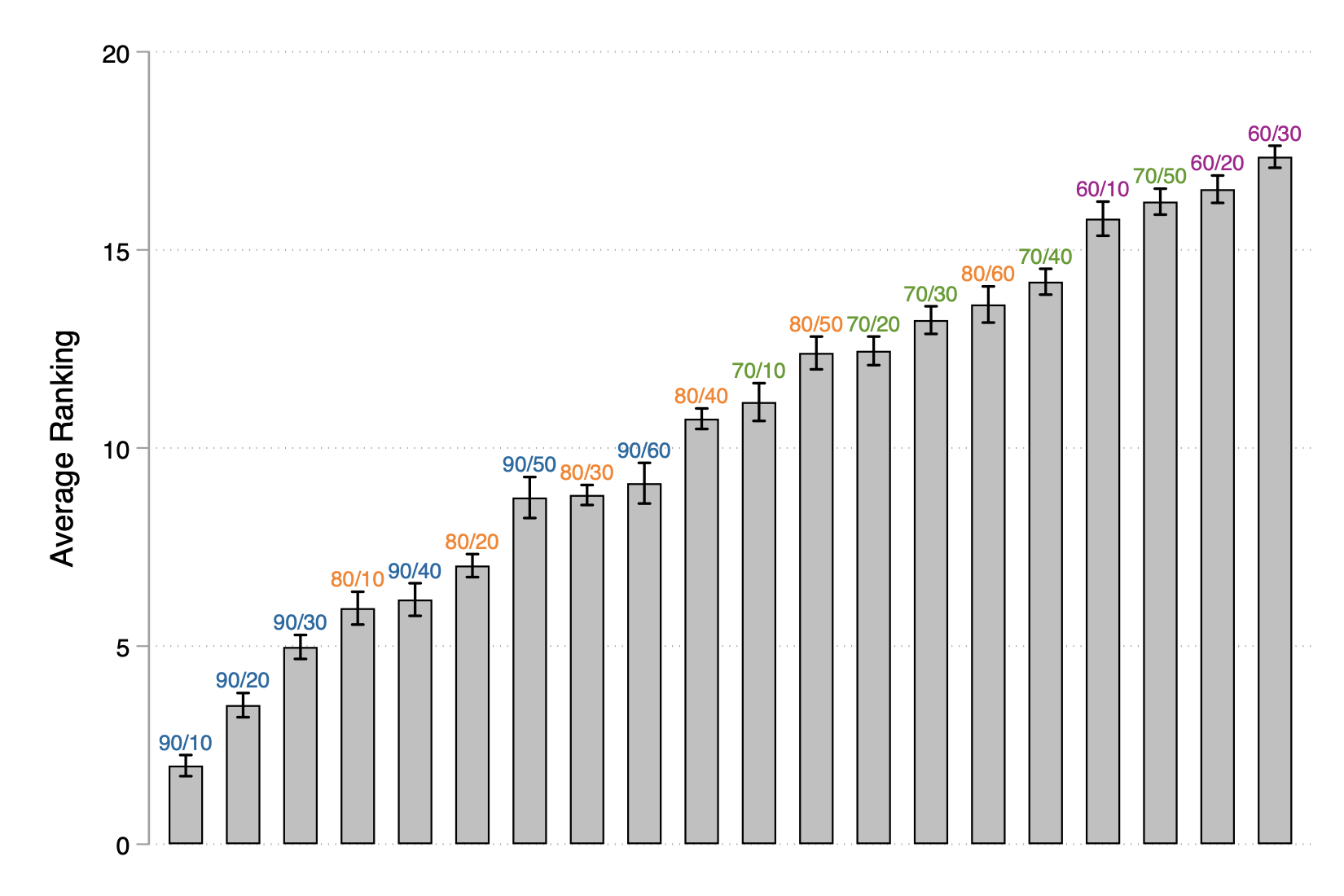}
\label{fig:Rankings_sorted}

    \vspace{0.5em}
    \begin{minipage}{0.9\textwidth} 
        \footnotesize \textit{Notes}: Each bar shows the average ranking of a profile with standard errors. Labels display the profile pair (Accuracy/AI reliance) and are color–coded according to accuracy.
    \end{minipage}
\end{figure}

I find robust evidence that workers penalize AI reliance when assuming the role of an HR manager and evaluating others’ profiles. To illustrate this, I examine how accuracy and AI reliance affect the scores and rankings assigned to profiles. Table \ref{tab:Evaluations} reports regression results where the dependent variable is either the score (ranging from 0 to 100) or the ranking of a profile (1 = best, 20 = worst), modeled as a function of the worker profile’s accuracy and their AI reliance. The specification includes evaluator fixed effects, and standard errors are clustered at the evaluator level. A one–percentage point increase in accuracy raises the score by about one point, whereas each additional percentage point of AI reliance reduced the score by 0.36 points. Both effects are statistically significant at the 1\% level. Because the accuracy coefficient is close to 1, the interpretation is straightforward: The weight placed on AI reliance is roughly 36\% of that placed on accuracy, but with the opposite sign. Put differently, evaluators penalize the adoption of three additional AI recommendations (–1.08 points) more than a single incorrect answer. The results for rankings closely mirror those for scores. Figure \ref{fig:Rankings_sorted} shows the 20 profiles sorted by average ranking. Although accuracy is the main driver, the figure illustrates that profiles with 10 percentage points lower accuracy can still achieve better rankings if they rely less on AI.\footnote{Figure \ref{fig:2Drankings} provides a two-dimensional, intuitive depiction of the average rankings.}

\begin{table}[h!]
    \centering
    \caption{Effect of Worker Accuracy and AI Reliance on Evaluations.}
    \label{tab:Evaluations}
    \sisetup{table-format=1.3}  
    \begin{tabular}{lSSS}
        \toprule
        & {Score} & & {Ranking} \\
        & {(1)} &  & {(2)} \\
        \midrule
        \textbf{Accuracy (pp)} & \textbf{1}*** & & \textbf{-0.43}***  \\\vspace{1em}
        & {(0.044)} & & {(0.015)} \\

         \textbf{AI reliance (pp)} & \textbf{-0.36}*** & & \textbf{0.142}*** \\\vspace{1em}
        & {(0.042)} & & {(0.013)}  \\   
        
        Individual FE & X & & X \\
        Constant & {-4.35} & & {39.4} \\
        Observations & {1,860} &  & {1,860}  \\
        \bottomrule
    \end{tabular}
    
    \vspace{0.5em}
    \begin{minipage}{0.9\textwidth} 
        \footnotesize \textit{Notes}: Each dependent variable is regressed on the evaluated worker’s accuracy and AI reliance rate, with evaluator fixed effects included and standard errors clustered at the evaluator level.\\
        * \(p<0.10\), ** \(p<0.05\), *** \(p<0.01\).
    \end{minipage}
\end{table}

There is also strong evidence at the individual level that evaluators penalize AI reliance. Following a similar empirical strategy as in the aggregate analysis, I regress each evaluator’s assigned scores on accuracy and AI reliance separately for 93 evaluators. Of these, 70 had a coefficient on AI reliance that was statistically significant at the 10\% level. Notably, 69 of the 70 significant coefficients were negative, indicating that nearly all evaluators with sufficient statistical power penalized AI reliance in their assigned scores. Figure \ref{fig:coefdist} shows the distribution of these statistically significant coefficients associated with AI reliance.

The evidence that workers penalize AI reliance when assuming the role of evaluators is particularly striking, given that these workers had already experienced what it feels like to be judged for using AI. Prior research shows that perspective-taking exercises can shift attitudes toward other groups \citep{Alan2021}, with recent evidence emphasizing relatability as a mechanism \citep{Andries2025}. Yet even the stronger intervention of directly experiencing the worker role did not diminish workers’ inclination to penalize AI reliance in others.

\section{Conclusion}\label{sec:conclusion}

This paper demonstrates that image concerns are a meaningful barrier to the adoption of AI in the workplace. In a field experiment on a large online labor marketplace, workers competing for contract extensions reduced their reliance on AI when its use was observable to an HR evaluator. This decline was not offset by greater effort or improved judgment about when to follow AI recommendations, leading to lower performance even though evaluators could directly observe these losses. As a result, the prospect of successful collaboration is diminished: by our benchmark (performing better than AI), one out of every four potential successful collaborations is lost when AI reliance is made visible.

Methodologically, the paper introduces a novel incentive-compatible elicitation based on platform feedback, offering a new way to study signaling motives and underlying mechanisms in digital labor markets. The results reveal that workers fear visible reliance on AI signals a lack of confidence in their judgment, a trait they view as more consequential than effort or skill when tasks involve AI assistance.

The broader implication is that the productivity promise of AI cannot be realized by improving algorithms alone. Institutions and organizations must also address the social meaning attached to using AI, whether by reframing reliance as a sign of adaptability, embedding AI more seamlessly into workflows, or reducing the visibility of individual choices. This paper provides evidence of how difficult it can be to mute these social channels in practice, and future research should seek solutions that do so. Without such efforts, workers may continue to underutilize AI not only because they doubt its accuracy or struggle to use it, but also because of what they fear its use reveals about them. Progress in developing better AI must therefore go hand in hand with careful attention to implementation if its full benefits are to be realized.

\clearpage
\bibliographystyle{plainnat} 
\bibliography{bibliography}  
\clearpage

\appendix
\counterwithin{figure}{section}
\counterwithin{table}{section}
\section{Additional Tables and Figures}

\begin{figure}[h!]
\centering
\caption{Illustrative Scenarios of AI-Human Collaboration}
\includegraphics[width=4.9in]{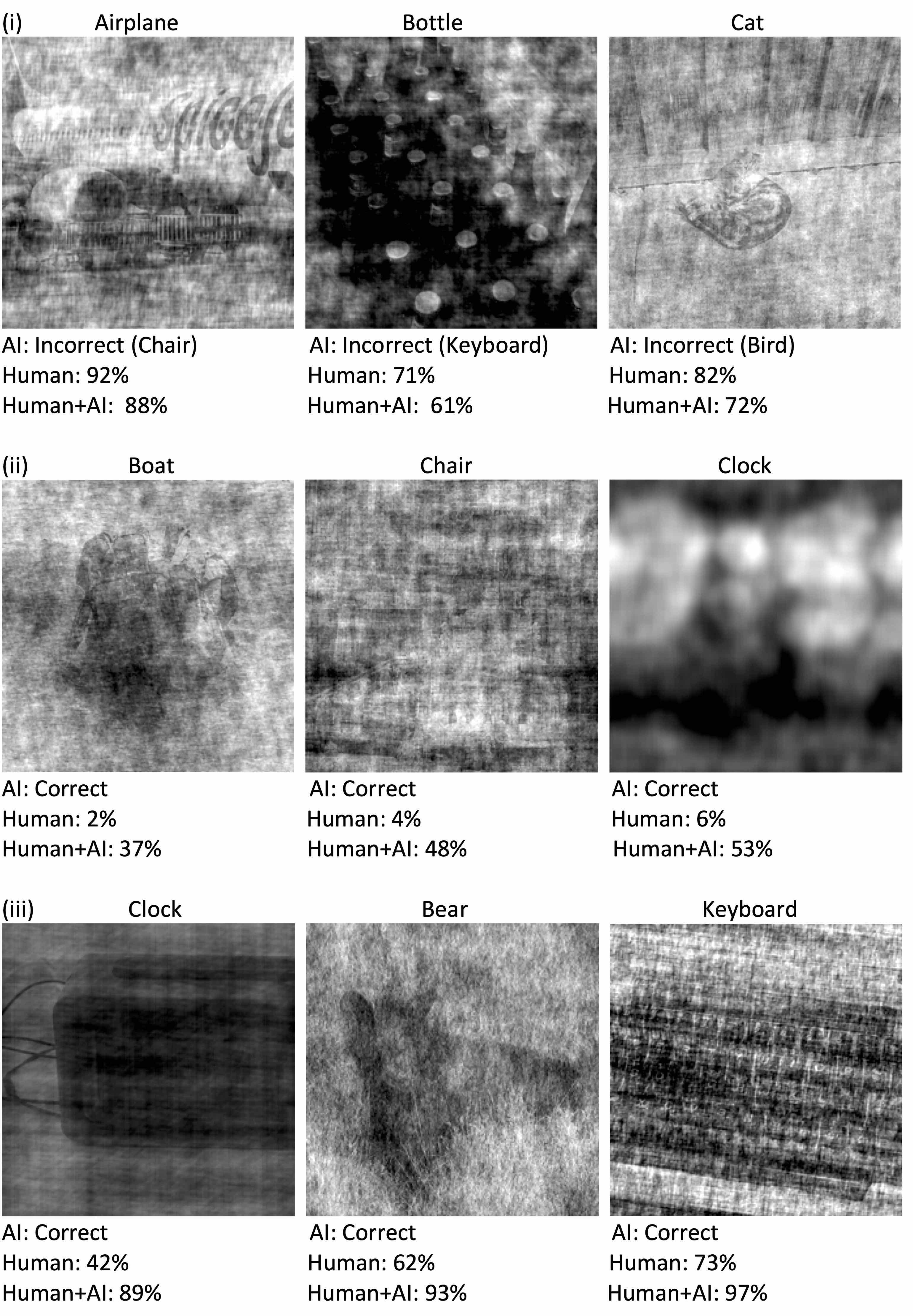}

    \vspace{0.5em}
    \begin{minipage}{0.9\textwidth} 
        \footnotesize \textit{Notes}: Images show the ground truth (top), the AI’s accuracy, and humans’ accuracy before and after seeing the AI recommendation. Each row illustrates a collaboration scenario: (i) Humans are mostly correct, but the AI is wrong. (ii) The AI is correct, and humans rarely are, so following it is essentially blind delegation. (iii) Humans are often correct, and many improve after reconsidering the image through the AI’s suggested lens.
    \end{minipage}
\label{fig:AppendixImages}
\end{figure}

\clearpage

\begin{figure}[h!]
\centering
\caption{Self-Reported First-Session Goal About HR Specialist’s Score}
\includegraphics[width=5in]{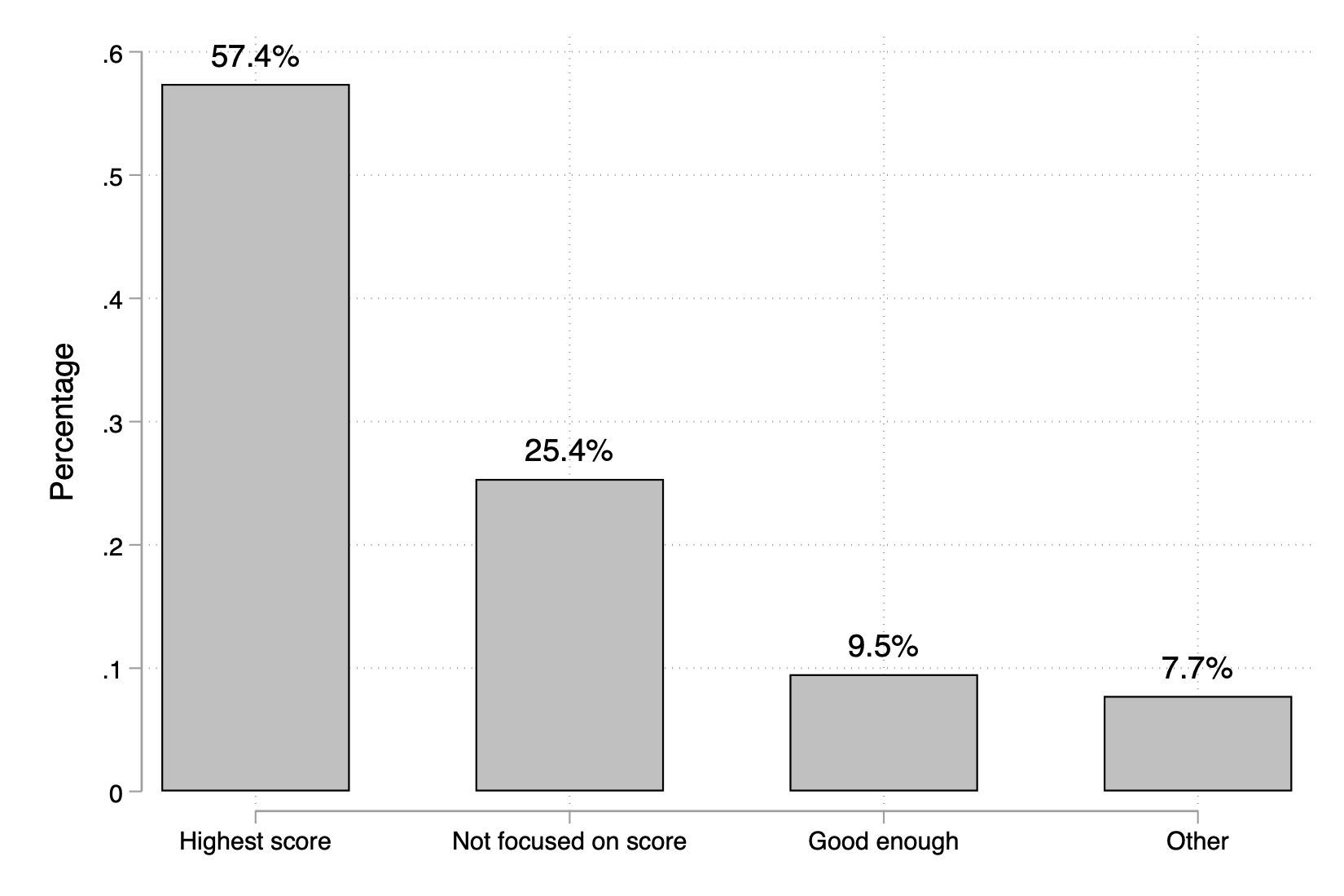}

    \vspace{0.5em}
    \begin{minipage}{0.9\textwidth} 
        \footnotesize \textit{Notes}: Based on answers from 284 workers. Among those who selected “not focused on score” or “other” and provided an explanation, 92\% gave a response consistent with maximizing (doing the best possible), while only 2\% mentioned a reason aligned with being good enough to get rehired.
    \end{minipage}
\label{fig:Maximize}
\end{figure}

\begin{figure}[htbp]
\centering
\caption{Treatment Effect by 10-Round Blocks.}
\begin{subfigure}[b]{.49\textwidth}
\centering
\includegraphics[width=3.1in]{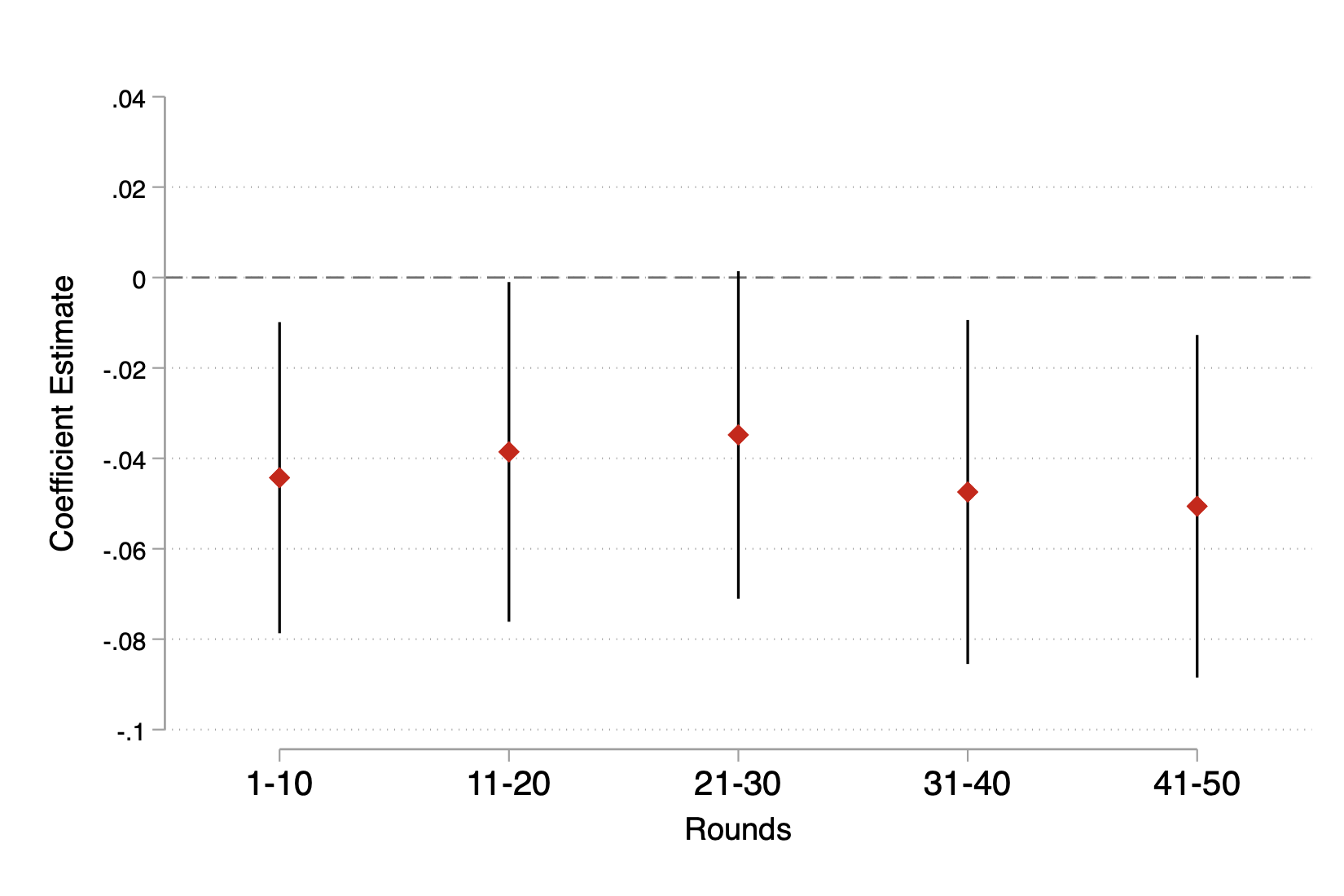}
\caption{AI Recommendation Reliance.}
\end{subfigure}
\begin{subfigure}[b]{.49\textwidth}
\centering
\includegraphics[width=3.1in]{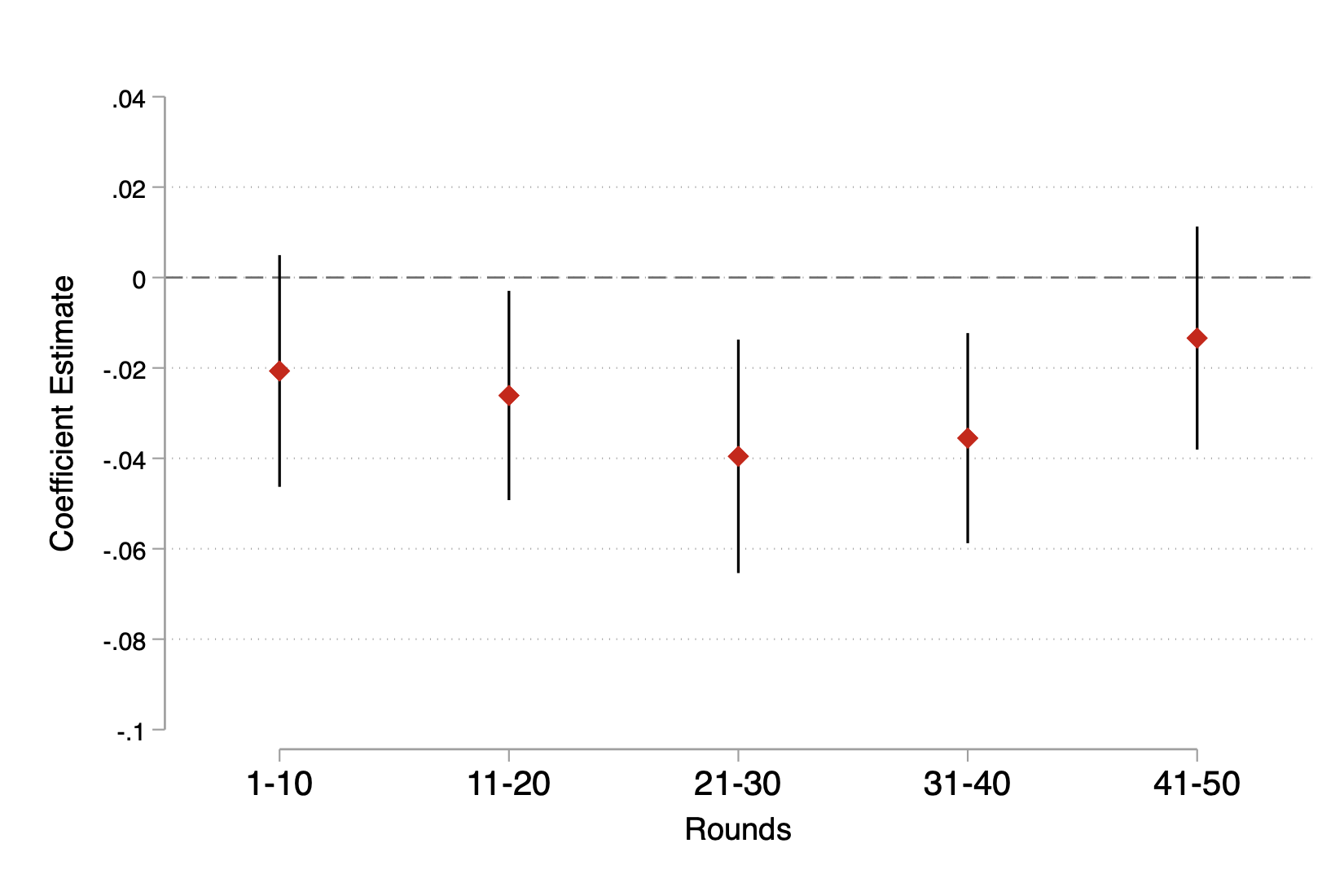}
\caption{Accuracy.}
\end{subfigure}

    \vspace{0.5em}
    \begin{minipage}{0.9\textwidth} 
        \footnotesize \textit{Notes}: Each dot represents the coefficient from an interaction between treatment status and a 10-round block indicator. Image-specific fixed effects are included. Standard errors are clustered at the worker level.
    \end{minipage}
\label{fig:RoundsTE}
\end{figure}

\begin{figure}[h!]
\centering
\caption{Accuracy When Following vs. Not Following AI Recommendations.}
\includegraphics[width=5in]{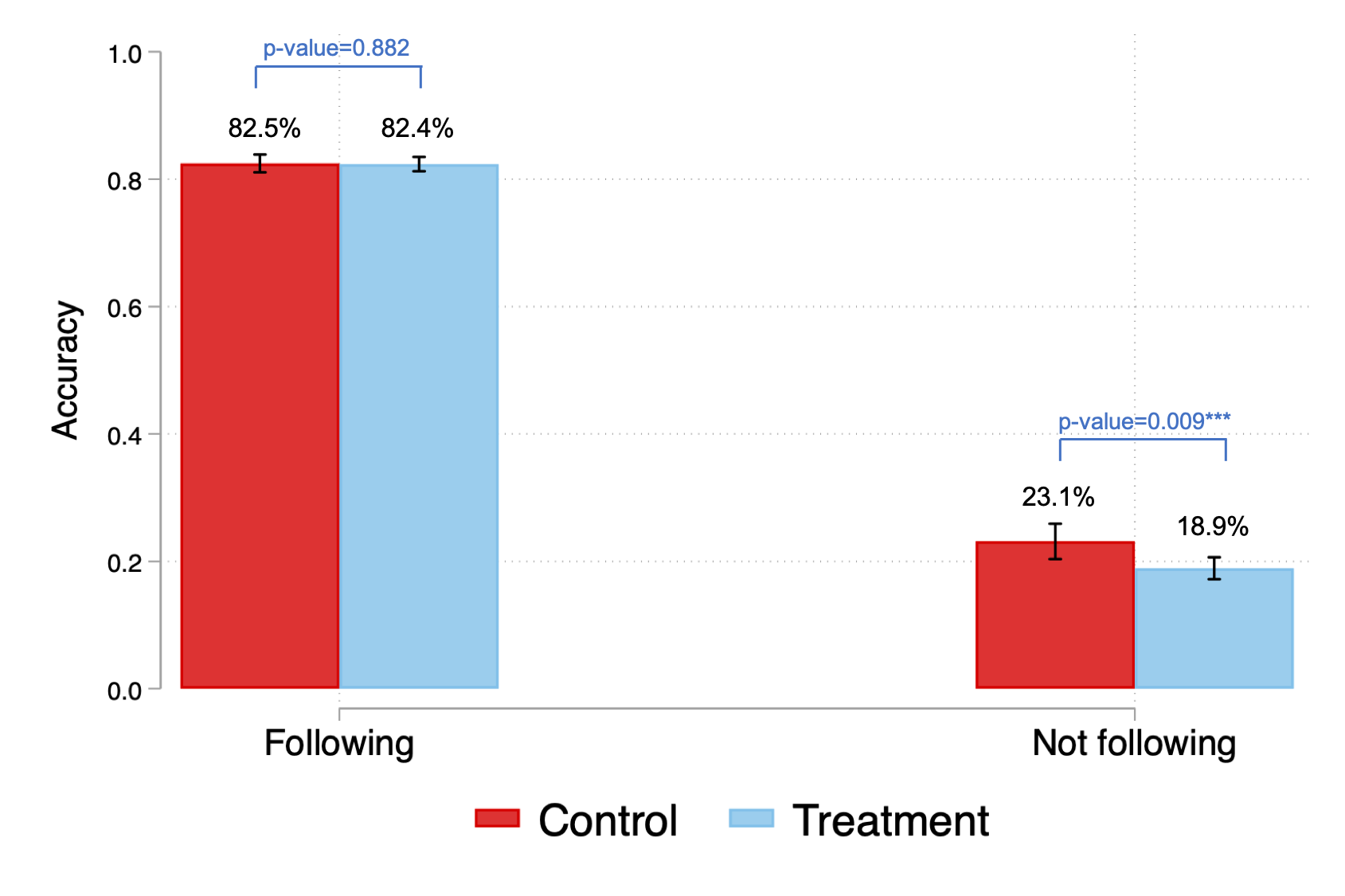}

    \vspace{0.5em}
    \begin{minipage}{0.9\textwidth} 
        \footnotesize \textit{Notes}: Image-specific fixed effects are included, and standard errors are clustered at the worker level.
    \end{minipage}
\label{fig:ByFollow}
\end{figure}

\begin{figure}[h!]
\centering
\caption{Gains from Working with AI.}
\includegraphics[width=5in]{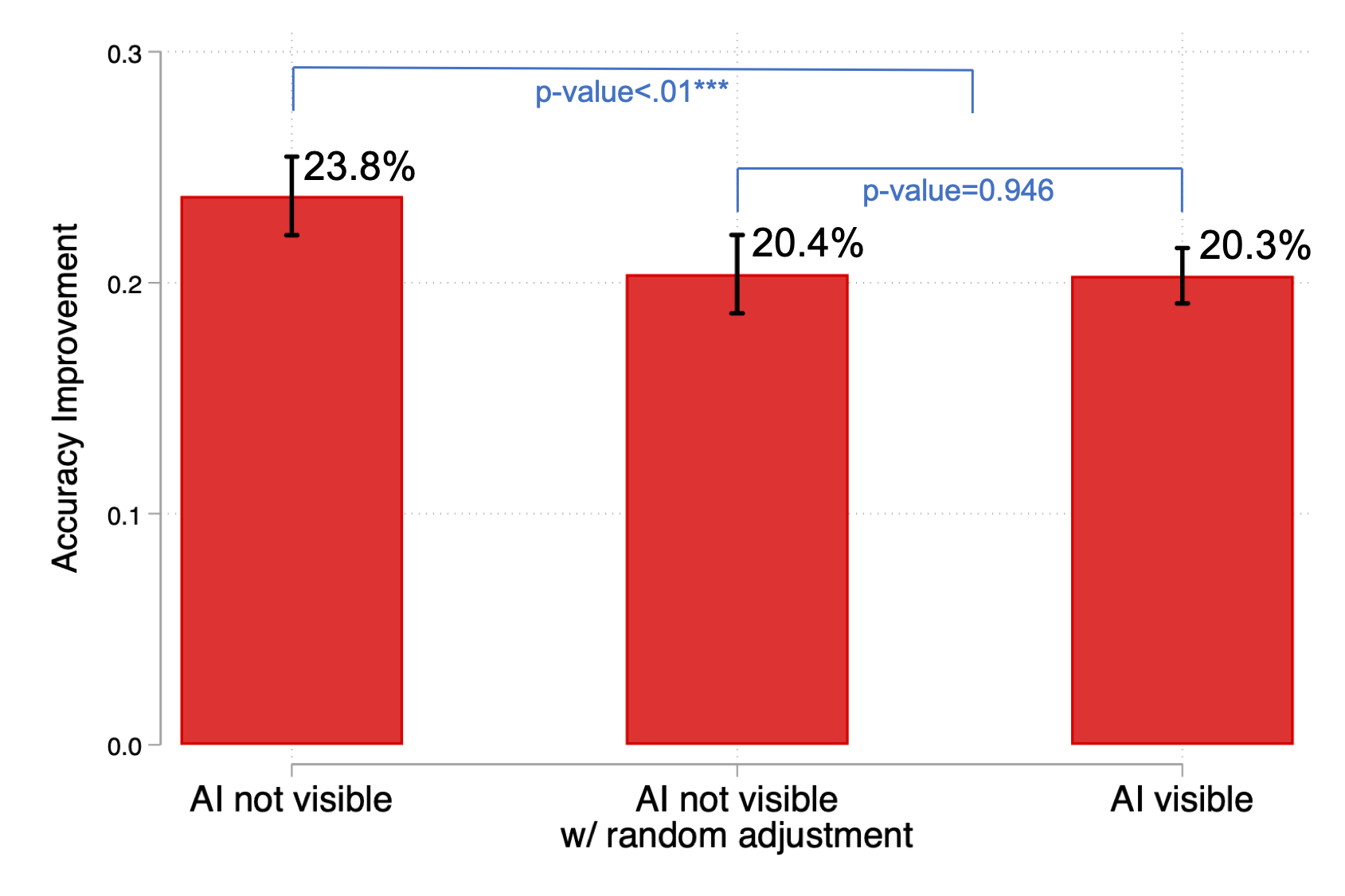}
    \vspace{0.5em}
    \begin{minipage}{0.9\textwidth} 
        \footnotesize \textit{Notes}: This graph plots the average accuracy gains per worker from having access to AI recommendations, measured as the improvement from workers’ initial choices to their final answers after seeing the AI suggestion. The first bar reports the average gain for the control group, in which AI reliance was not visible to the HR evaluator. The second bar reports the same control group after a random adjustment that lowered their AI reliance to match the treatment group: specifically, 14\% of accepted AI recommendations were randomly overruled. The final bar shows the average gain for the treatment group, where AI reliance was visible to the HR evaluator.
    \end{minipage}
\label{fig:Random}
\end{figure}

\begin{figure}[htbp]
\centering
\caption{Heterogeneous Treatment Effects by Demographics}
\begin{subfigure}[b]{.49\textwidth}
\centering
\includegraphics[width=3.1in]{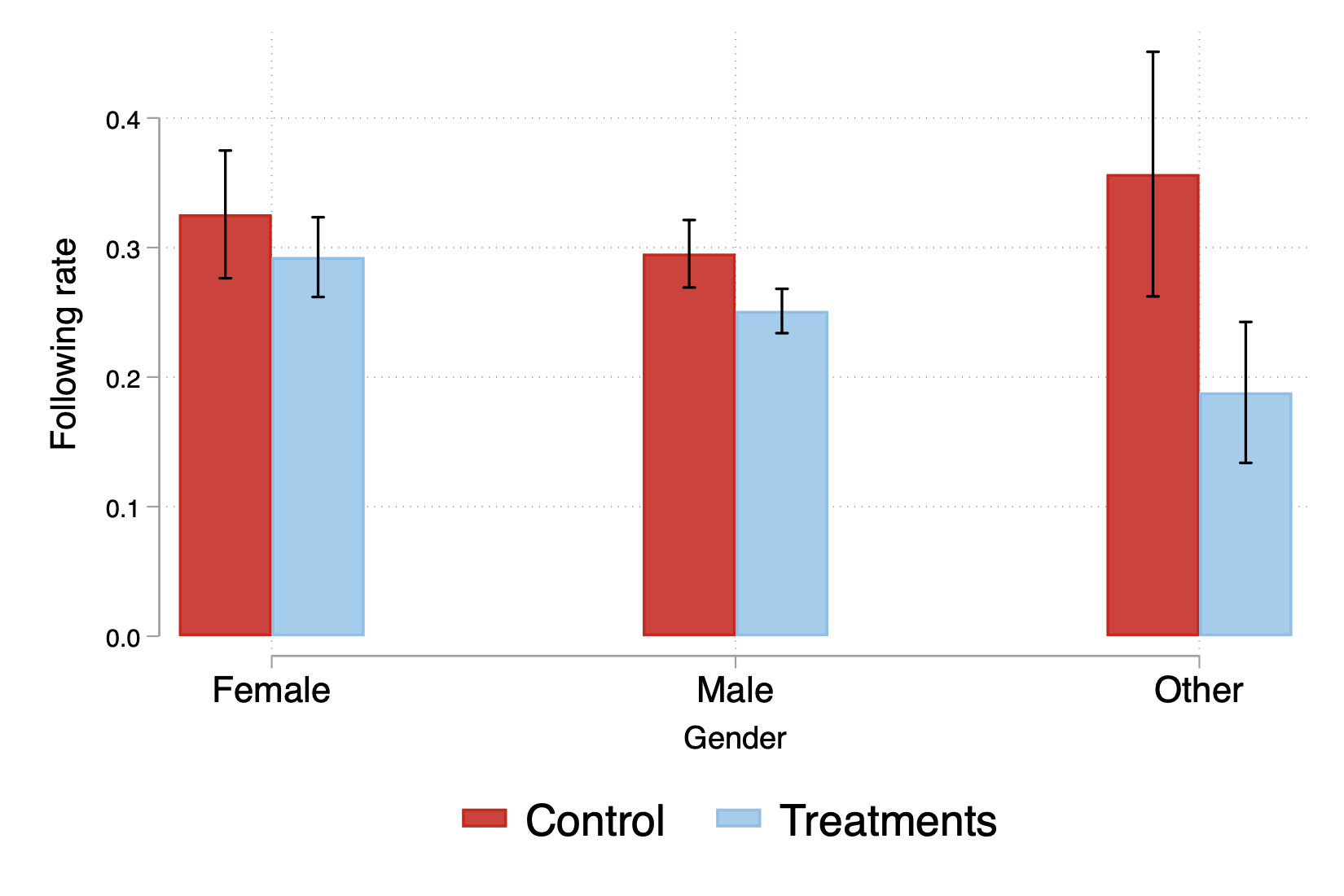}
\caption{Gender: AI Recommendation Reliance.}
\end{subfigure}
\begin{subfigure}[b]{.49\textwidth}
\centering
\includegraphics[width=3.1in]{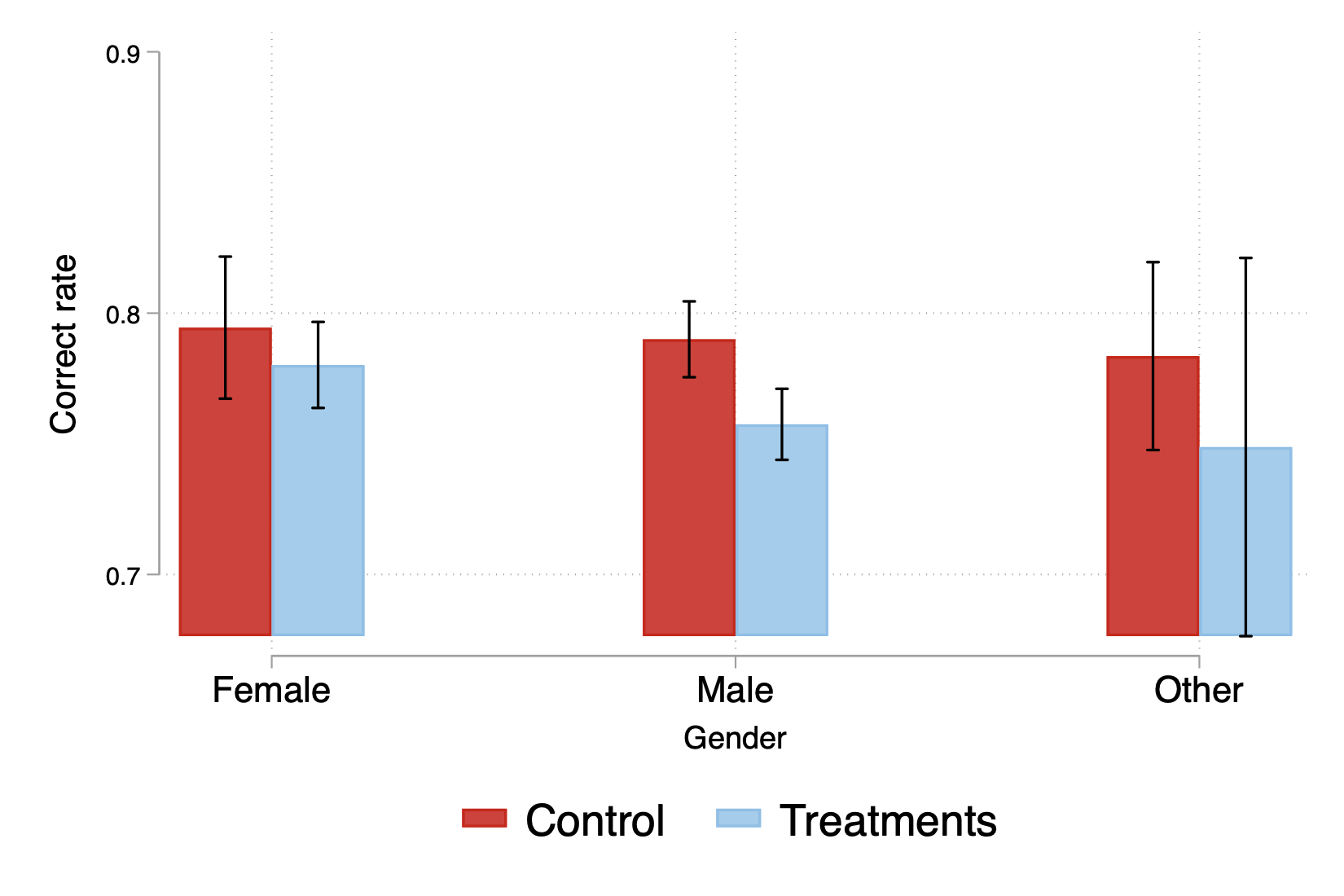}
\caption{Gender: Accuracy.}
\end{subfigure}

\vspace{1.5em} 

\begin{subfigure}[b]{.49\textwidth}
\centering
\includegraphics[width=3.1in]{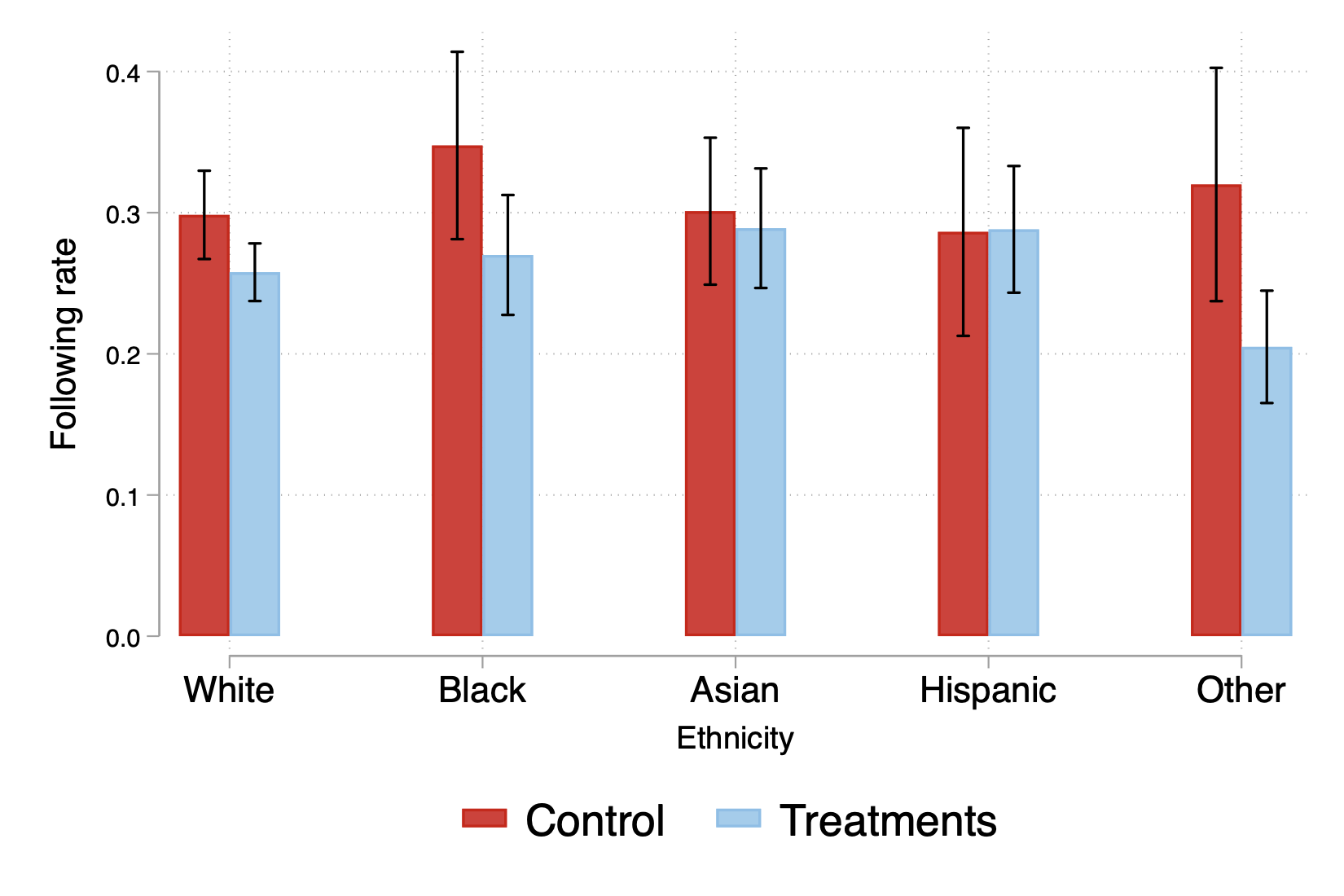}
\caption{Ethnicity: AI Recommendation Reliance.}
\end{subfigure}
\begin{subfigure}[b]{.49\textwidth}
\centering
\includegraphics[width=3.1in]{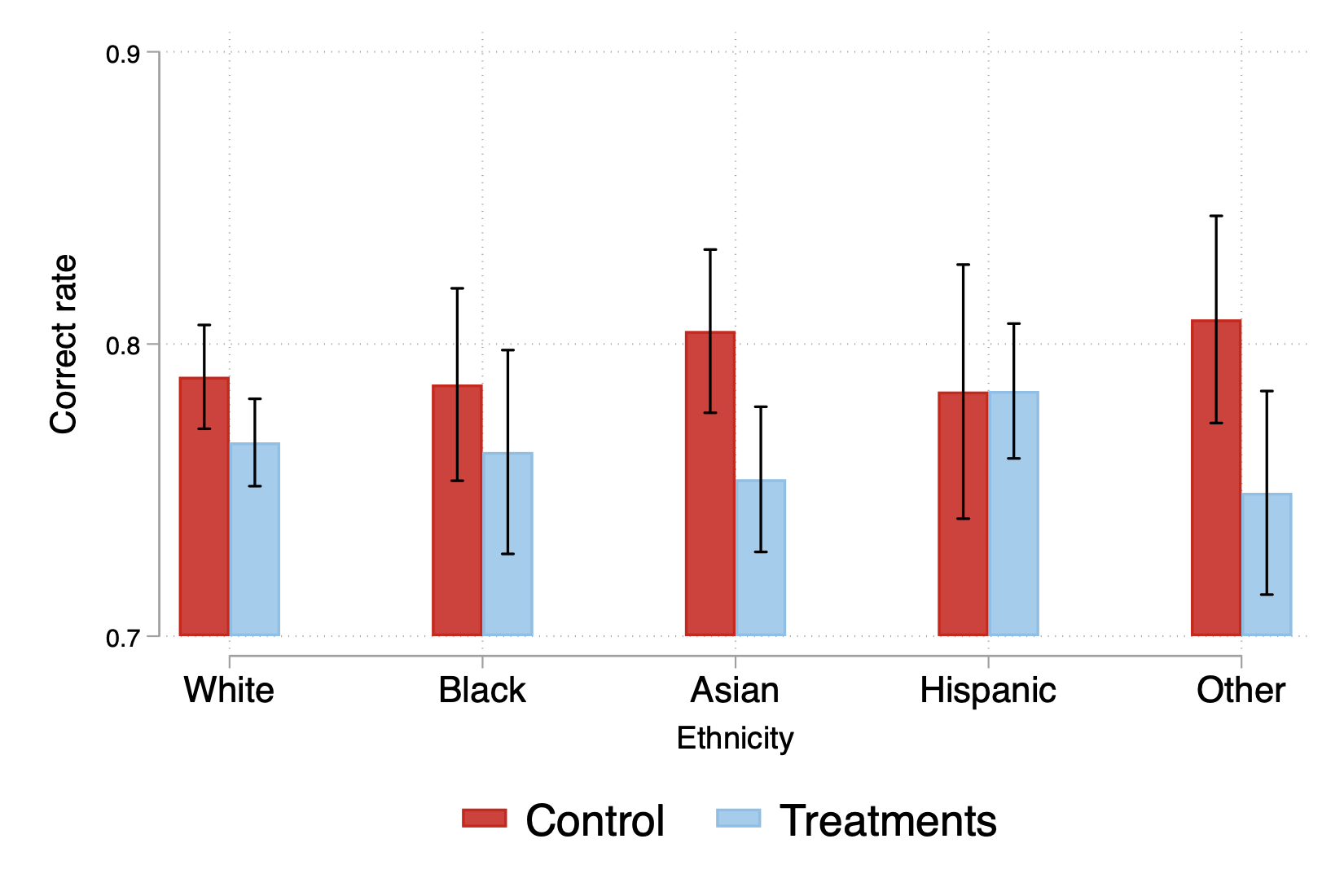}
\caption{Ethnicity: Accuracy.}
\end{subfigure}

\vspace{1.5em} 

\begin{subfigure}[b]{.49\textwidth}
\centering
\includegraphics[width=3.1in]{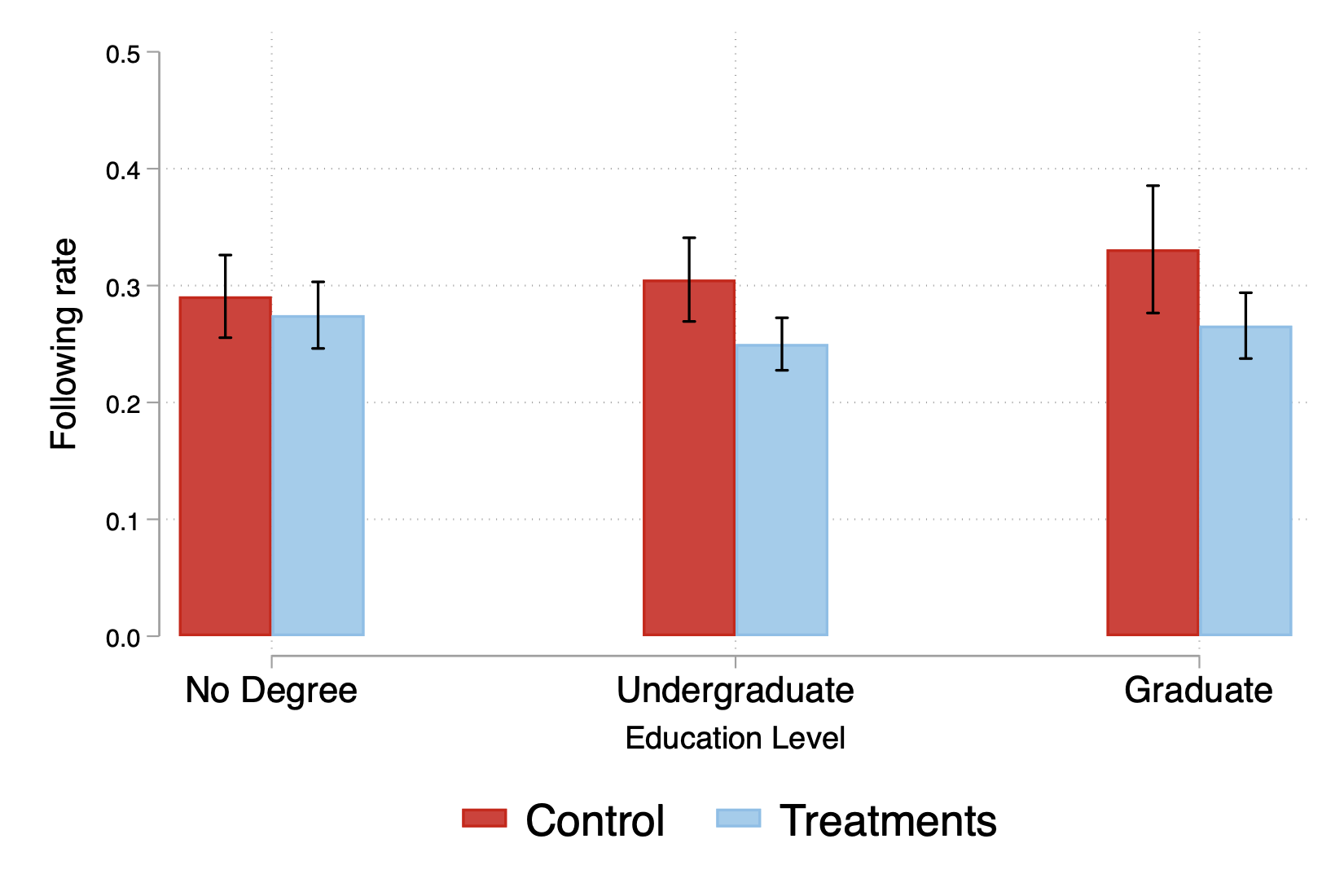}
\caption{Education: AI Recommendation Reliance.}
\end{subfigure}
\begin{subfigure}[b]{.49\textwidth}
\centering
\includegraphics[width=3.1in]{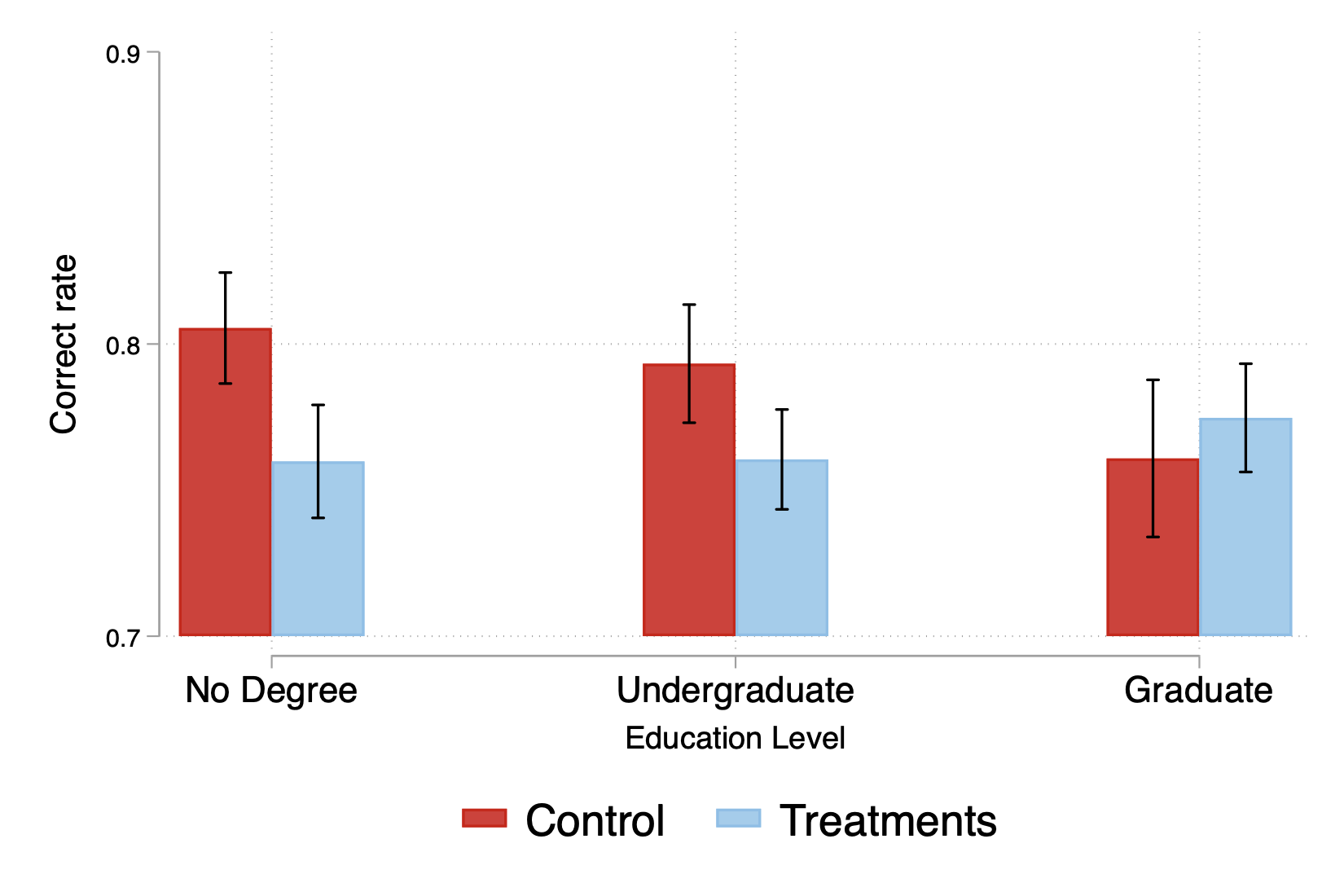}
\caption{Education: Accuracy.}
\end{subfigure}

\vspace{0.5em}
\begin{minipage}{0.9\textwidth}
\footnotesize \textit{Notes}: Panels (a) and (b) present treatment effects on AI reliance and accuracy by gender, while panels (c) and (d) do the same by ethnicity. Panels (e) and (f) replicate the analysis by workers’ highest level of education. Image-specific fixed effects are included, and standard errors are clustered at the worker level.
\end{minipage}
\label{fig:Het_dem}
\end{figure}

\begin{figure}[htbp]
\centering
\caption{Heterogeneous Treatment Effects by Platform Experience.}
\begin{subfigure}[b]{.49\textwidth}
\centering
\includegraphics[width=3.1in]{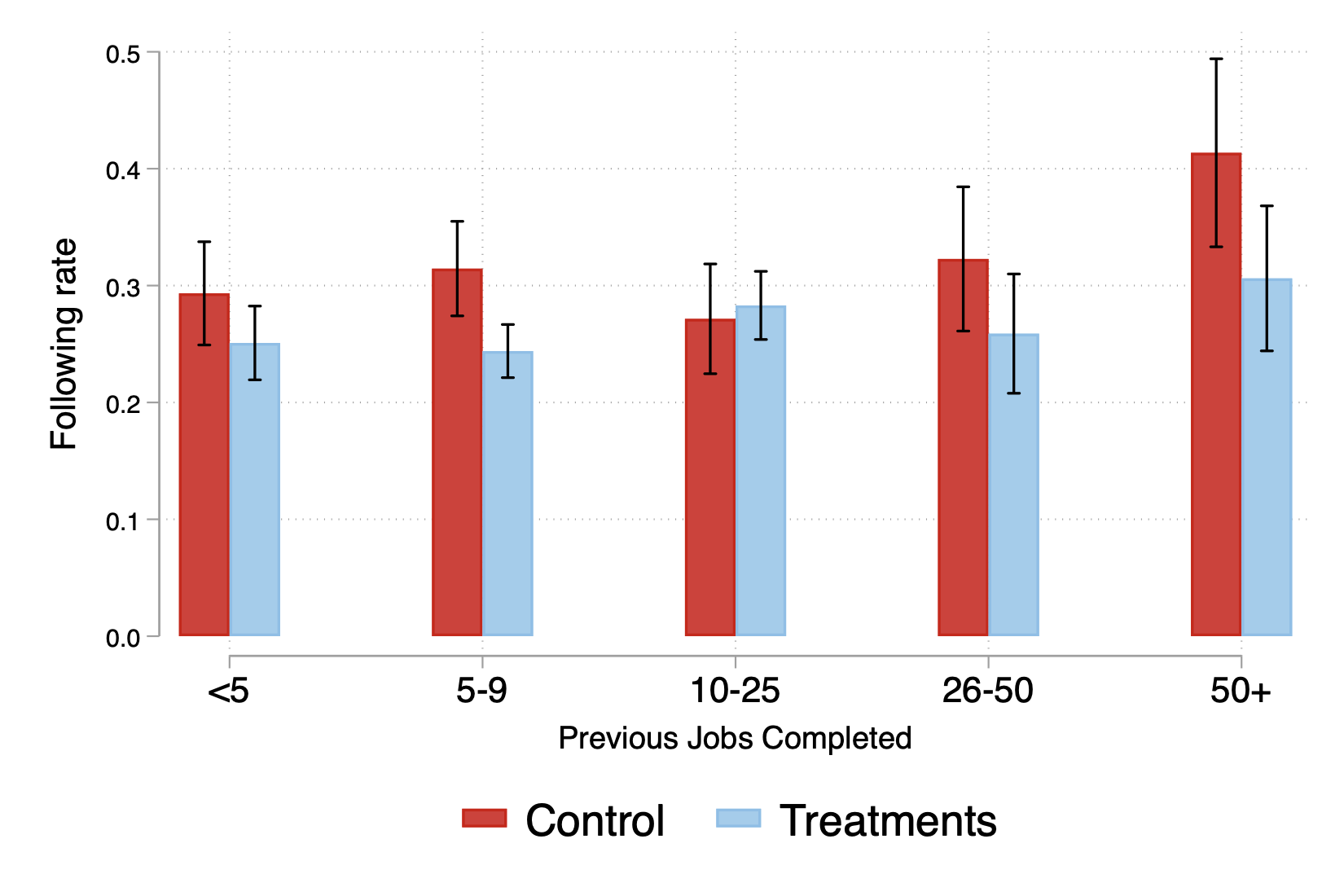}
\caption{Jobs: AI Recommendation Reliance.}
\end{subfigure}
\begin{subfigure}[b]{.49\textwidth}
\centering
\includegraphics[width=3.1in]{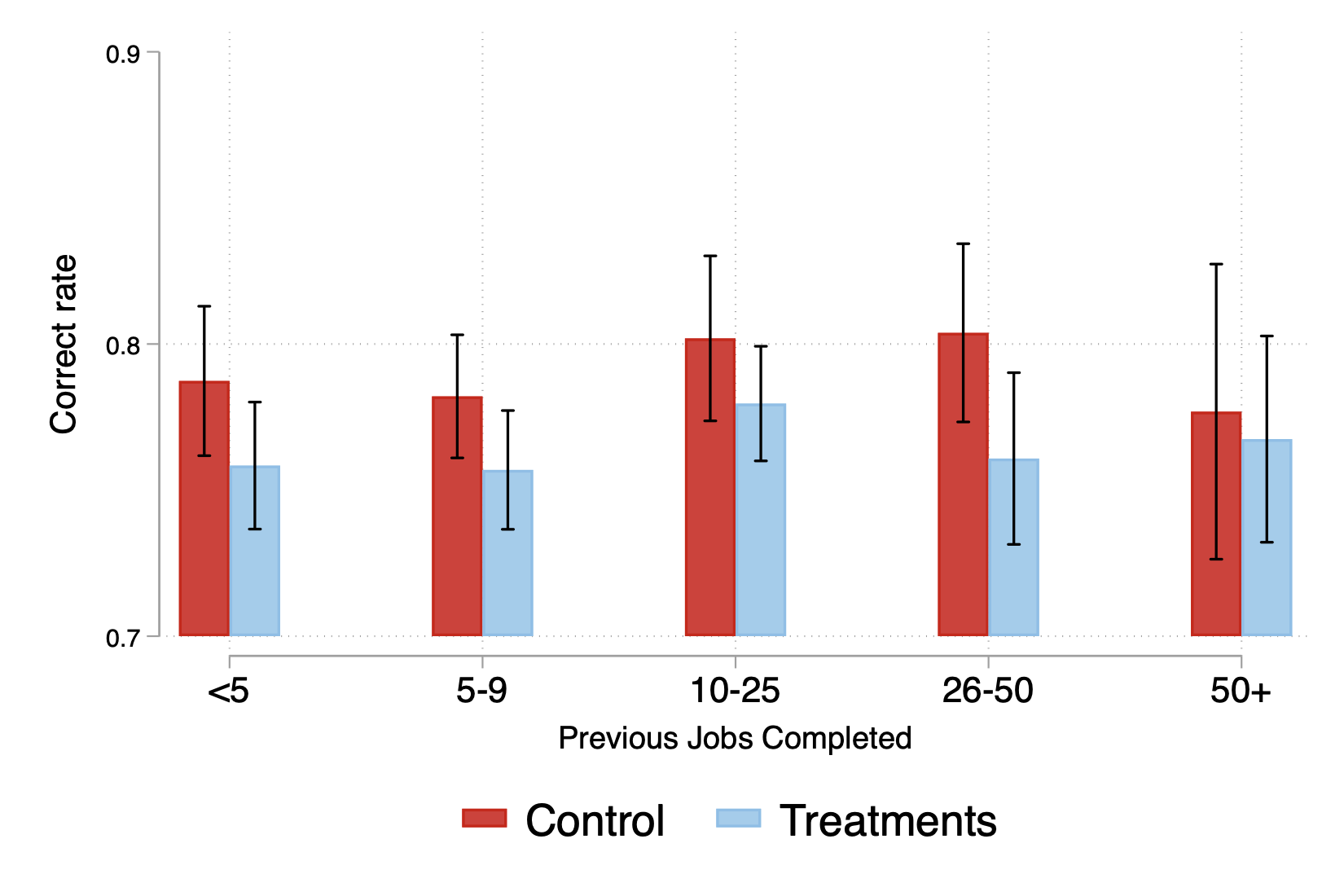}
\caption{Jobs: Accuracy.}
\end{subfigure}

\vspace{1.5em} 

\begin{subfigure}[b]{.49\textwidth}
\centering
\includegraphics[width=3.1in]{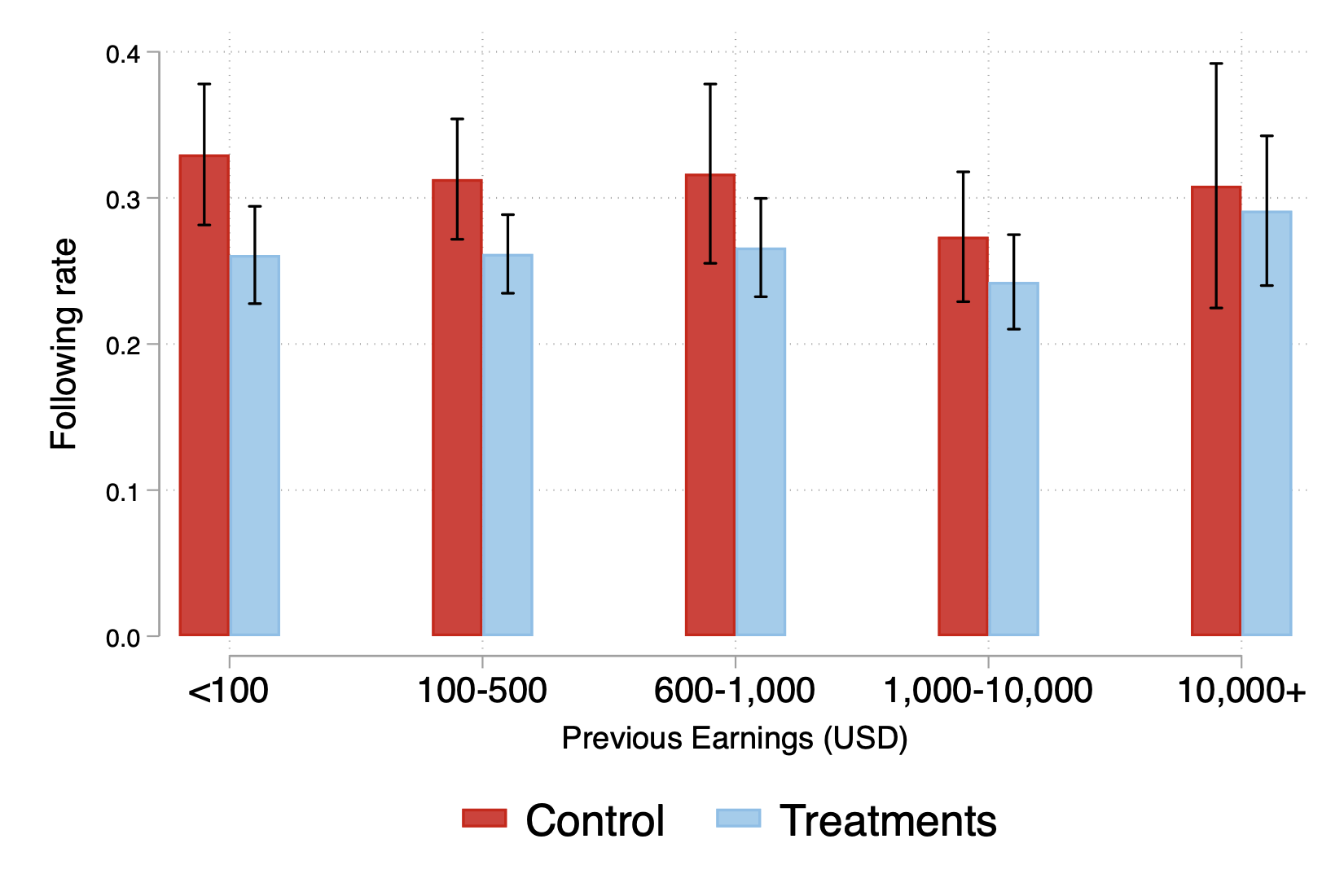}
\caption{Earnings: AI Recommendation Reliance.}
\end{subfigure}
\begin{subfigure}[b]{.49\textwidth}
\centering
\includegraphics[width=3.1in]{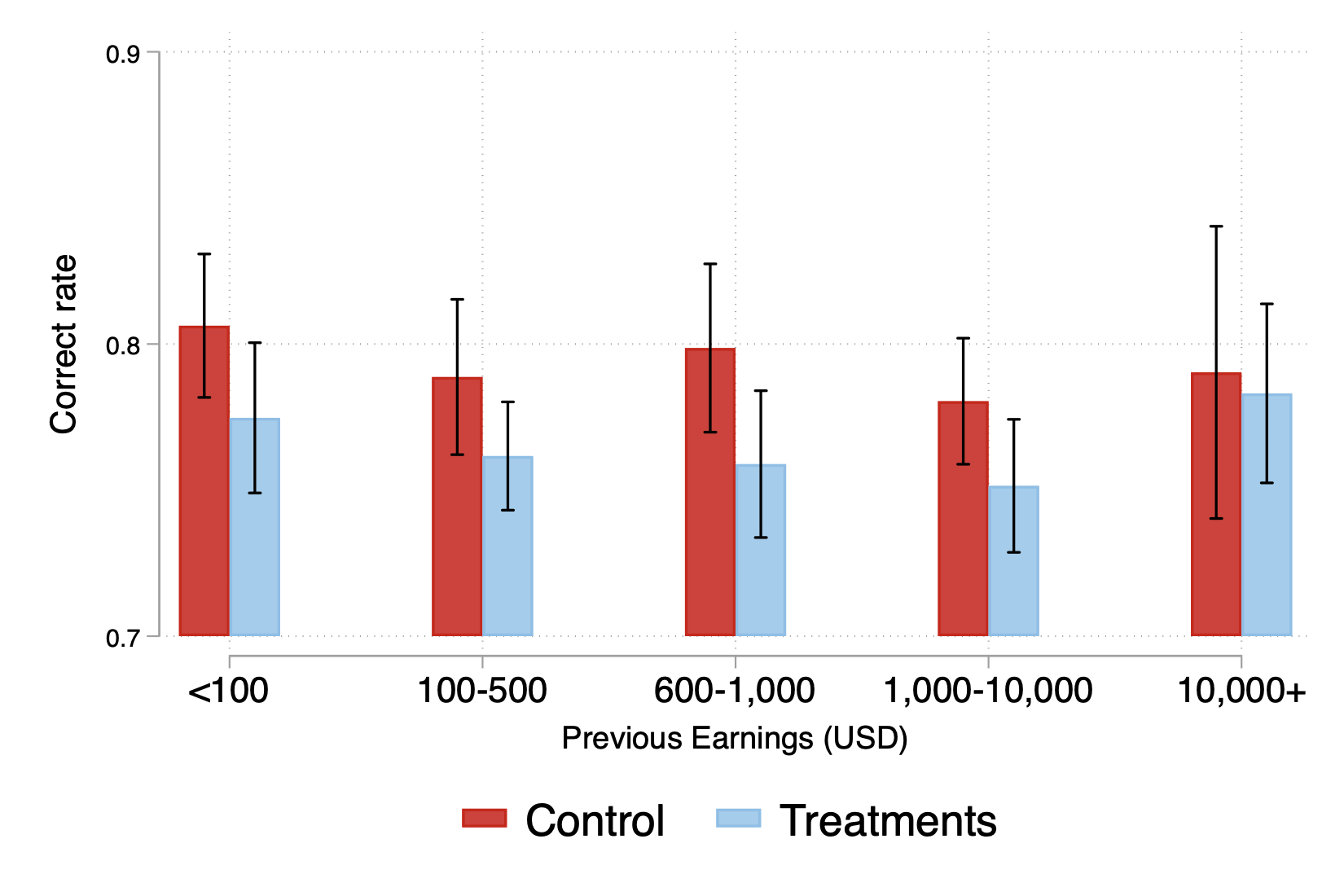}
\caption{Earnings: Accuracy.}
\end{subfigure}

\vspace{0.5em}
\begin{minipage}{0.9\textwidth}
\footnotesize \textit{Notes}: Panels (a) and (b) present treatment effects on AI reliance and accuracy by buckets of the number of previously completed jobs on the platform, while panels (c) and (d) do the same by grouping workers according to their previous earnings. The correlation between earnings and jobs is 0.55—positive but far from one—so examining both dimensions separately is not redundant. Image-specific fixed effects are included, and standard errors are clustered at the worker level.
\end{minipage}
\label{fig:Het_platform}
\end{figure}

\begin{figure}[h!]
\centering
\caption{Share of Workers Outperforming AI}
\includegraphics[width=5in]{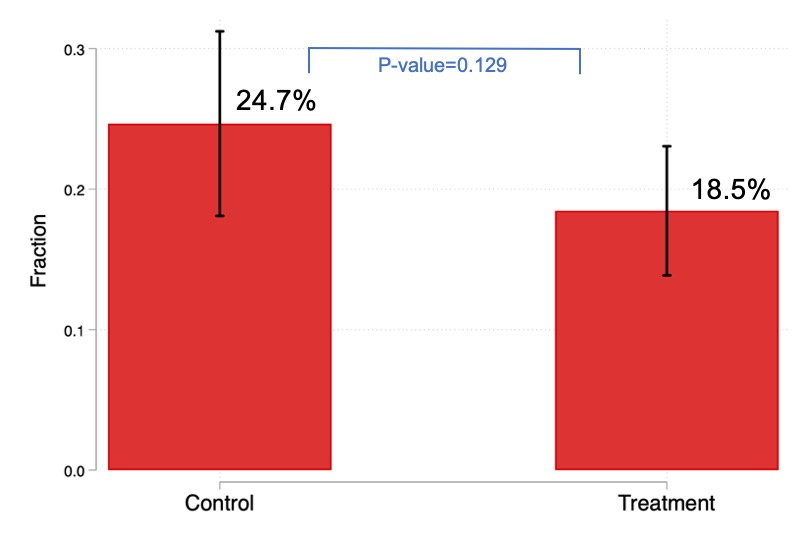}
\label{fig:Share}
\end{figure}

\begin{table}[h!]
    \centering
    \caption{Treatment Effects with Separate Treatment Groups.}
    \label{tab:Results_Separate}
    \sisetup{table-format=1.3}  
    \begin{tabular}{lSSSSSSSSS}
        \toprule
        & \multicolumn{2}{c}{AI recommendation reliance} & & \multicolumn{2}{c}{Correct answer} & & \multicolumn{2}{c}{Response time} \\
        \cmidrule(lr){2-3} \cmidrule(lr){5-6} \cmidrule(lr){8-9}
        & {All} & {Conditional} & & {Initial} & {Final} & & {Initial} & {Rec. stage}  \\
        & {(1)} & {(2)} & & {(3)} & {(4)} & & {(5)} & {(6)}   \\
        \midrule
        \textbf{Public} & \textbf{-0.040}** & \textbf{-0.080}*** & &\textbf{0.005} & \textbf{-0.028}***  &  & \textbf{2.14} & \textbf{-0.82} \\\vspace{1em}
        & {(0.016)} & {(0.025)} & & {(0.014)} & {(0.010)}  & & {(1.33)}  & {(0.62)}   \\

         \textbf{Public w/ info.} & \textbf{-0.046}*** & \textbf{-0.081}*** & &\textbf{0.012} & \textbf{-0.026}**  &  & \textbf{2.05} & \textbf{-0.61} \\\vspace{2em}
         & {(0.016)} & {(0.026)} & & {(0.014)} & {(0.010)}  & & {(1.28)}  & {(0.64)}   \\   

         \midrule
Equality test 
                   & {0.67} & {0.97} & & {0.57} & {0.88}  & & {0.95}  & {0.71}  \\
    ($p$-value) 
                   &  &  & &  &  & &   &   \\
        \midrule

        Constant & 0.305 & 0.640 & & 0.553 & 0.791  & & {21.3} & {10.1} \\
        Observations & {22,398} &  {10,554} &  & {22,398} & {22,398} & &{22,398} &  {10,554} \\
        \bottomrule
    \end{tabular}
    
    \vspace{0.5em}
    \begin{minipage}{0.9\textwidth} 
        \footnotesize \textit{Notes}: Separating treatment groups changes the empirical specification to: $Y_{ij}= \alpha + \beta_1{T1_i} + \beta_2{T2_i} + \gamma X_j + \epsilon_{ij}$. Where each outcome variable is regressed on indicators for treatment assignment. Specifically, $T1_i$ equals 1 if worker $i$ was assigned to the \textit{Public} treatment group, and $T2_i$ equals 1 if assigned to the \textit{Public with Information} group. All regressions include image-specific fixed effects, and standard errors are clustered at the worker level. Equality test reports the $p$-value from a Wald test of equal coefficients for the \textit{Public} and \textit{Public with Information} treatments. \\
        * \(p<0.10\), ** \(p<0.05\), *** \(p<0.01\). 
    \end{minipage}
\end{table}

\begin{figure}[htbp]
\centering
\caption{Average Time Spent on Instruction Screens}
\begin{subfigure}[b]{.49\textwidth}
\centering
\includegraphics[width=3.1in]{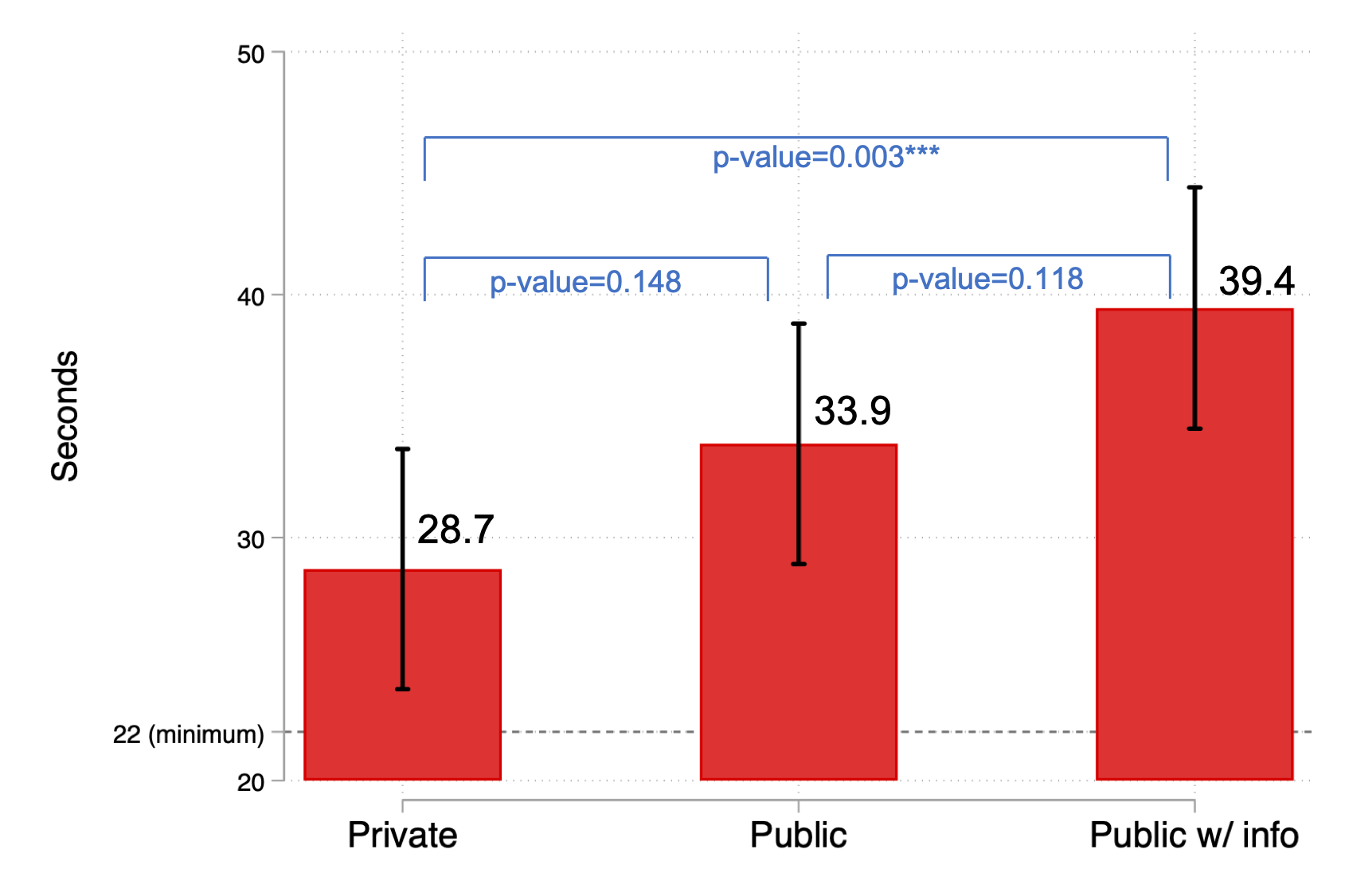}
\caption{Screen with Evaluation Explanation}
\end{subfigure}
\begin{subfigure}[b]{.49\textwidth}
\centering
\includegraphics[width=3.1in]{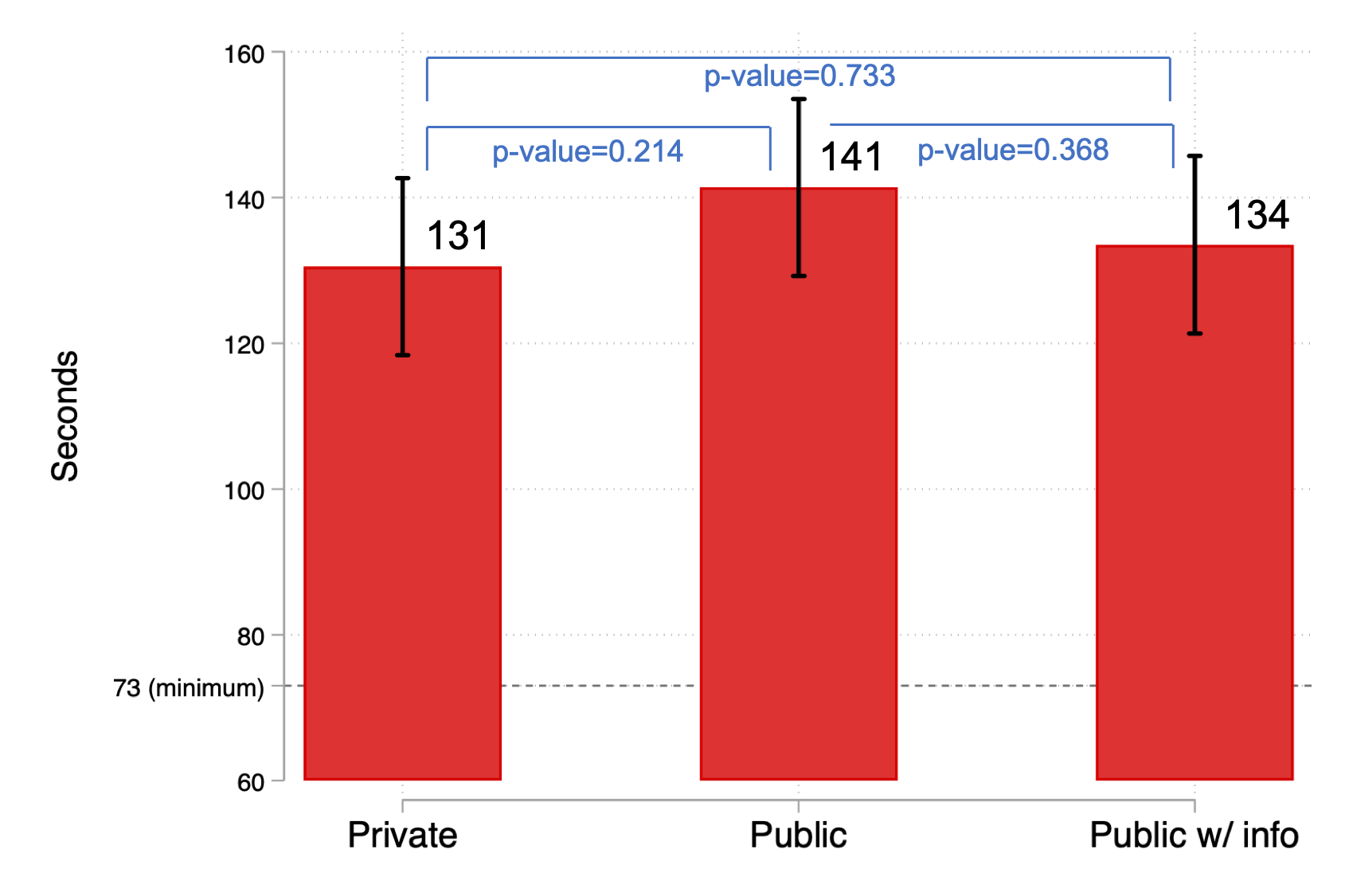}
\caption{Remaining Screens.}
\end{subfigure}

    \vspace{0.5em}
    \begin{minipage}{0.9\textwidth} 
        \footnotesize \textit{Notes}: Minimum time requirements were imposed to ensure careful reading; horizontal dashed lines denote the shortest allowable time per screen. Panel (a) reports time spent on the instruction screen containing the treatment variation, while Panel (b) shows the total time across all other screens.
    \end{minipage}
\label{fig:Time_Instructions}
\end{figure}

\begin{figure}[h!]
\centering
\caption{Distribution of Significant AI Recommendation Coefficients.}
\includegraphics[width=5in]{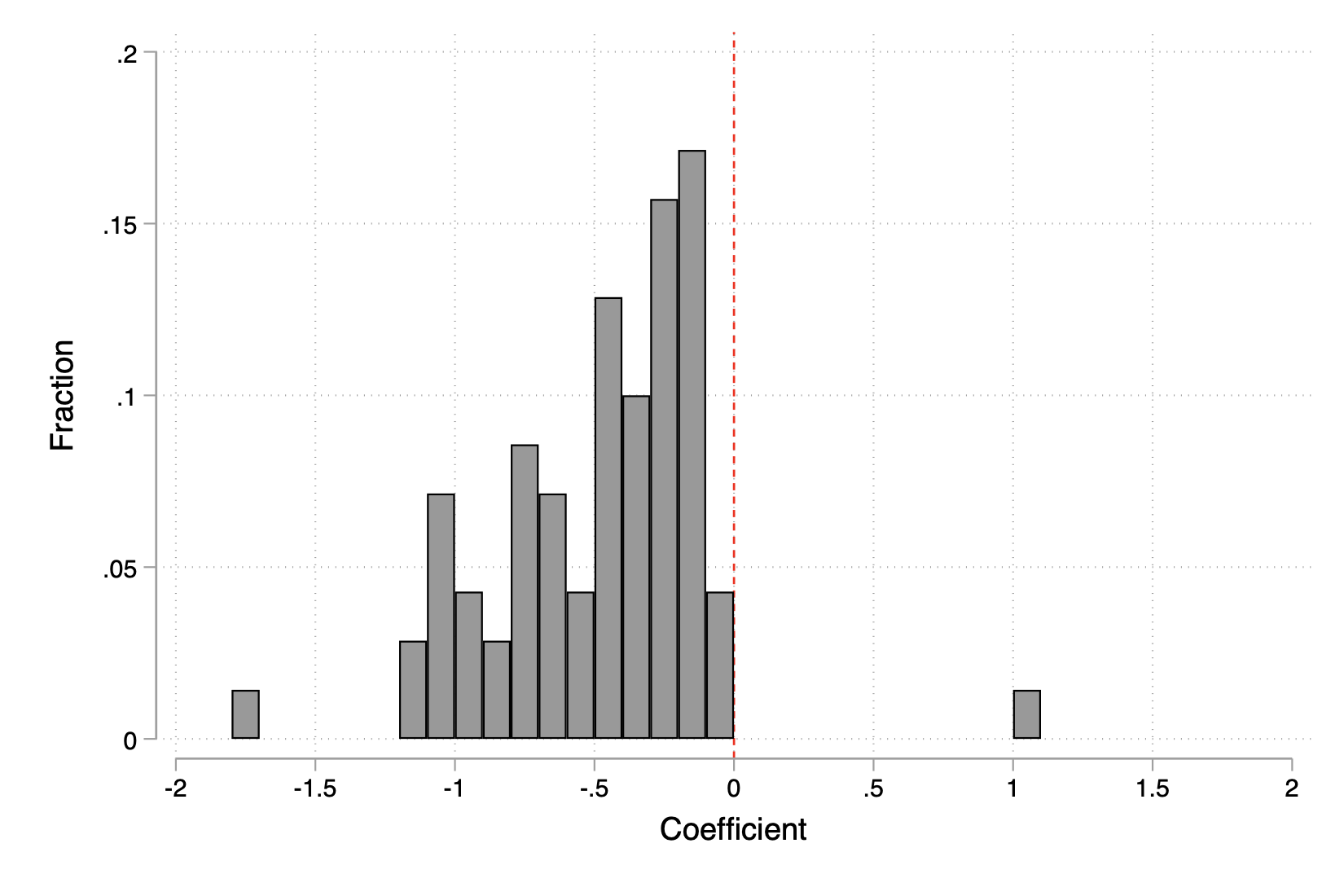}

    \vspace{0.5em}
    \begin{minipage}{0.9\textwidth} 
        \footnotesize \textit{Notes}: We estimate individual regressions for each of the 93 returning workers who evaluated 20 worker profiles. The figure shows the distribution of the coefficients on AI reliance for the 70 evaluators with coefficients significant at the 10\% level.
    \end{minipage}
\label{fig:coefdist}
\end{figure}

\begin{figure}[h!]
\centering
\caption{2-D Visualization of Rankings.}
\includegraphics[width=5in]{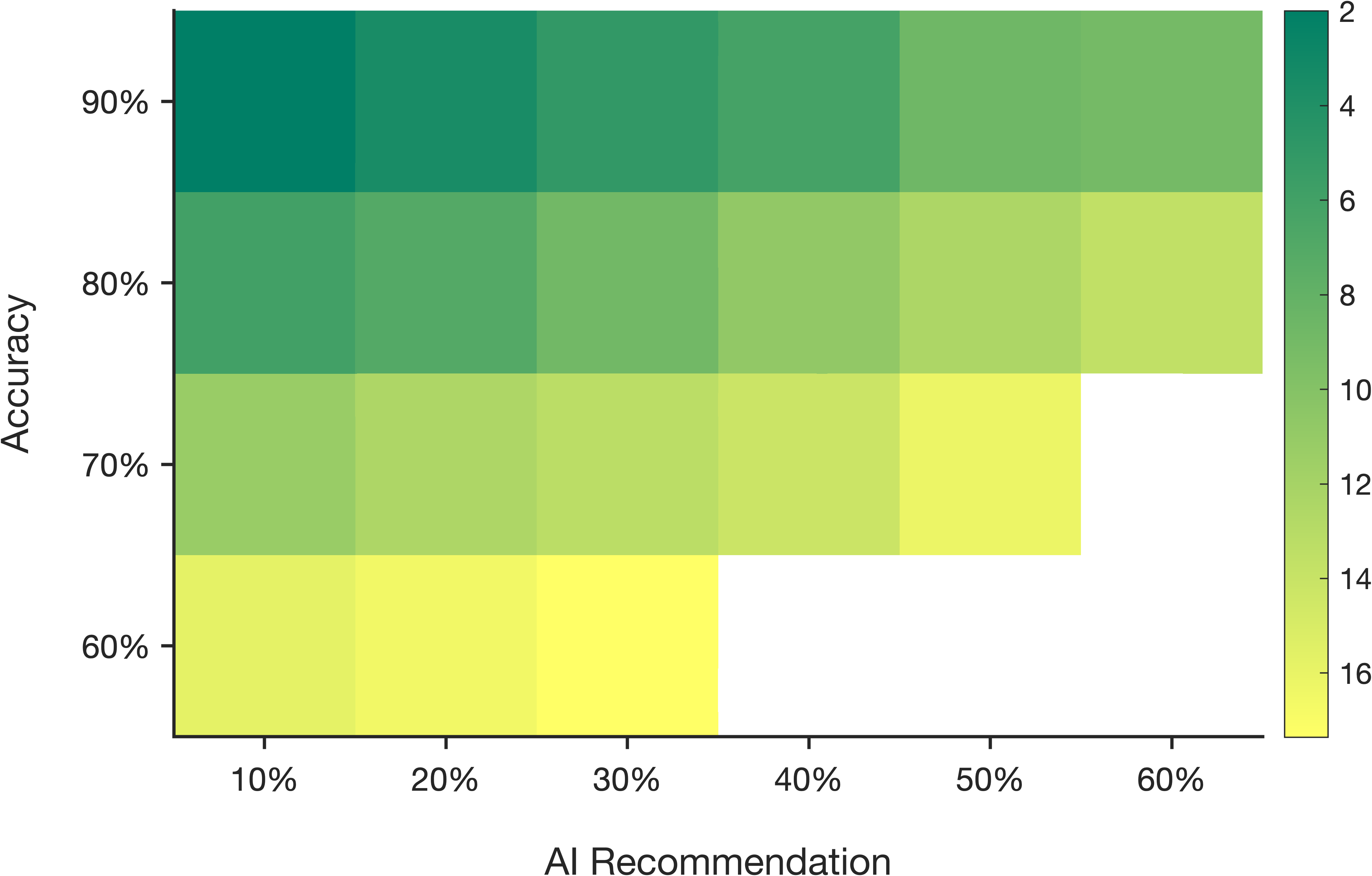}
\label{fig:2Drankings}
\end{figure}

\clearpage
\counterwithin{figure}{section}
\counterwithin{table}{section}

\section{Experiment Implementation}

\subsection{Experimental Instructions for First Job}

\begin{figure}[h!]
\centering
\fbox{\includegraphics[width=4.8in]{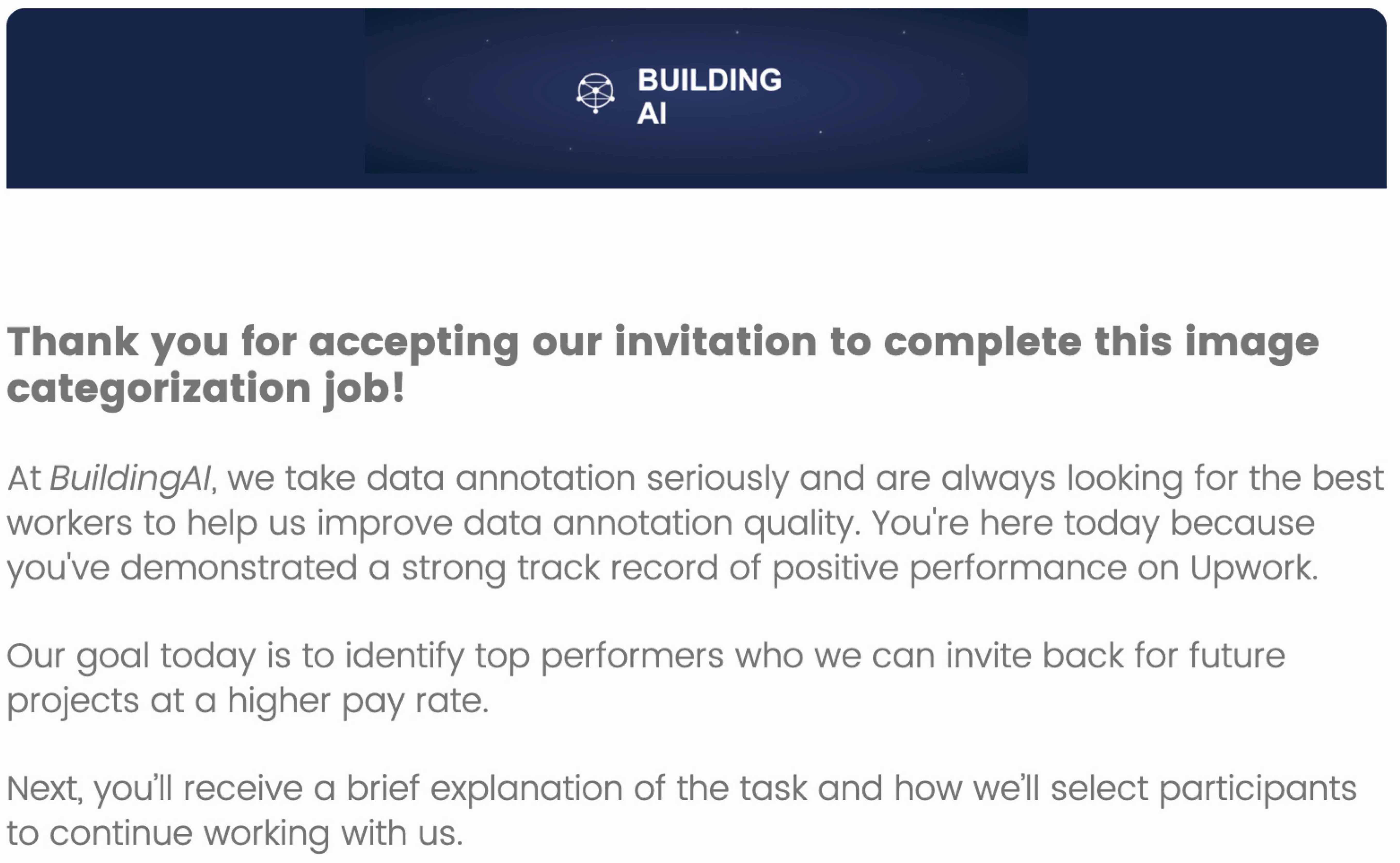}}
\end{figure}

\begin{figure}[h!]
\centering
\fbox{\includegraphics[width=4.8in]{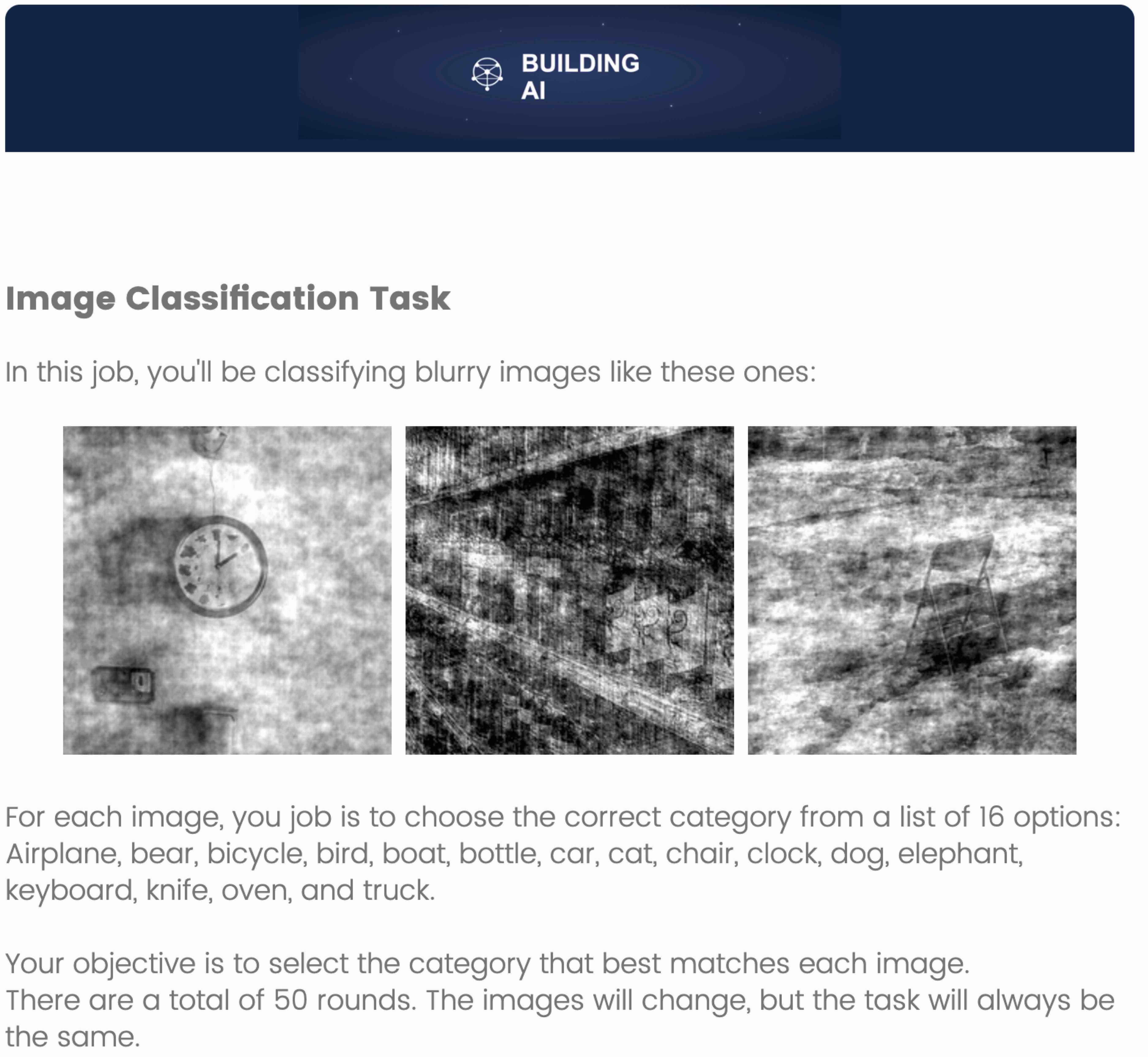}}
\end{figure}

\begin{figure}[h!]
\centering
\fbox{\includegraphics[width=5.5in]{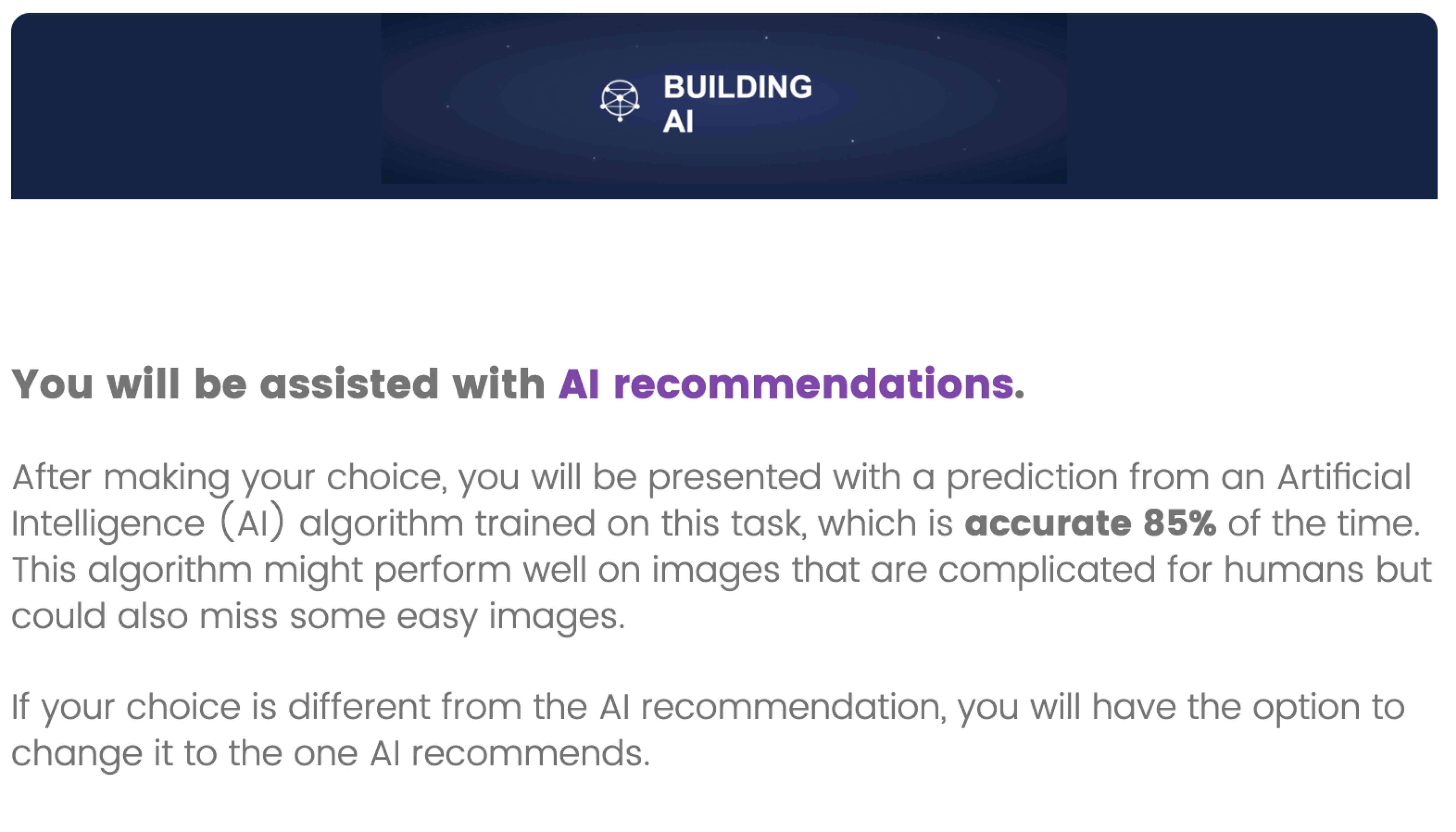}}
\end{figure}

\begin{figure}[h!]
\centering
\fbox{\includegraphics[width=5.5in]{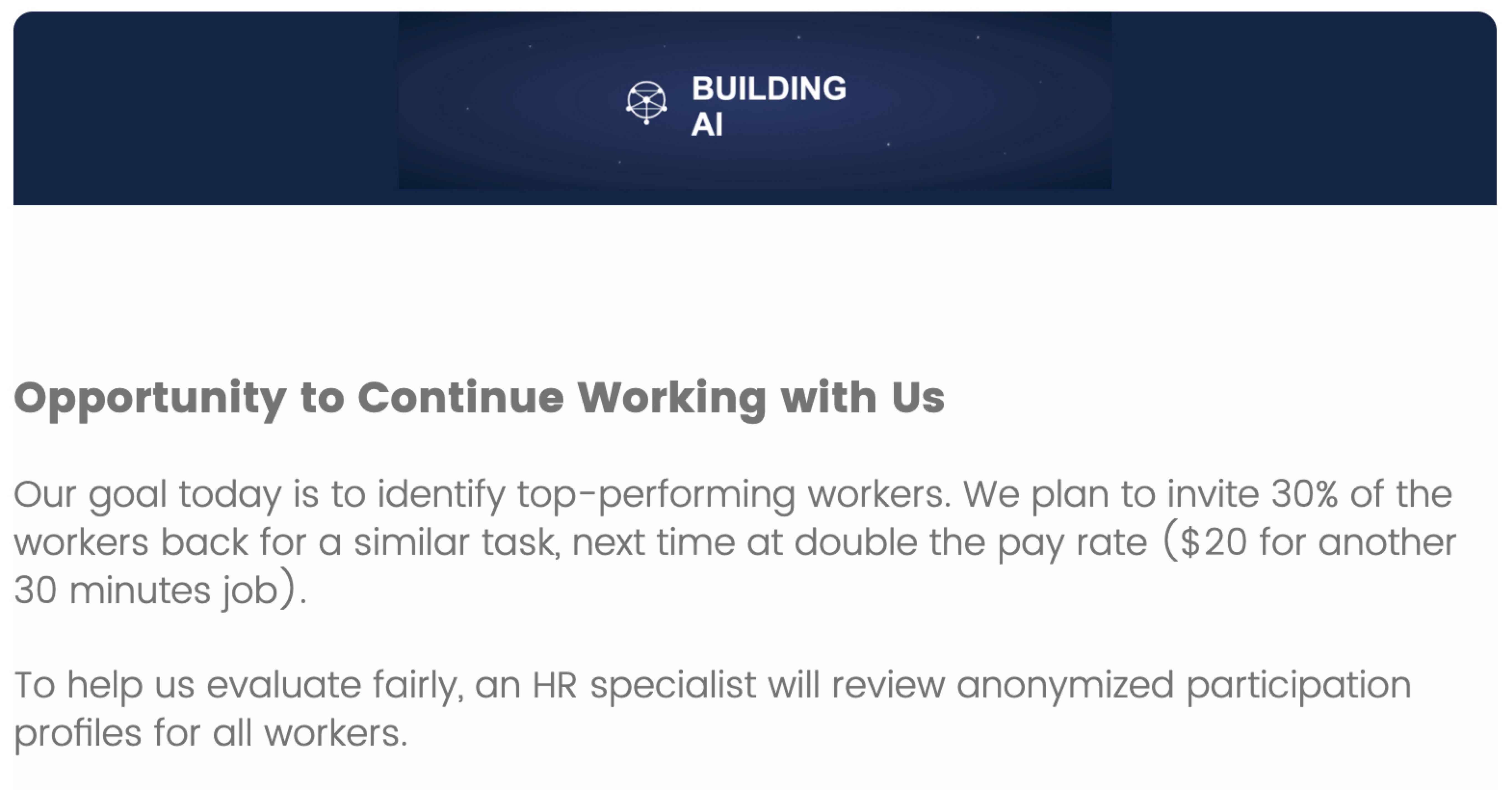}}
\end{figure}

\begin{figure}[h!]
    \centering
    \caption*{HR Evaluation Process Explained in Control}
    \fbox{\includegraphics[width=5in]{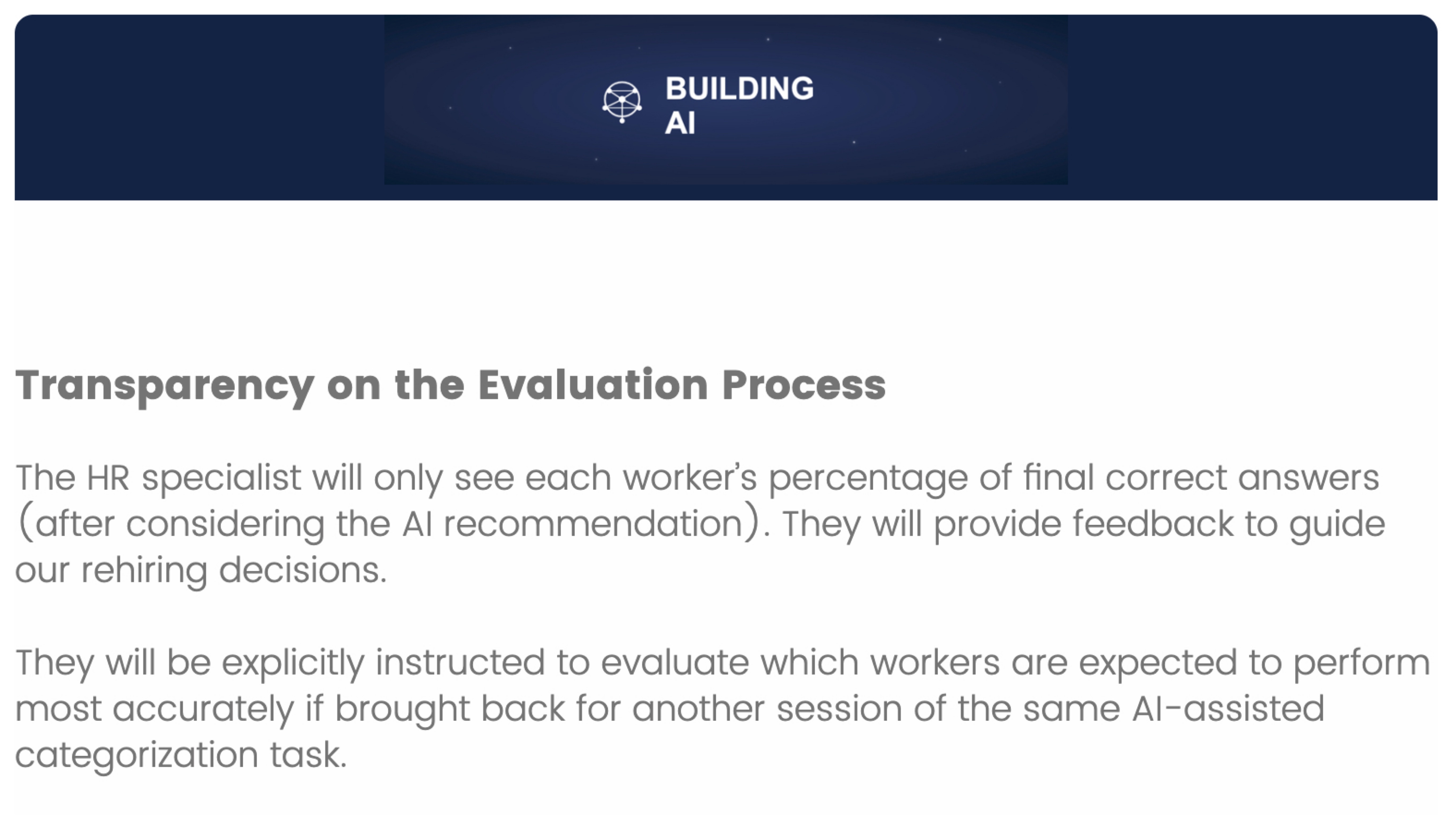}}\\[3em]
   \fbox{\includegraphics[width=5in]{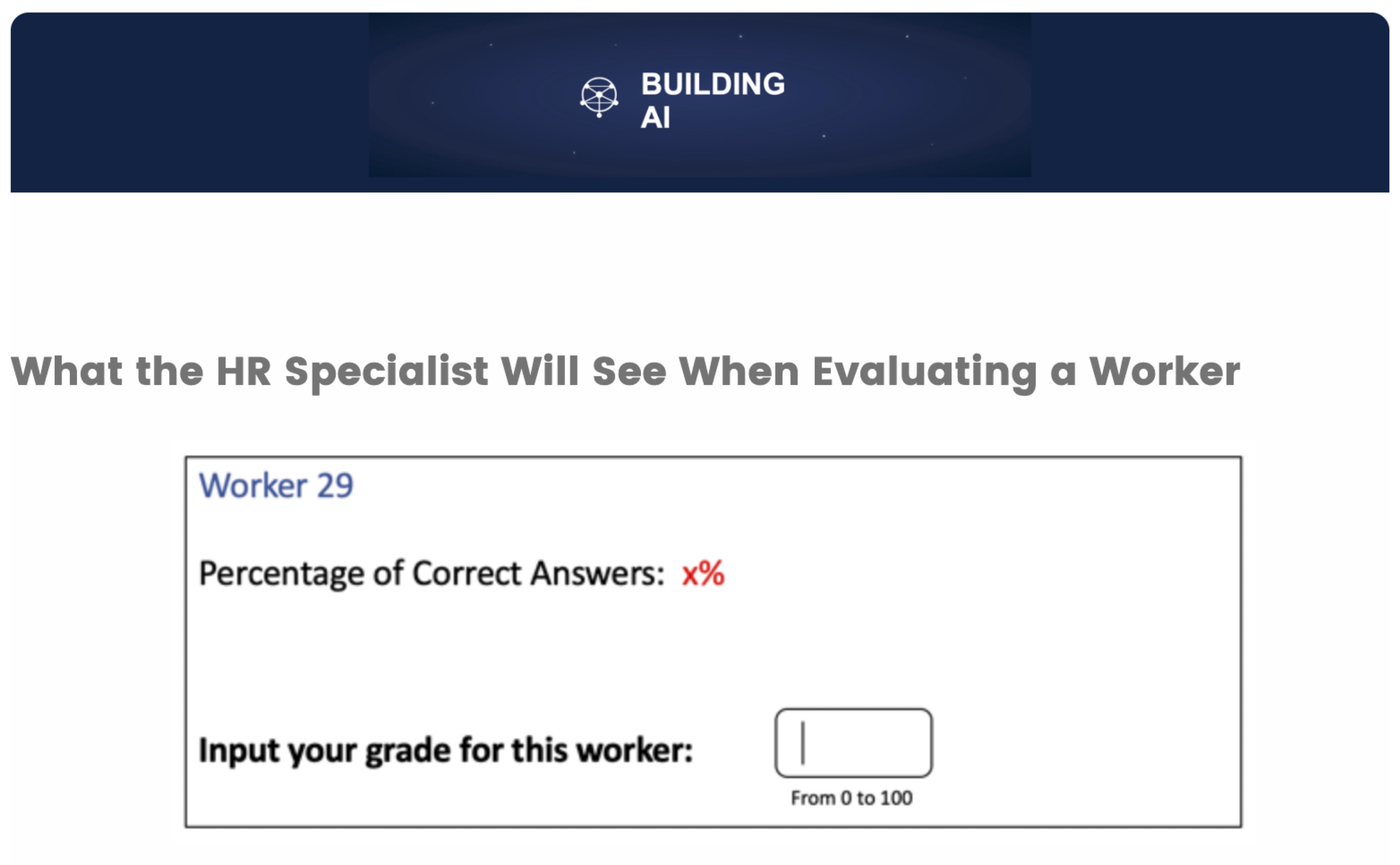}}
\end{figure}

\begin{figure}[h!]
    \centering
    \caption*{HR Evaluation Process Explained in Treatment 1}
    \fbox{\includegraphics[width=5in]{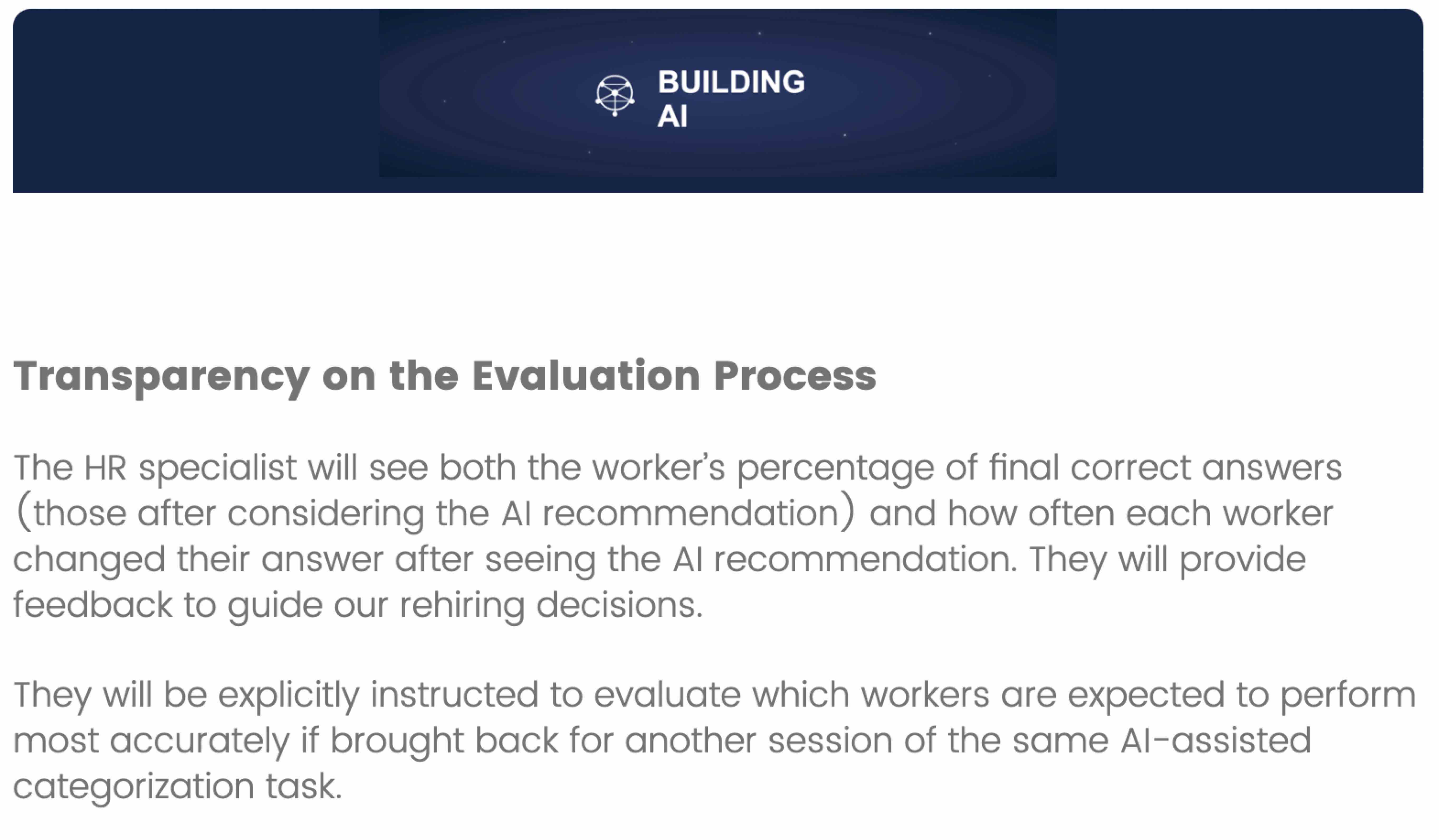}}\\[3em]
    \fbox{\includegraphics[width=5in]{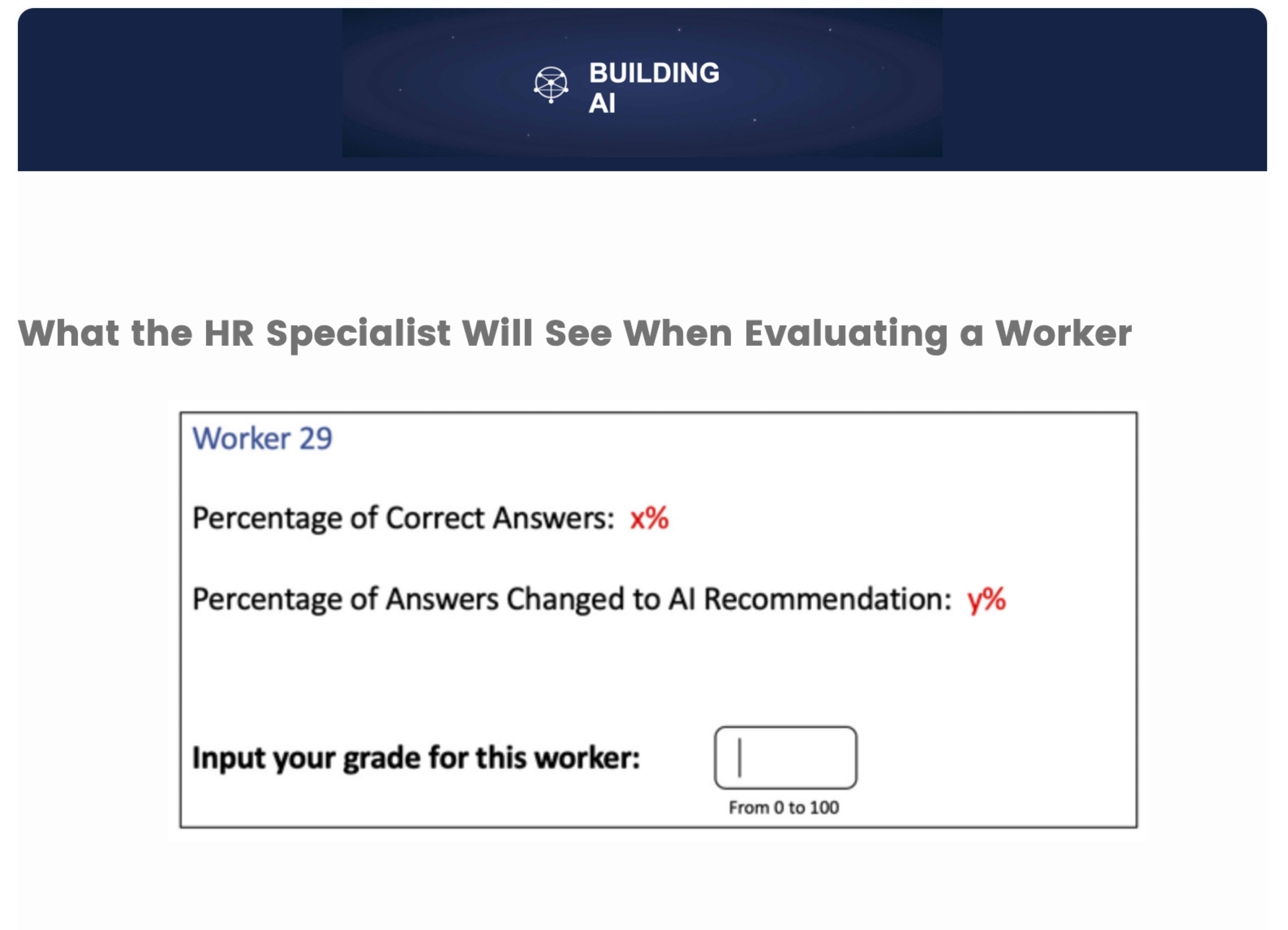}}
\end{figure}

\begin{figure}[h!]
    \centering
    \caption*{HR Evaluation Process Explained in Treatment 2}
    \fbox{\includegraphics[width=5in]{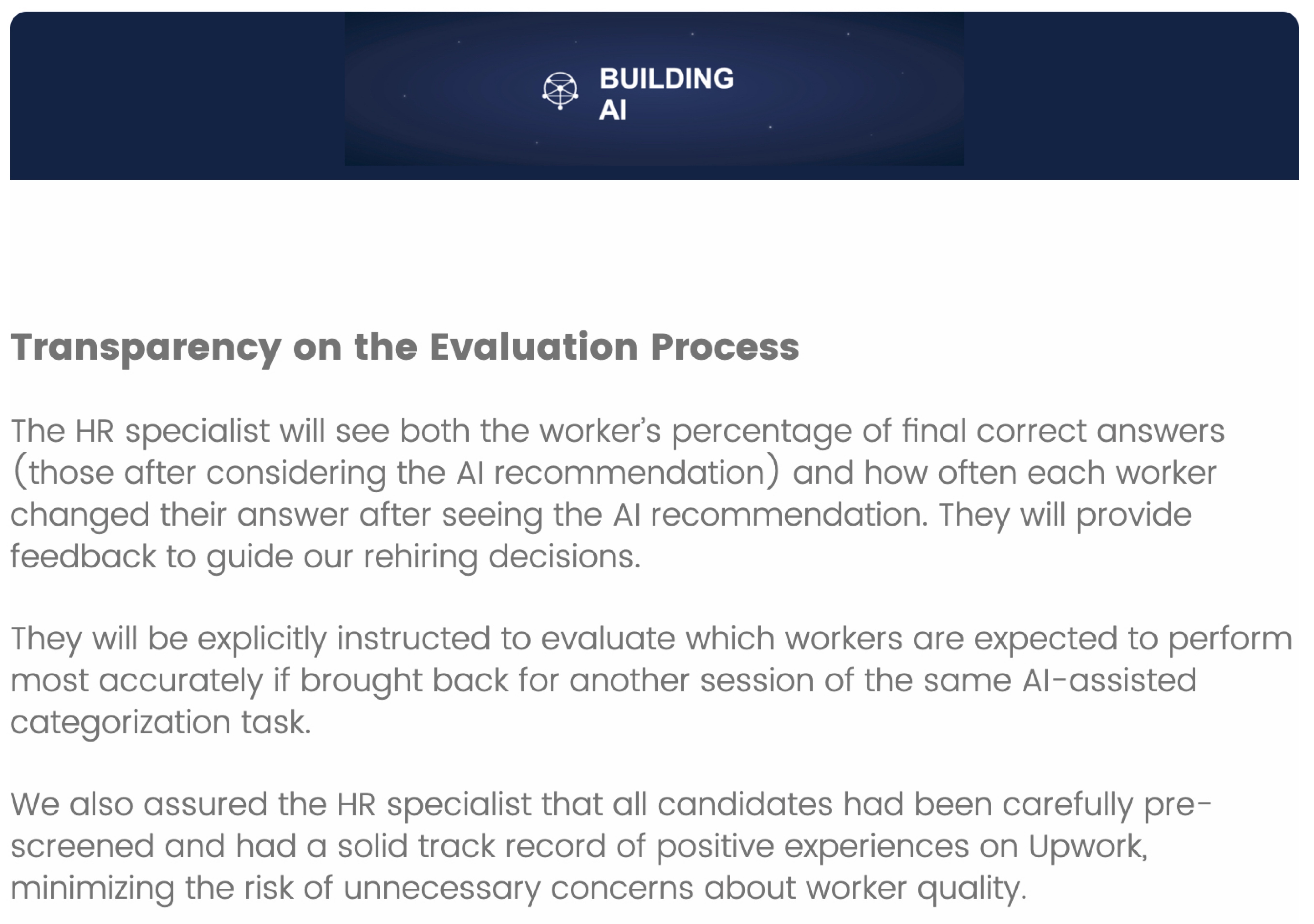}}\\[3em]
    \fbox{\includegraphics[width=5in]{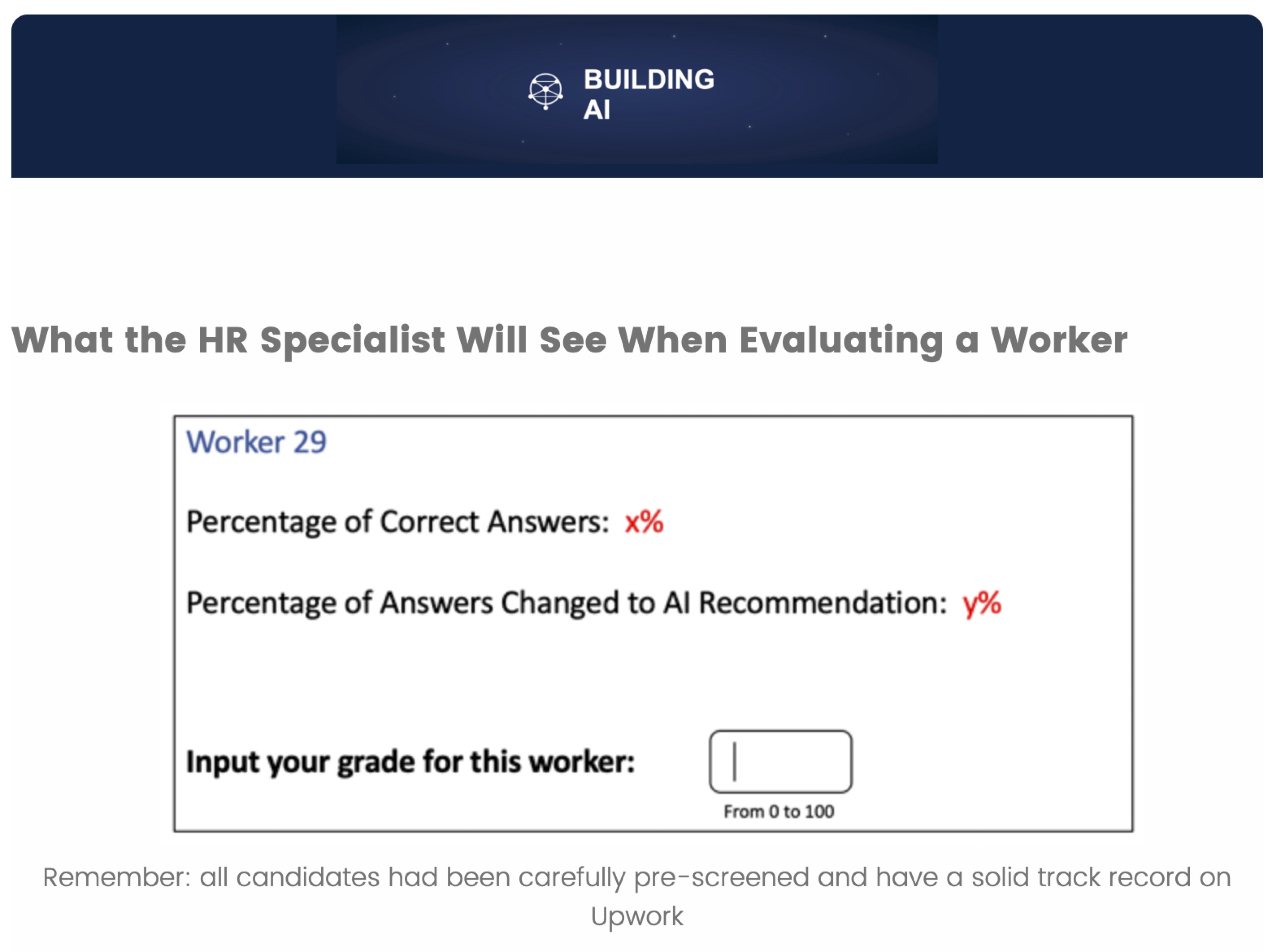}}
\end{figure}

\clearpage
\subsection{Additional Questions in Second Job}

Instructions shown are for returning treated workers (either in the \textit{public} or \textit{public with information} conditions). Returning workers from the control group received similar instructions, except that the “Worker Selection” segment was omitted.

\begin{figure}[h!]
\centering
\fbox{\includegraphics[width=5.5in]{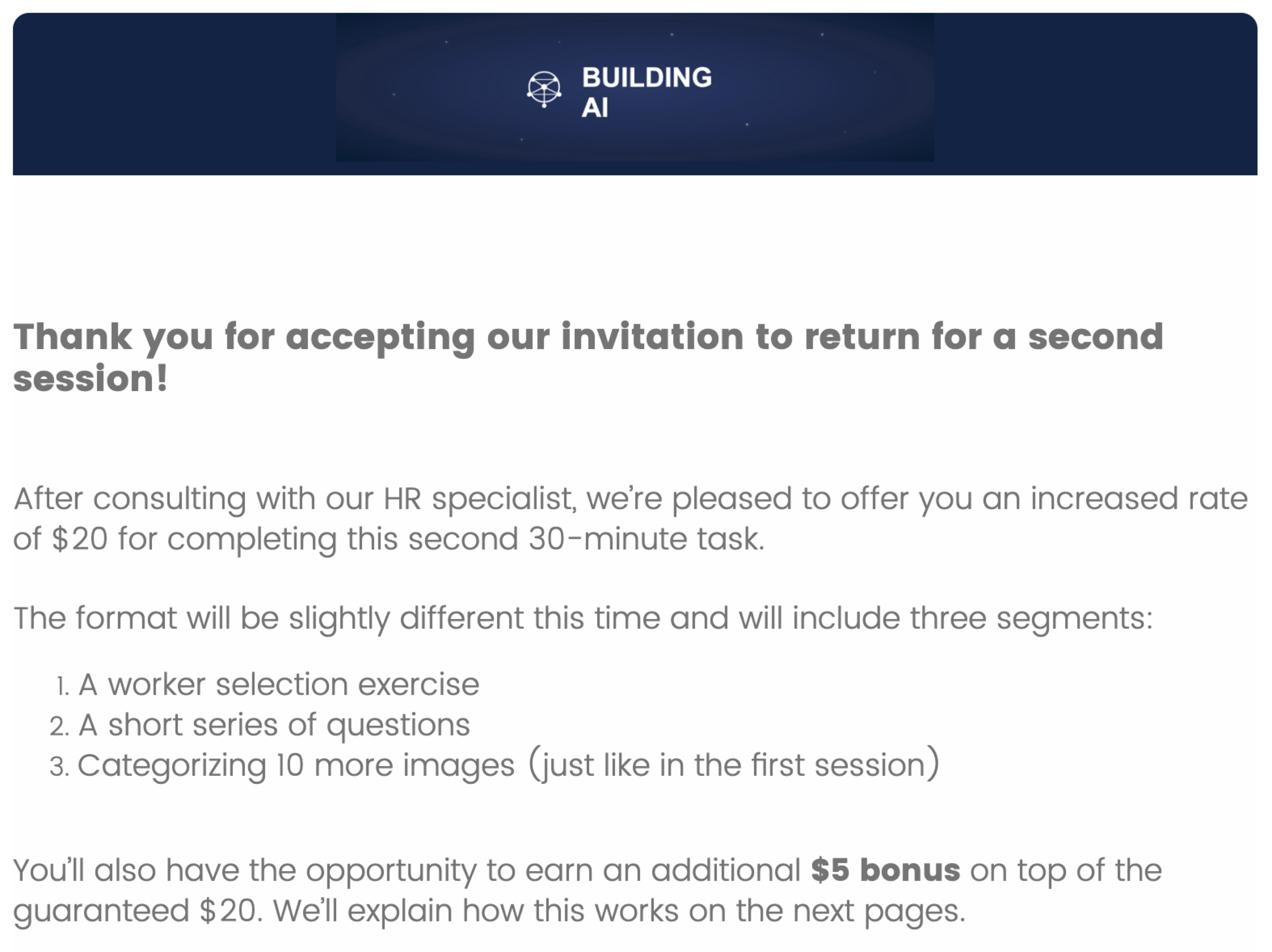}}
\end{figure}

\begin{figure}[h!]
\centering
\fbox{\includegraphics[width=5.5in]{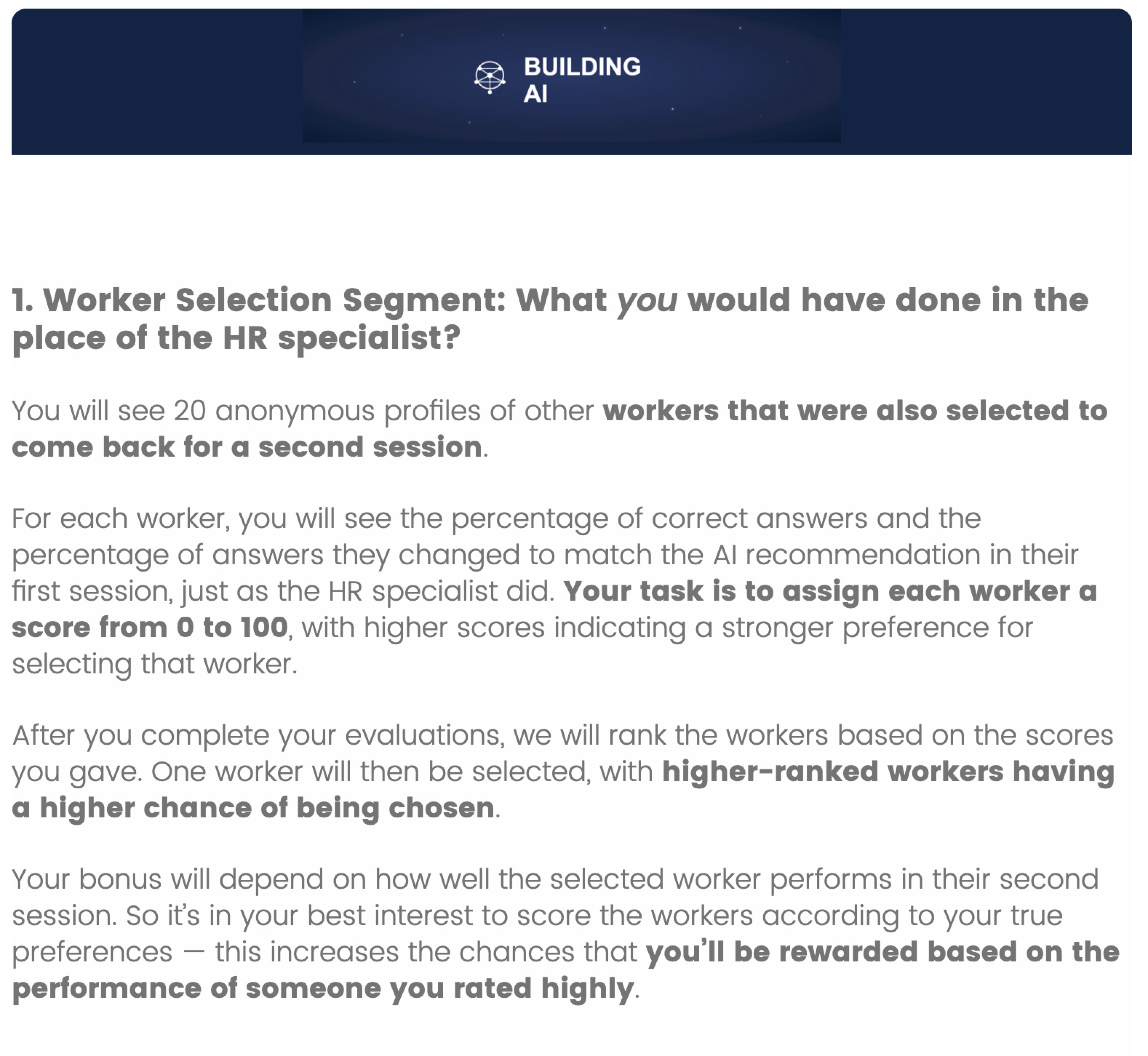}}
\end{figure}

\begin{figure}[h!]
\centering
\fbox{\includegraphics[width=5.5in]{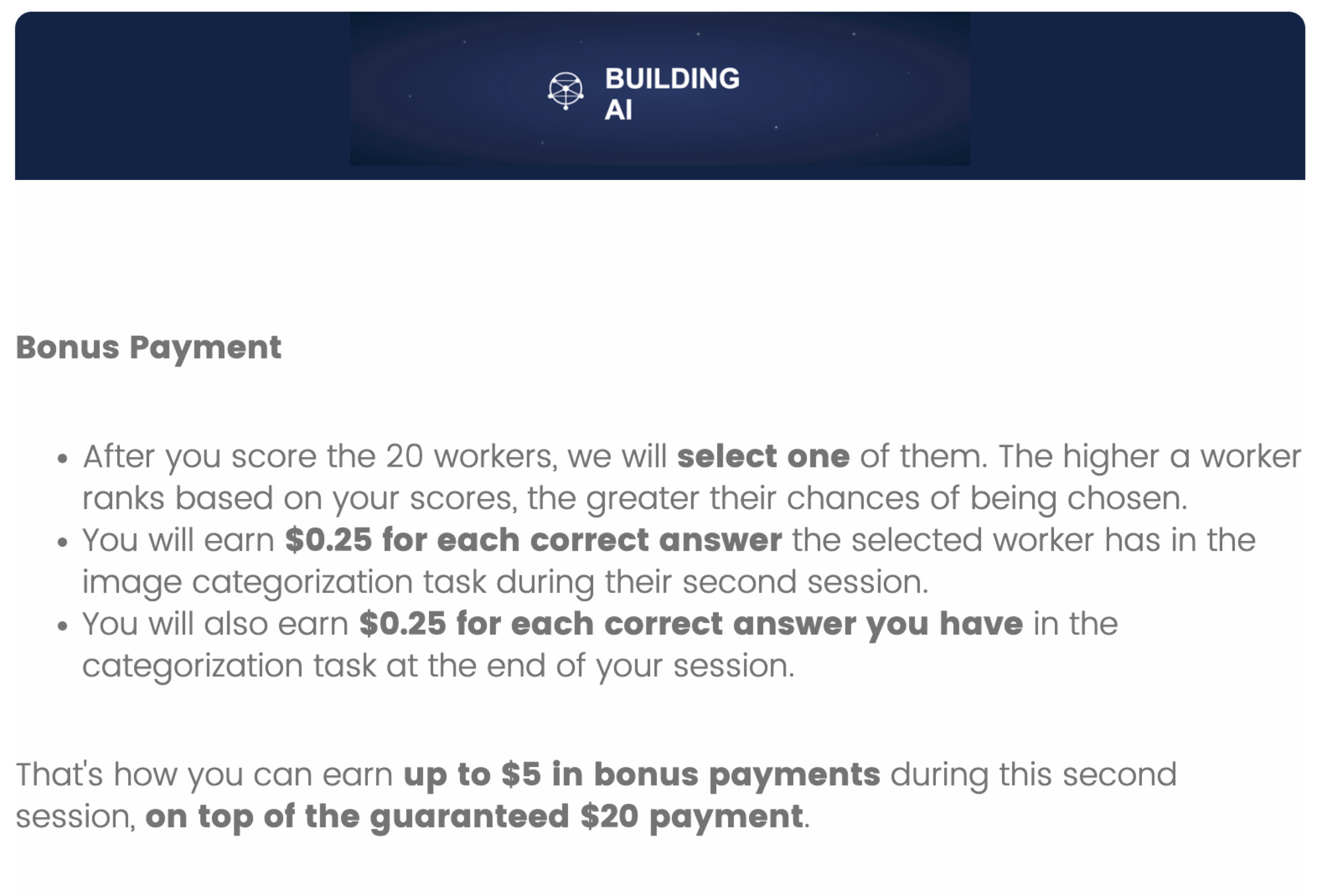}}
\end{figure}

\begin{figure}[h!]
\centering
\fbox{\includegraphics[width=5.5in]{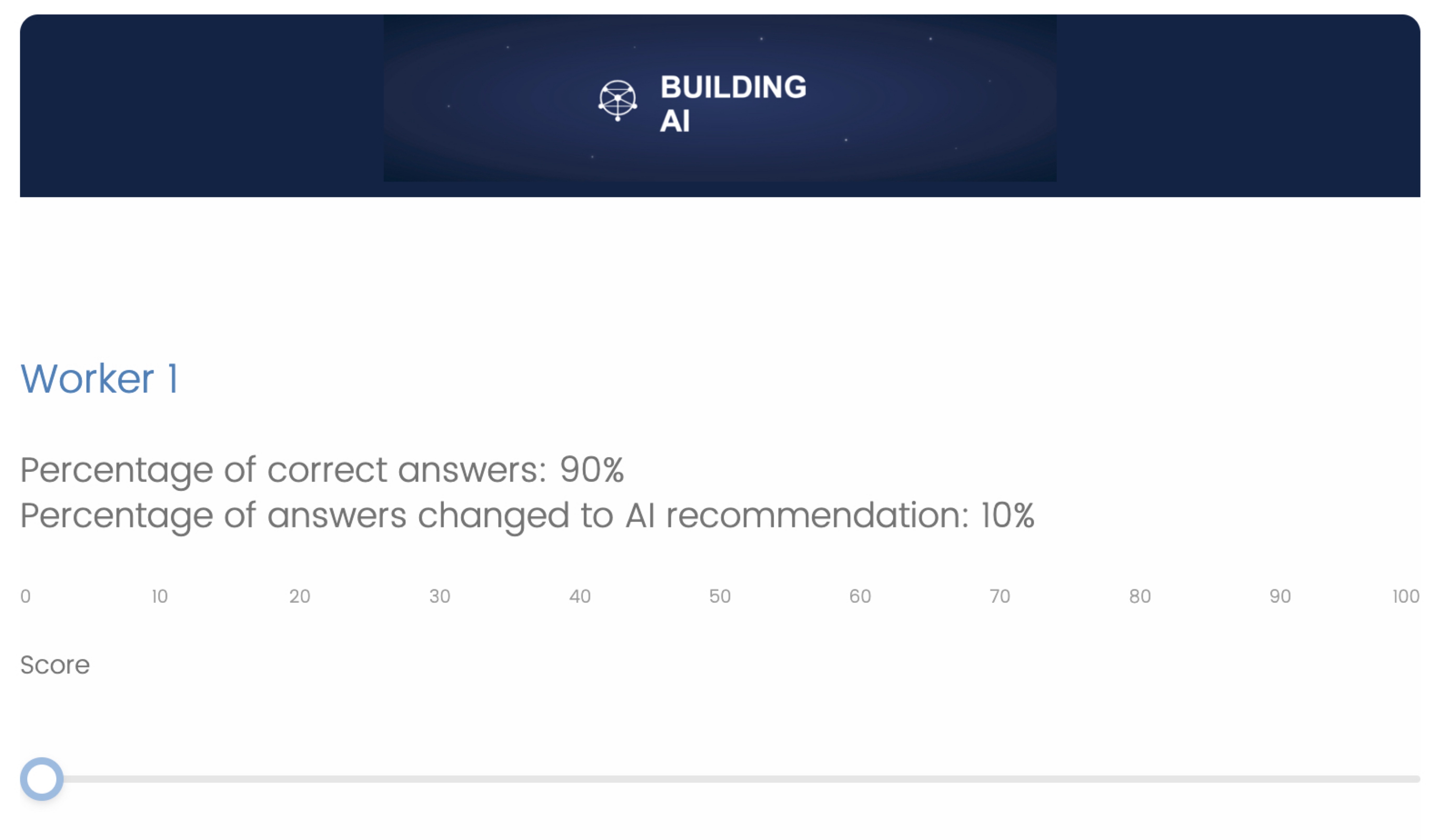}}
\end{figure}

\begin{figure}[h!]
\centering
\fbox{\includegraphics[width=5.5in]{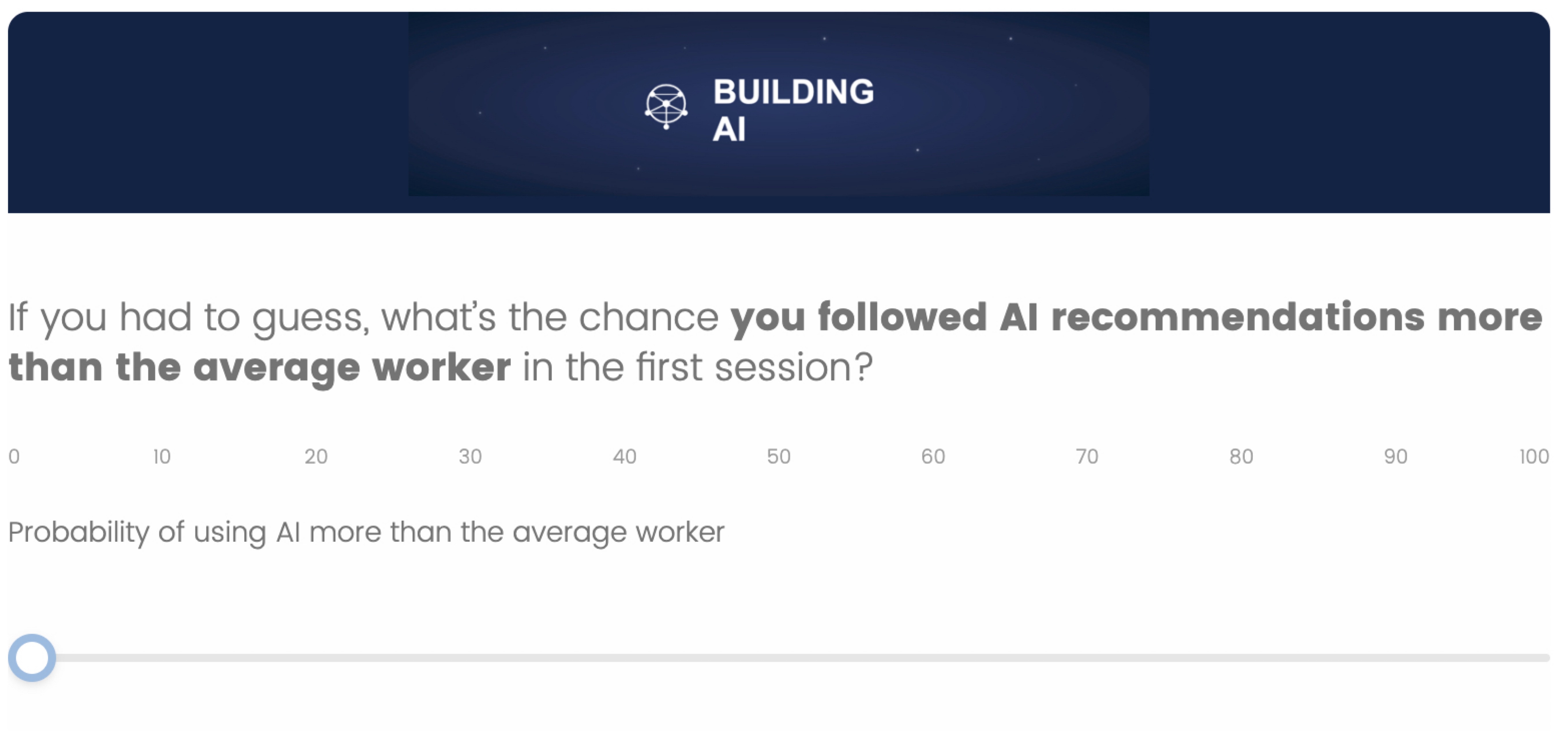}}
\end{figure}

\begin{figure}[h!]
\centering
\fbox{\includegraphics[width=5.5in]{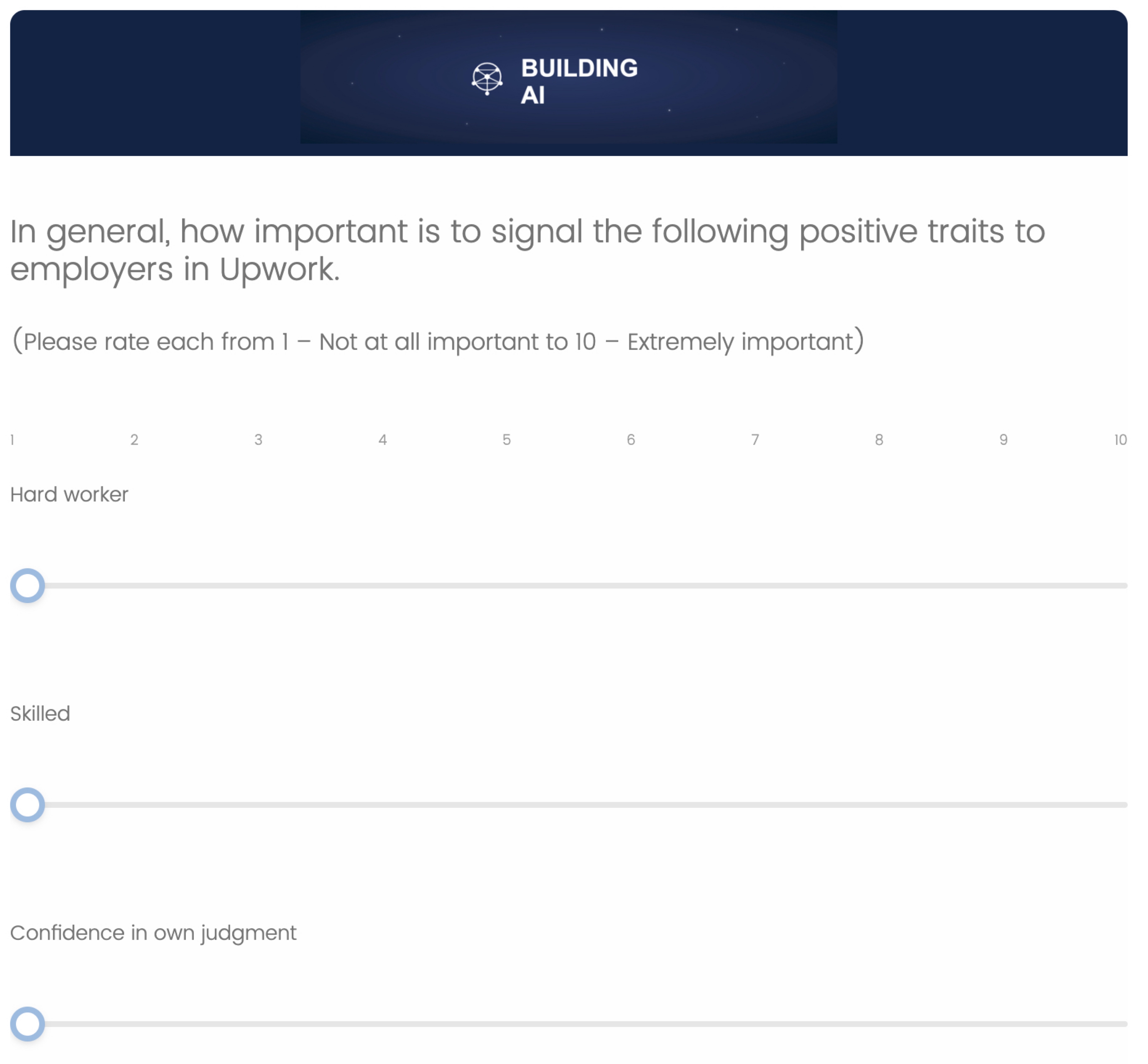}}
\end{figure}

\begin{figure}[h!]
\centering
\fbox{\includegraphics[width=5.5in]{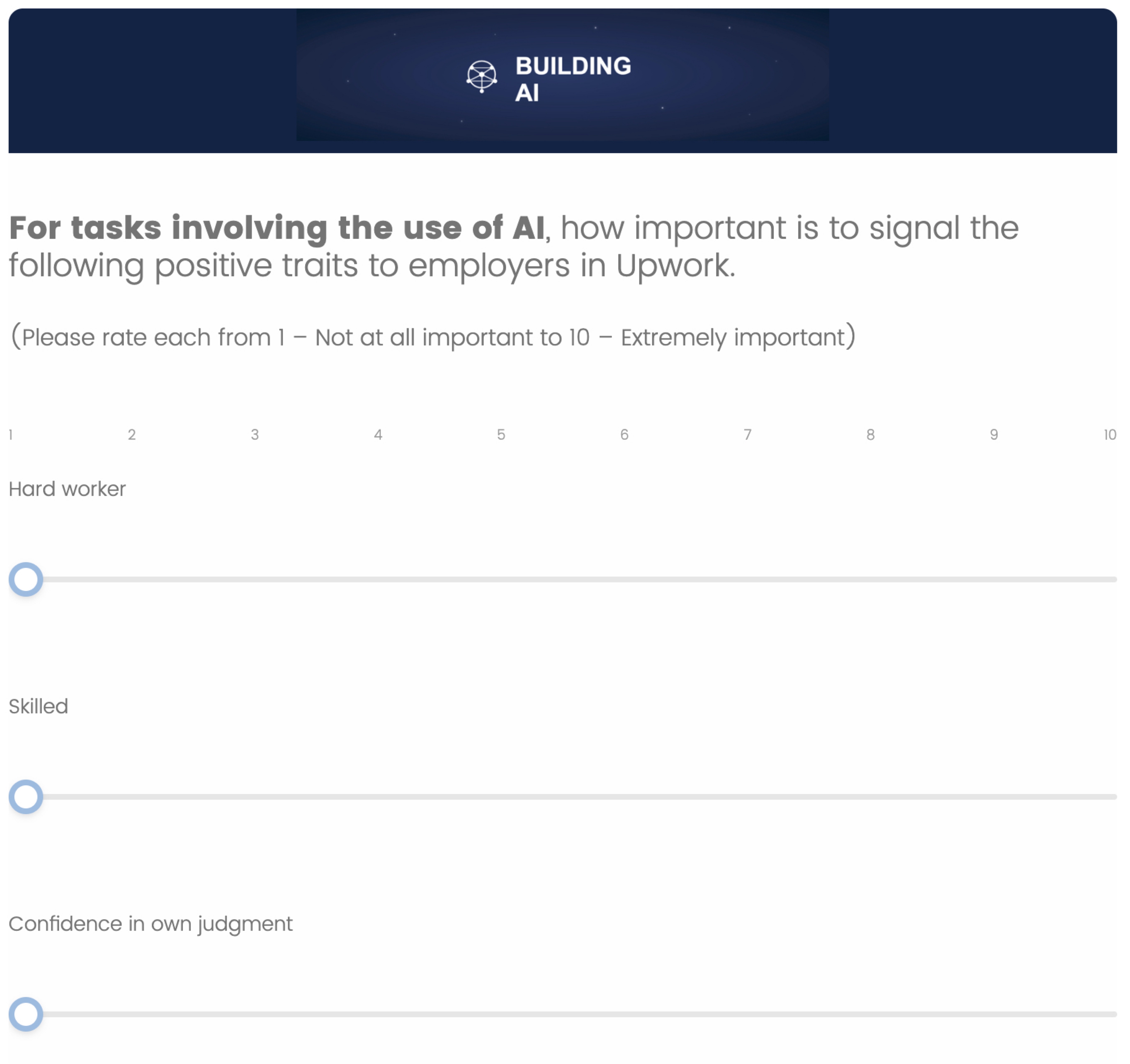}}
\end{figure}

\clearpage
\subsection{\textit{BuildingAI} Webpage}\label{sec:Webpage}

\begin{figure}[h!]
\centering
\caption*{Welcome Screen}
\fbox{\includegraphics[width=5.5in]{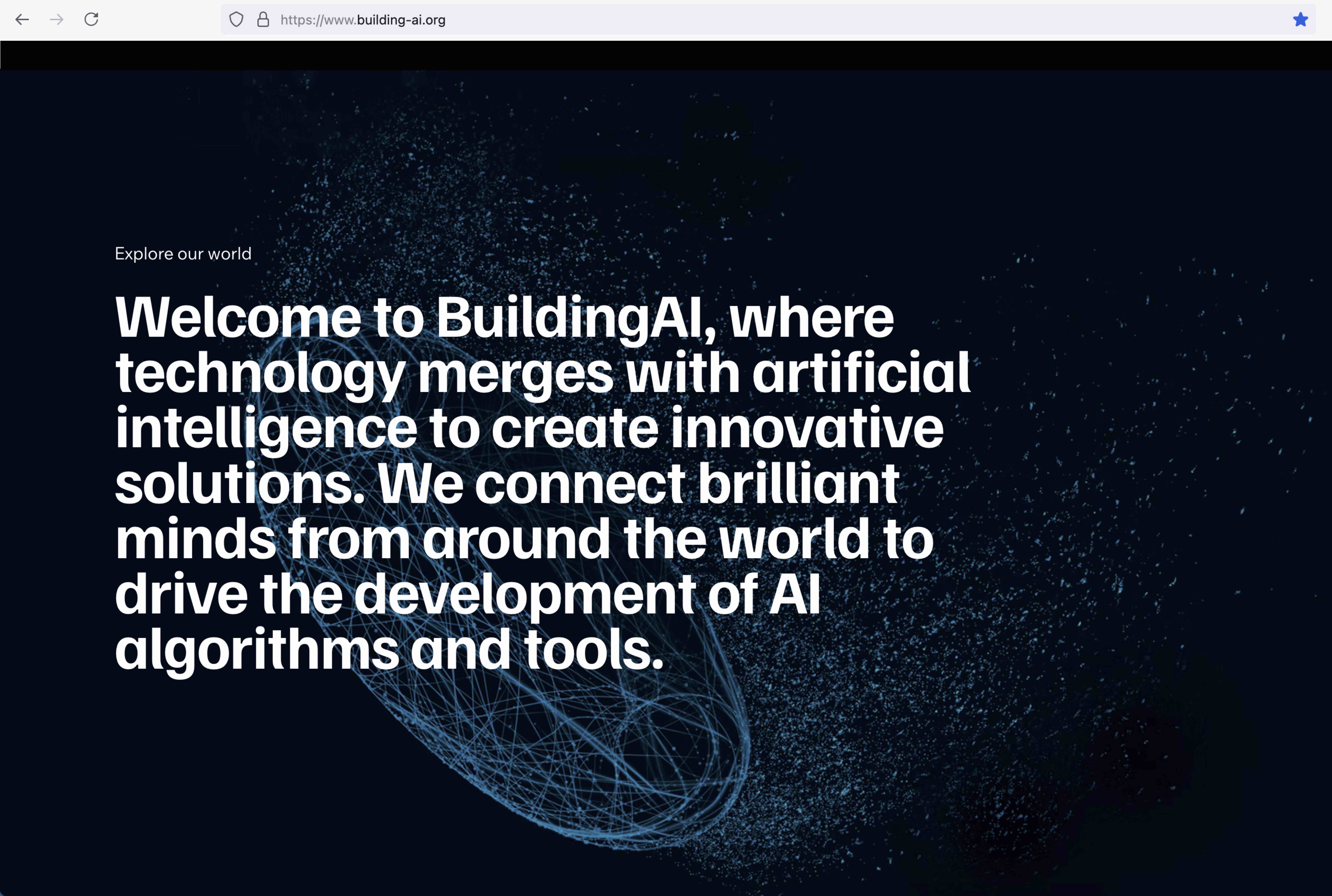}}
\label{fig:Page1}
\end{figure}

\begin{figure}[h!]
\centering
\caption*{Additional Content}
\fbox{\includegraphics[width=5.5in]{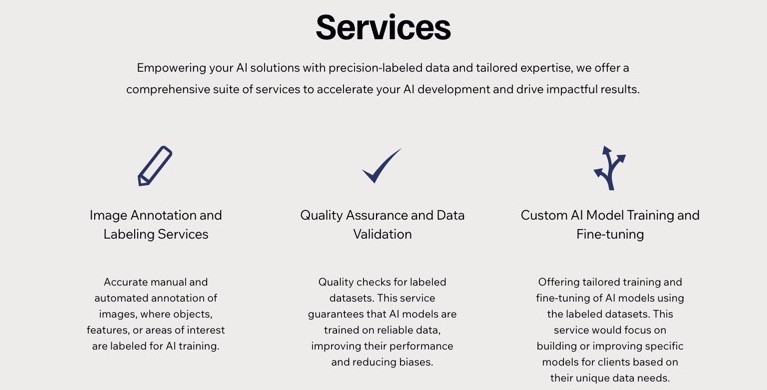}}
\label{fig:Page2}
\end{figure}

\clearpage

\subsection{Task Design}\label{sec:TaskAppendix}
I retrieved 80 images from the dataset made publicly available  by \cite{Steyvers2022}. These images were selected through multiple rounds of piloting to ensure they were diverse in difficulty yet sufficiently challenging.

\begin{figure}[h!]
\centering
\caption{Initial Choice Accuracy.}\label{fig:TaskDifficulty}
\includegraphics[width=4.2in]{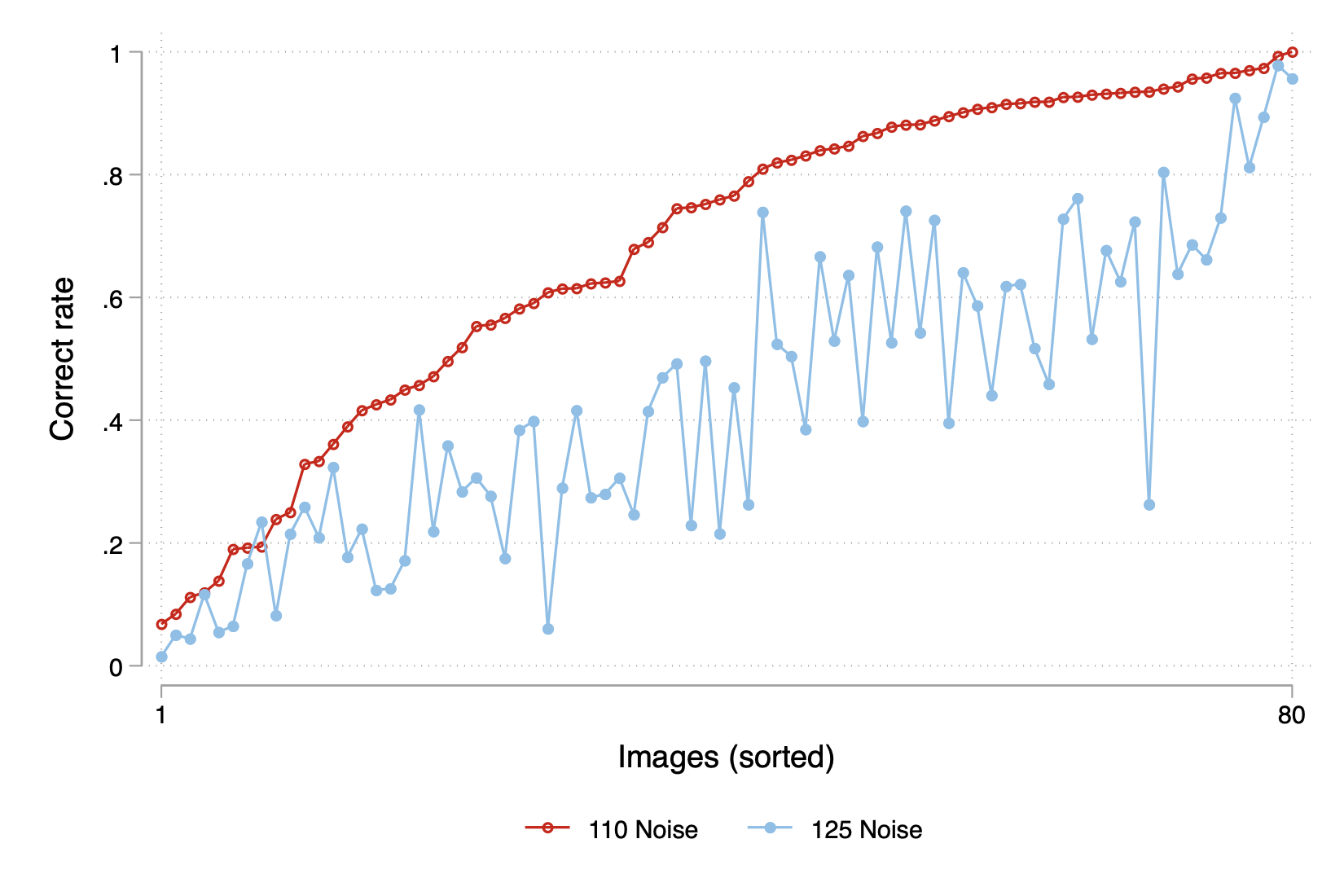}
\vspace{0.5em}
    \begin{minipage}{0.9\textwidth} 
        \footnotesize \textit{Notes}: I compute the accuracy in the first choice as a proxy for image difficulty. Images are then sorted based on this difficulty metric using the 110 noise level (shown in red). The overlaid blue line connects the corresponding accuracy values for each image at the 125 noise level.
    \end{minipage}
\end{figure}

To vary the degree of difficulty in the human and machine classifier experiments, \cite{Steyvers2022} applied four levels of phase noise distortion to the images (80, 95, 110, and 125), which determined the degree of blurriness. This introduced an objective source of variation in image difficulty. In my experiment, I used only the two highest noise levels (110 and 125), with the noise level randomized before an image was shown. Figure \ref{fig:TaskDifficulty} illustrates the variation in task difficulty and highlights a clear within-image increase in difficulty as noise levels rise.

A major advantage of this dataset is that it includes predictions from five pre-trained ImageNet models, along with performance evaluations in collaborative settings: human–machine, two humans, and two machines. This feature enables a non-deceptive experimental design and allows the experiment to draw on algorithms previously tested in studies of human–AI complementarity. In the experiment, I provided participants with predictions from the VGG-19 model, a convolutional neural network with 19 layers. This model was selected because, according to the results in \cite{Steyvers2022}, it exhibited the highest complementarity potential among the available options. However, none of the findings in this paper hinge on the use of this specific model.

\section{Proofs for Section \ref{sec:CF}}\label{app:proofs}

\subsection{Proof of Proposition \ref{prop1}}\label{proof:threshold}
We treat (i) and (ii) in turn. Throughout, consider only tasks with \(y^0\neq \hat y\);
on agreement tasks the choice to stay with their initial choice is straightforward.

\smallskip\noindent\emph{(i) Control scoring \(S_c(a)\).}
Under calibration, if the worker keeps \(y^0\) her expected correctness on that task equals \(p\);
if she follows the AI it equals \(\kappa\). Since \(S_c\) is increasing in accuracy, the task-level optimal action is to pick the option with the higher correctness probability.
Thus, follow the AI iff \(\kappa\ge p\), i.e., use the cutoff \(\tau^\ast=\kappa\). 

\smallskip\noindent\emph{(ii) Treated scoring \(S_t(a,r)\).}
Consider the effect on the overall score of switching from \(y^0\) to \(\hat y\) on a single
disagreement task with private confidence \(p\). This marginal change is
\[
\Delta S_t(p)\;=\;\underbrace{\frac{\partial S_t}{\partial a}}_{>0}\cdot (\kappa-p)
\;+\;\frac{\partial S_t}{\partial r}
\]
The first term is the gain in expected accuracy \((\kappa-p)\) scaled by the marginal score on accuracy; the second term is the marginal score impact of one more unit of AI reliance.
Because \(\partial S_t/\partial a\) and \(\partial S_t/\partial r\) depend only on the
aggregate \((a,r)\), \(\Delta S_t(p)\) is strictly \emph{decreasing} in \(p\).
Therefore, among all measurable selection rules that specify which \(p\)-tasks to switch on,
the score is maximized by switching on the \emph{lowest} \(p\)'s first. Hence the optimal rule is a lower-threshold set
\(\{p\le \tau^\ast\}\). The cutoff is determined by the margin where \(\Delta S_t(\tau^\ast)=0\):
\[
\frac{\partial S_t}{\partial a}\,(\kappa-\tau^\ast)
\;+\;\frac{\partial S_t}{\partial r}\;=\;0
\quad\Longrightarrow\quad
\tau^\ast \;=\; \kappa \;-\; \Big(-\frac{\partial S_t/\partial r}{\partial S_t/\partial a}\Big)
\;=\; \kappa-\lambda,
\]
If \(\partial S_t/\partial r=0\), this reduces to \(\tau^\ast=\kappa\).

\subsection{The Accuracy–Reliance Curve Exhibits an Inverse-U Shape}
\label{subsec:invU}

Under any cutoff \(\tau\in[1/n,1]\), the AI reliance level is:
\[
r(\tau)\;=\;\Pr(p\le \tau)\, D\;=\;F(\tau)\, D,
\]
which is strictly increasing because \(f>0\) on \([1/n,1]\). Hence \(r(\tau)\) is invertible; write its inverse as \(\tau(r)=F^{-1}(r)\).

\paragraph{Accuracy as a function of reliance.}
From the solution reported in Section \ref{sec:CF},
\[
A(\tau)\;=\;\left[\theta+\int_{1/n}^{\kappa-\lambda}(\kappa-p)\,f(p)\,dp\;\right]\, D \,+\, (\theta\kappa)\,(1-D)
\quad\Longrightarrow\quad
\frac{dA(\tau)}{d\tau}=f(\tau)\big(\kappa-\tau\big)\, D.
\]
By the chain rule, the \emph{accuracy--reliance} curve
\(A(r)\equiv A(\tau(r))\) satisfies
\begin{equation*}
\label{eq:dAdr}
\frac{dA(r)}{dr}
=\frac{dA/d\tau}{dr/d\tau}
=\big(\kappa-\tau(r)\big)\, D
=\big(\kappa-F^{-1}(r)\big)\, D,
\end{equation*}
and
\begin{equation*}
\label{eq:d2Adr2}
\frac{d^2A(r)}{dr^2}
=-\,\frac{d\tau(r)}{dr}\, D
=-\,\frac{\, D}{f(\tau(r))}< 0 \quad \text{(since $D>0$ and $f>0$)}.
\end{equation*}

\begin{figure}[h!]
    \centering
\caption{Expected Accuracy as a Function of AI Reliance.}
    \begin{tikzpicture}
    \begin{axis}[
        width=12cm,
        height=8cm,
        xmin=0, xmax=1,
        ymin=0.68, ymax=0.92,
        axis lines=left,
        xlabel={AI reliance},
        ylabel={Expected Accuracy},
        samples=200,
        domain=0:1,
        ticklabel style={/pgf/number format/fixed},
        xtick={0,0.7,1},
        xticklabels={$0$,$F(\kappa) $,$\, 1$},
        ytick={0.7,0.85},
        yticklabels={$\theta$,$\kappa$},
    ]
        \addplot[thick]
            {-0.375*x^2 + 0.525*x + 0.7};

        \addplot[densely dotted] coordinates {(0,0.70) (1,0.70)};
        \addplot[densely dotted] coordinates {(0,0.85) (1,0.85)};

        \addplot[dashed] coordinates {(0.7,0.68) (0.7,0.92)};
        \node[anchor=south] at (axis cs:0.7,0.92)
            {$\tau = \kappa$};

        \addplot[only marks, mark=*] coordinates {(0,0.70) (1,0.85)};
        \node[anchor=east] at (axis cs:0,0.70) {$\theta$};
        \node[anchor=west] at (axis cs:1,0.85) {$\kappa$};
    \end{axis}
    \end{tikzpicture}

    \label{fig:accuracy_threshold}
    
        \vspace{0.5em}
    \begin{minipage}{0.9\textwidth} 
        \footnotesize \textit{Notes}: When $\tau = 0$, workers never use AI and expected accuracy is $\theta$; when $\tau = 1$, workers always use AI and expected accuracy is $\kappa$. For intermediate levels of AI reliance, the system can perform strictly better, with maximal expected accuracy at $\tau = \kappa$, which yields a reliance level of $F(\kappa)$. Without lost of generality, the figure focuses on the case $\theta < \kappa$.
    \end{minipage}
\end{figure}
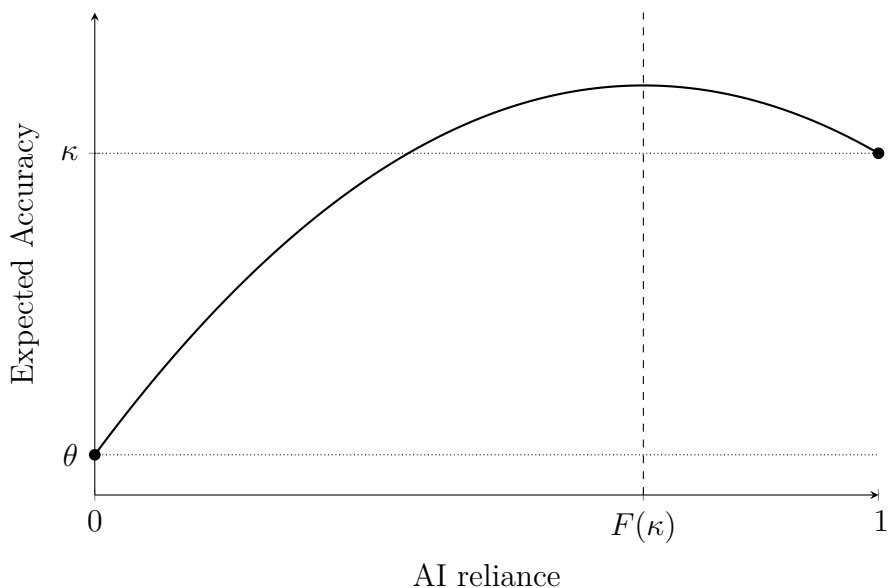

Thus \(A(r)\) is \emph{strictly concave} in \(r\), implying that accuracy initially increases with AI reliance, reaches a maximum at \(\tau=\kappa\), and then decreases. In other words, the accuracy--reliance curve takes an inverse-U shape. Figure \ref{fig:accuracy_threshold} illustrates the inverse-U-shaped relationship between expected accuracy and AI reliance, focusing for simplicity only on situations in which the worker and the AI disagree.

\end{document}